\documentclass[12pt,a4paper]{article}
\pdfoutput=1
\usepackage[T1]{fontenc}
\usepackage[nomath]{lmodern}
\usepackage{jheppub,microtype,amsmath,amsfonts,amssymb,mathrsfs,bbm, physics,graphicx,enumitem,framed,float,booktabs,array,soul,multirow}
\usepackage{tikz}
\usetikzlibrary{calc,arrows,decorations.markings}
\usepackage[prependcaption,textsize=scriptsize]{todonotes}

\newcommand{\pluslash}{\ensuremath\diagup\!\!\!\!\!{+}}
\newcommand{\SL}{\mathrm{SL}}
\newcommand{\Z}{\mathbb{Z}}

\newcolumntype{L}[1]{>{\raggedright\let\newline\\\arraybackslash\hspace{0pt}}m{#1}}
\newcolumntype{C}[1]{>{\centering\let\newline\\\arraybackslash\hspace{0pt}}m{#1}}
\newcolumntype{R}[1]{>{\raggedleft\let\newline\\\arraybackslash\hspace{0pt}}m{#1}}

\def\t{{\tau}}

\title{S-foldings of 5d SCFTs}

\author[a,b]{Hee-Cheol Kim,}
\author[c]{Sung-Soo Kim,}
\author[d]{Kimyeong Lee}

\affiliation[a]{Department of Physics, POSTECH, Pohang 37673, Korea}
\affiliation[b]{Asia Pacific Center for Theoretical Physics, Postech, Pohang 37673, Korea}
\affiliation[c]{School of Physics, University of Electronic Science and Technology of China,\\
Chengdu, Sichuan 611731, China}
\affiliation[d]{School of Physics, Korea Institute for Advanced Study,  
	  Seoul 02455, Korea}

\emailAdd{heecheol1@gmail.com}
\emailAdd{sungsoo.kim@uestc.edu.cn}
\emailAdd{klee@kias.re.kr}

\abstract{We explore the $\mathbb{Z}_{2,3,4,6}$ S-foldings of some 5d superconformal field theories  from the $(p,q)$ 5-brane web perspective. The    S-folding   involves both a spatial quotient and an $\mathrm{SL}(2,\mathbb{Z})$ transformation on 5-branes simultaneously. The  $\mathbb{Z}_{2,3,4,6}$ S-foldings  are achieved by  the insertion of the  $D_4, E_6, E_7, E_8$ 7-branes, respectively. The deficit angles and monodromies of these 7-branes are exactly those necessary for the  S-foldings.  We explore the details of the S-folding process, especially   the enhancement of  global flavor symmetry   in various simple cases. The characteristic of the S-folding depends sharply on  whether the fixed point of the discrete symmetry is at the center of a compact face (or surface), at a 5-brane, or at a crossing point of 5 branes. The analysis of the prepotential greatly supports this view of the discrete gauging.}

\begin{document}
\preprint{KIAS-21060}
\maketitle
\bigskip

%---------- Introduction -------------------
\section{Introduction}

 5d superconformal field theories (SCFTs)  manifest  as a complete UV description of some 5d quantum field theories \cite{Seiberg:1996bd,Morrison:1996xf,Intriligator:1997pq}. The planar  web diagrams of $(p,q)$ 5-branes  in Type IIB string theory  provide a simple approach to the Coulomb  branch physics of these SCFTs \cite{Aharony:1997ju,Brunner:1997gf,Aharony:1997bh,Intriligator:1997pq}. The 5-brane web diagrams  for some 5d superconformal theories   can have  some discrete rotational symmetries on the $(p,q)$-plane if we ignore the 5-brane charge.  For example,  the $E_1$ theory, that is  the $SU(2)$ gauge theory with trivial discrete theta parameter, %
 arises as the intersection of $(1,1)$ and $(1,-1)$ 5-branes. This theory has a $\mathbb{Z}_4$ discrete  symmetry when $(1,1)$ and $(1,-1)$ 5-branes have the same tension. The $E_3$ theory arises as the intersection of $(1,0),(0,1),(1,1)$ 5-branes and is described by the $SU(2)$ Yang-Mills theory with two fundamental hypermultiplets.  When $(1,0),(0,1),(1,1)$ 5-branes take the same tension, the theory has  a $\mathbb{Z}_6$ rotational symmetry. When we put 5-brane charges under consideration, the true discrete symmetry is a rotational  symmetry   accompanied by  a specific monodromy transformation on the brane charges by an $\mathrm{SL}(2, \mathbb{Z})$ transformation.

Recently, the gauging of  such $\mathbb{Z}_3$ rotational symmetry of  5d $T_N$ theories   has been explored in \cite{Acharya:2021jsp}.
This gauging 
can be thought of as an S-folding made of both a spatial $\mathbb{Z}_3$ rotation of the $(p,q)$-plane and a $\mathbb{Z}_3$ subgroup  of the $\mathrm{SL}(2,\mathbb{Z})$ transformation. 
The same $\mathbb{Z}_3$ quotient could be achieved by an insertion of a  specific 7-brane  configuration for the $E_6$ symmetry  which has a deficit angle $4\pi/3$ and a simple monodromy. Another analysis of discrete quotients of 5d SCFTs has been carried out in geometric context recently in \cite{Tian:2021cif}.   

In this work, we  generalize the above S-folding to more general cases. Some  5d SCFTs  can be described by $(p,q)$  5-brane web diagrams with $\mathbb{Z}_{n}$ symmetry generated by  both spatial rotation on the $(p,q)$-plane around a fixed point  and $\SL(2,\Z)$ transformation on 5-brane charges.
The 5-brane web can be cut into $n$ slices of the same angle $2\pi/n$ under the $\mathbb{Z}_n$ symmetry. 
The discrete quotient or S-folding of the $\mathbb{Z}_n$ symmetry keeps only one slice of the $(p,q)$-plane, and then coalesce smoothly the 5-brane webs along two boundaries of the slice by a necessary $\SL(2,\Z)$ transformation $K_{\mathbb{Z}_n}$.  In addition, it requires an insertion of a specific type of 7-brane configuration
at the fixed point. The 7-brane configuration should have   a specific deficit angle bigger than or equal to $\pi$ with a constant axio-dilaton field  and a specific monodromy matrix. There are exactly only four such  7-brane configurations  of type $D_4,E_6,E_7,E_8$ \cite{Sen:1996vd, Dasgupta:1996ij}, which  would correspond  to  $\mathbb{Z}_2,\mathbb{Z}_3,\mathbb{Z}_4,\mathbb{Z}_6 $ quotients, respectively. These 7-brane configurations for the $D_4,E_6,E_7,E_8$ symmetry before we resolve them are precisely   kind of  S-folds of \cite{Garcia-Etxebarria:2015wns,Aharony:2016kai} (see also \cite{Apruzzi:2020pmv,Giacomelli:2020jel,Heckman:2020svr,Giacomelli:2020gee} for 4d  $\mathcal{N}=2$ S-folds), which are generalizations of ordinary orientifolds. In this work, we  study these S-foldings of 5d  SCFTs  and some 6d SCFTs. 
The prepotential analysis for 5d SCFTs before and after the $\mathbb{Z}_n$ quotients would support the above prescription of inserting special 7-brane configurations for the  S-folds.

The massive particle-like objects in the Coulomb phase of 5d SCFTs could be described by the finite $(p,q)$ string webs in the 5-brane webs. They are roughly dyonic instantons in the field theory description~\cite{Lambert:1999ua}. The discrete symmetry we are considering is acting on these particle-like states non-trivially. 
The S-folding keeps only the $\mathbb{Z}_n$ invariant states, which reduces the number of degrees of freedom. On the other hand, the insertion of the $D_4,E_{6},E_{7},E_{8}$ 7-branes introduces additional degrees of freedom accounting for    the twisted sectors  at the $\mathbb{Z}_n$ singularity.  
Accordingly, the S-folding process decreases the flavor symmetry and as well as the dimension of the Coulomb branch in the original brane web.
However, since the additional degrees of freedom at the singularity carry flavor symmetry,  the full flavor symmetry after the S-folding might increase. The details depend on whether the fixed point of the $\mathbb{Z}_n$ action lies in the center of a face or on a 5-brane or at a crossing point of  5-branes.   
To be more specific,  in the SCFT of a local $\mathbb{P}^2$, a state with electric charge $3$ on the Coulomb branch is represented by a Y-shaped string web made of, say, (1,0),(0,1), and (1,1) strings ending on the 5-branes, and is also invariant under $\mathbb{Z}_3$ symmetry. Under the $\mathbb{Z}_3$ S-folding the theory turns to the rank 1 SCFT with $E_6$ flavor symmetry and this state gets transformed into a state carrying electric charge $1$ and the $E_6$ fundamental charge which is represented by, say a (1,0) string, connecting a 5-brane and the $E_6$ 7-branes at the fixed point. For the SCFT of a local $\mathbb{P}^1\times\mathbb{P}^1$ with $\mathbb{Z}_4$ symmetry, the $\mathbb{Z}_4$ S-folding leads to the rank 1 SCFT with $E_7$ flavor symmetry. The W-boson of charge $2$ on the Coulomb branch of the progenitor SCFT is truncated under the S-folding as it is not invariant under the $\mathbb{Z}_4$ symmetry. On the other hand, a state with charge $4$ made of intersecting (1,0) and (0,1) strings is invariant under the $\mathbb{Z}_4$ and gets transformed in the resulting theory into a unit charge state, represented by a fundamental string connecting a 5-brane to $E_7$ 7-branes.  
We elaborate on the details of these S-foldings in many simple examples, which leads to a general perspective.

The 1-loop corrected prepotential of a 5d SCFT in the Coulomb branch provides the exact tensions of magnetic monopole strings \cite{Seiberg:1996bd,Intriligator:1997pq}. A magnetic monopole string is  represented  in a 5-brane web diagram  by a D3 brane  covering a Coulomb branch face with string tension   given by the   area of the face \cite{Aharony:1997ju}. When we mod out $\mathbb{Z}_n$ of the 5-brane web, Coulomb branch faces as well as flavor 5- and 7-branes related by the symmetry are identified. This leads us to propose that the prepotential for the S-folded theory is obtained from the original prepotential  in such a way that one divides  
it by $n$ after taking into account of the identifications of K\"ahler parameters related by the symmetry.

The prepotential computation gives us a useful consistency check for our S-folding prescriptions.  For example, for the  $SU(2)$ gauge theory with $N_f$ hypermultiplets in the fundamental representation (flavors), the cubic prepotential is given by $(8-N_f)\phi^3/3!$ with a Coulomb parameter $\phi$ when all mass parameters vanish \cite{Seiberg:1996bd}. The $\mathbb{Z}_n$ S-folding is possible only when $8-N_f$ is divisible by $n$. The prepotentials divided by $n$ are exactly what one would get by S-foldings with an insertion of 7-branes which increases the flavor symmetry in this case.

We check the consistency of our  prescription for the discrete quotient or S-folding in numerous cases, by comparing the results from the brane web approach and those from the prepotential analysis. Interestingly, we find that the prediction for $\mathbb{Z}_3$ folding of the $T_4$ theory in \cite{Acharya:2021jsp} does not match our proposal.

The plan of this work is as follows: In section \ref{sec:2}, we first review 5d $\mathcal{N}=1$ superconformal field theories. Especially we review  the prepotential on the Coulomb branch,   geometric descriptions and 5-brane webs for the 5d SCFTs, and then  the role of  7-branes and the appearance of  discrete symmetries. In section \ref{sec:3}, we study the S-folding in the cases where the fixed point is on a face of the 5-brane webs. We first provide a general S-folding prescription in terms of the 5-brane web diagrams and insertion of the relevant 7-branes.   Then we carry out a detailed study of rank-1   and some  higher rank examples. In section \ref{sec:4}, we study the S-folding when the fixed point is on a line or at a vertex of the 5-brane web. We provide a prescription and check their consistency with a few examples. In section \ref{sec:5}, we study various examples of the discrete quotient of the 6d superconformal field theories. In section \ref{sec:6}, we conclude with some remarks.

%----------------------------------
\section{On 5d SCFTs}\label{sec:2}
In this section, we first review 5d $\mathcal{N}=1$ supersymmetric field theories from various perspectives in relation to   prepotential,     geometry and 5-brane webs.

\subsection{Prepotential on the Coulomb branch}
$\mathcal{N}=1$ superconformal theories in five dimensions are supersymmetric theories of eight supercharges with an $SU(2)_R$ symmetry \cite{Seiberg:1996bd,Intriligator:1997pq}. A large class of these theories admits gauge theory description at low energies. Given a gauge group $G$, supersymmetric multiplets of the $\mathcal{N}=1$ gauge theory are made of the vector multiplet and charged matter hypermultiplets. The vector multiplet consists of a vector field $A_\mu$ and a real scalar field $\phi$. The hypermultiplets consist of complex scalar fields forming an $SU(2)_R$ doublet which transform in a representation $\bf R$ of $G$. Fermionic fields are also presented as superpartners in each multiplet. On the Coulomb branch of the moduli space, %which is our interest, 
the scalar field $\phi$ in the vector multiplet gets expectation values in the Cartan of the gauge group $G$, which completely breaks $G$ to $U(1)^{\text{rank}(G)}$. The Coulomb branch is hence parameterized by the scalar expectation values $\phi_i$ (or dynamical K\"ahler parameters in the associated Calabi-Yau geometry) where $i=1,\cdots, \text{rank}(G)$. The corresponding low energy theory is then described by an effective theory of the Abelian gauge groups.

This effective Abelian theory is characterized by a cubic prepotential $\mathcal{F}(\phi)$ which is a 1-loop exact cubic polynomial of K\"aher parameters taking the form of~\cite{Seiberg:1996bd, Intriligator:1997pq}
\begin{align}\label{eq:prepotential}
    \mathcal{F}(\phi)= \frac{m_0}2 h_{ij}\phi_i\phi_j+\frac{\kappa}{6}d_{ijk}\phi_i\phi_j\phi_k+\!\frac1{12}\bigg( \!
    \sum_{e\in\bf R}|e\cdot \phi|^3-\sum_f\!\sum_{w\in {\bf w}_f}|w\cdot\phi+m_f|^3\!
    \bigg)\!,
\end{align}
where $m_0=1/{g_0^2}$ is the inverse of the gauge coupling $g_0$ squared,  $h_{ij}={\rm Tr}(T^G_iT^G_j)$ is the Killing form of $G$ with the generators $T^G_i$ in the fundamental representation of $G$, and 
$\kappa$ is the classical Chern-Simons (CS) level and $d_{ijk}=\frac{1}{2}{\rm Tr}(T^G_i\{T^G_j,T^G_k\})$ is non-zero only for $G=SU(N)$ with $N\ge 3$. The remaining two terms with the parentheses are the one-loop contributions where $\mathbf{R}$ are the roots of the Lie algebra of $\mathfrak{g}$ associated with $G$ and $\mathbf{w}_f$ are the weights of the $f$-th hypermultiplet with mass $m_f$. The Coulomb branch is divided into sub-chambers distinguished by the signs of masses appearing in the prepotential~\eqref{eq:prepotential}. For example, the prepotential for the 5d   $SU(3)$ gauge theory without flavor at the CS level 0 is given by
    \begin{align}\label{eq:SU3_0}
  \mathcal{F}_{SU(3)_0}=m_0\big(\phi_1^2-\phi_1\phi_2+\phi_2^2\big)+\frac16\big( 8\phi_1^3-3\phi_1^2\phi_2-3\phi_1\phi_2^2+8\phi_2^3 \big)\ ,
\end{align}
in a Weyl chamber where $2\phi_1>\phi_2>\phi_1/2$.

We note that the multiple derivatives of prepotentials with respect to $\phi$ give rise to some quantum observables. The first derivative yields the magnetic monopole string tension $T_i=\partial_i \mathcal{F}$ and the second derivative determines the gauge kinetic terms of the effective action, yielding the effective gauge coupling
\begin{align}
    \tau_{ij}=(1/g_{\rm eff}^2)_{ij}=\partial_i\partial_j \mathcal{F}\ , 
\end{align}
which is the metric on the Coulomb branch $ds^2= \tau_{ij}d\phi^id\phi^j$. The third derivatives lead to  cubic Chern-Simons couplings $C_{ijk}=\partial_i\partial_j\partial_k\mathcal{F}$ which are quantized as  $C_{ijk}\in \mathbb{Z}$ due to gauge invariance.

%At the infinite coupling limit, many theories flow to UV fixed points where  superconformal theories (SCFTs) arise \cite{Seiberg:1996bd, Intriligator:1997pq}. 
Different gauge theories can arise from different relevant deformations of the same SCFT~\cite{Seiberg:1996bd, Intriligator:1997pq}, which is often referred to as ``UV duality,'' \cite{Gaiotto:2015una}
or equivalently, it can be stated that an SCFT may have several different relevant deformations which yield different gauge theory descriptions at low energies. At the fixed point, global symmetry of the theory can be enhanced to a bigger symmetry  $G_F \supset G_{\text{hyper}}\times U(1)_I$, where $G_{\text{hyper}}$ is the flavor symmetry from hypermultiplets and $U(1)_I$ is a topological symmetry from instantonic conserved currents. One of interesting examples is the 5d $SU(2)$ gauge theory with $N_f$ hypermultiplets in the fundamental representation, which has an enhanced symmetry $E_{N_f+1}\supset SO(2N_f)\times U(1)_I$ in the UV fixed point, where $G_{\text{hyper}}=SO(2N_f)$ comes from $N_f\le 7$ flavors~\cite{Seiberg:1996bd}.

Another interesting aspect is that there exists a marginal number of hypermultiplets or a marginal CS level where the theory with this marginal hypermultiplets/CS level admits a UV completion as a 6d theory rather than a 5d theory. Such a theory is called a Kaluza-Klein (KK) theory which is in fact a 6d theory put on a circle with or without a twist. For instance, the 5d $SU(2)$ gauge theory with 8 fundamental hypers is a KK theory arising from the 6d rank-1 $E$-string theory on a circle without any twist \cite{Witten:1995gx,Ganor:1996mu,Seiberg:1996vs,Kim:2014dza}. The 5d  $SU(3)$ gauge theory without flavor at the Chern-Simons level 9, which we denote by $SU(3)_9$, is another example of KK theory arising from the 6d minimal $SU(3)$ SCFT on a circle with a $\mathbb{Z}_2$ twist \cite{Jefferson:2017ahm,Jefferson:2018irk}.

So far, we have discussed 5d or 6d SCFTs described by gauge theories at low energy after mass deformations. There are, in fact, many 5d theories without any Lagrangian descriptions. The $E_0$-theory \cite{Morrison:1996xf} would be a well-known non-Lagrangian theory, which can arise by decoupling instantonic hypermultiplet from the 5d $SU(2)$ theory at discrete theta angle $\theta=\pi$~\cite{Witten:1982fp}. 

The 5d $\mathcal{N}=1$ SCFTs are closely related to M-theory compactification on local Calabi-Yau 3-folds and 5-brane webs of Type IIB string theory. We shall discuss this in the following subsections in more detail.

\subsection{Geometry and 5-brane web }
Now we briefly introduce geometric engineering in M-theory and also Type IIB 5-brane constructions of 5d SCFTs.

\paragraph{Geometric constructions.}

Let us first review some basic features of the geometric construction of 5d $\mathcal{N}=1$ SCFTs. We consider M-theory compactified on a smooth non-compact Calabi-Yau threefold $X_6$. In a singular limit where all the K\"ahler parameters are turned off, this compactification gives rise to a 5d SCFT \cite{Witten:1996qb}.

The Calabi-Yau 3-fold $X_6$ associated to a 5d SCFT has a description as a local neighborhood of a collection of K\"ahler surfaces $\cup_{i=1}^r S_i \in X_6$ glued each other. Here, $r$ is the rank of the Calabi-Yau 3-fold corresponding to the dimension of the Coulomb branch in the 5d field theory. The properties of the K\"ahler surfaces and their gluing rules have been extensively studied in \cite{Jefferson:2018irk,Bhardwaj:2019fzv}. See also \cite{Xie:2017pfl,Bhardwaj:2018yhy,Bhardwaj:2018vuu,Apruzzi:2018nre,Apruzzi:2019vpe,Apruzzi:2019enx,Bhardwaj:2019jtr,Apruzzi:2019opn,Apruzzi:2019kgb,Bhardwaj:2020gyu,Bhardwaj:2020kim}.

The M2- and M5-branes in M-theory can wrap around compact holomorphic 2- and 4-cycles respectively in the 3-fold. An M2-brane wrapping a holomorphic 2-cycle provides a charged BPS particle state in the 5d field theory and the mass of the BPS particle is proportional to the volume of the 2-cycle. On the other hand, an M5-brane wrapping a compact surface (or 4-cycle) leads to a BPS monopole string state, the tension of which is proportional to the volume of the surface.

The volumes of the compact $p$-cycles embedded in $X_6$ are measured by their intersections with the K\"ahler class $J$ defined as
\begin{align}
  J = \sum_{i} \phi_i S_i + \sum_{f}m_f N_f \ ,
\end{align}
where $\phi_i$'s denote the dynamical K\"ahler parameters and $m_f$ are mass parameters (or non-dynamical K\"ahler parameters) associated to the non-compact K\"ahler surfaces $N_f$.
One can compute the total volume of the 3-fold as
\begin{align}
  {\rm vol}(X_6) = \frac{1}{3!}\int_{X_6} J^3 \ .
\end{align}
Upon the M-theory compactification on $X_6$, this volume of the 3-fold is mapped to the prepotential $\mathcal{F}$ on the Coulomb branch of the 5d field theory, i.e. $\mathcal{F}={\rm vol}(X_6)$. This implies that the cubic Chern-Simons coefficients in the low energy 5d theory are determined by the triple intersection numbers of surfaces in the 3-fold
\begin{align}
  C_{IJK} = S_I \cdot S_J \cdot S_K \ ,
\end{align}
where the index $I$ here labels both the compact surfaces $S_i$ and the non-compact surfaces $N_f$.
The volume of a curve $C$ is computed by
\begin{align}
  {\rm vol}(C) = -J\cdot C \ ,
\end{align}
which sets the mass of the BPS state arising from an M2-brane wrapping the curve $C$. Similarly, the volume of a surface $S_i$ is
\begin{align}
  {\rm vol}(S_i) = J\cdot J\cdot S_i=\frac{1}{2} \int_{X_6}J^2 \wedge S_i \ ,
\end{align}
which can equally be obtained by the first derivative of the total volume of the 3-fold with respect to $\phi_i$, i.e. $\partial_i {\rm vol}(X_6)=\partial_i \mathcal{F}$. This will determine the tension of the monopole string arising from an M5-brane wrapping the surface $S_i$.

We are interested in Calabi-Yau 3-folds whose local singularities have dual realizations in Type IIB $(p,q)$ 5-brane webs. The duality maps the compact holomorphic 2-cycles and 4-cycles in the 3-folds to $(p,q)$ 5-branes and compact faces, respectively, in the 5-brane webs.
Such geometries have a natural $T^2$ fibration over K\"ahler surfaces and the modular group of this $T^2$ is identified with the $\mathrm{SL}(2,\mathbb{Z})$ transformation in Type IIB theory. 
In this case, the 3-fold can have, at special points of the K\"ahler moduli space, discrete symmetries acting on the K\"ahler surfaces accompanied by $\mathrm{SL}(2,\mathbb{Z})$ transformations.

For example, the Calabi-Yau threefold with a local $\mathbb{P}^2$ has a toric realization in terms of $(p,q)$ 5-branes. M-theory compactified on this 3-fold in the decoupling limit gives rise to the simplest rank-1 5d SCFT without flavor symmetry. The prepotential on the Coulomb branch of the SCFT is determined by the volume of the Calabi-Yau 3-fold as
\begin{align}
  6\cdot {\rm vol}\left(X_6(\mathbb{P}^2)\right)=6\cdot\mathcal{F}_{\mathbb{P}^2} = 9\phi^3 \ ,
\end{align}
where $\phi$ is the K\"ahler parameter for the local $\mathbb{P}^2$.
This geometry has a discrete $\mathbb{Z}_3$ symmetry which permutes three fixed points in the $\mathbb{P}^2$ under the $T^2$ action. In addition, the combination of this $\mathbb{Z}_3$ action on $\mathbb{P}^2$ with a $\mathbb{Z}_3$ subgroup of the $\mathrm{SL}(2,\mathbb{Z})$ action also becomes another discrete symmetry of the geometry. We will discuss how to gauge the later $\mathbb{Z}_3$ symmetry in the next section.

Another simple but interesting example is the Calabi-Yau threefold corresponding to the 5d $SU(3)$ gauge theory at the Chern-Simons level $0$. The 3-fold is now a rank-2 geometry engineered by gluing two Hirzebruch surfaces like $\mathbb{F}_1\cup \mathbb{F}_1$ by identifying two curves $e_1$ and $e_2$ with $e_1^2=e_2^2=-1$ where the subscript $1,2$ denotes respectively the first and the second $\mathbb{F}_1$ surface. The volume of this 3-fold is 
\begin{align}\label{eq:F1F1}
  6\cdot{\rm vol}\left(X_6(\mathbb{F}_1\cup \mathbb{F}_1)\right) = 8\phi_1^3-3\phi_1^2\phi_2-3\phi_1\phi_2^2+8\phi_2^3 \ ,
\end{align}
where $\phi_{1,2}$ are the K\"ahler parameters for two $\mathbb{F}_1$'s. Here we switched off the non-dynamical parameter (or the gauge coupling of the $SU(3)$ gauge group). The volume of the 3-fold agrees with \eqref{eq:SU3_0} when $m_0=0$. 
This geometry has a $\mathbb{Z}_2$ symmetry exchanging two Hirzebruch surfaces, thus exchanging two K\"ahler parameters as $\phi_1\leftrightarrow \phi_2$, combined with a $\mathbb{Z}_2$ action in $\mathrm{SL}(2,\mathbb{Z})$. We remark that this symmetry is unbroken even after turning on the gauge coupling of the $SU(3)$ gauge group. S-fold of this $\mathbb{Z}_2$ symmetry will also be discussed later.

There are other approaches to calculate the triple intersection numbers of Calabi-Yau 3-folds. For example, the gauged linear sigma model (GLSM) approach to the toric Calabi-Yau 3-fold has been extensively studied to calculate the prepotential in  \cite{Closset:2018bjz}. Their method can be employed to reproduce the prepotential obtained from the field theory in simple examples.

The discrete symmetries of 3-folds we discussed above are more evident in the associated 5-brane webs. So let us now move on to the $(p,q)$ 5-brane webs in Type IIB string theory.

\paragraph{Type IIB 5-brane webs.} A large class of 5d $\mathcal{N}=1$ theories can be constructed from 5-branes of Type IIB string theory. Here, we discuss salient features of the 5-brane construction of 5d SCFTs \cite{Aharony:1997bh,Aharony:1997ju,Leung:1997tw,Benini:2009gi}. 
To begin with, recall the complexified coupling of the IIB string theory $\tau$ is given by
\begin{align}
    \tau= \chi+ i/g_s, 
\end{align}
where $\chi$ is the axion and $g_s$ is the string coupling. The tension of a 5-brane with charge $(p, q)$ is then expressed as 
\begin{align}
\label{eq:bpstension}
    T_{(p,q)} = |p-\tau q|T_{\rm D5}, 
\end{align}
where $T_{\rm D5}\sim 1/g_s $ is the D5-brane tension. Note that the tension  is invariant  under the $\mathrm{SL}(2,\mathbb{Z})$ transformation 
$({{}_a\ {}_b\atop {}^c\ {}^d})\in \mathrm{SL}(2,\mathbb{Z})$ 
of the axio-dilaton field $\tau$ and the 5-brane charge $[p,q]$,
\begin{align}
    \tau\rightarrow \frac{a\tau+b}{c\tau+d}\ , \qquad  \left({p\atop q}\right) \rightarrow \left({a \ b \atop c\ d}\right) \left({p\atop q}\right) \ .
\end{align} 
We take the convention of $(p,q)$-brane charge as follows: $(1,0)$ denotes the charge of a D5-brane, while $(0,1)$ denotes that of an NS5-brane. 5d theories are realized as a charge conserving configuration of $(p,q)$ 5-branes on a two-dimensional plane called a $(p,q)$-plane where D5- and NS5-branes are extended. On the $(p,q)$-plane, a $(p_1,q_1)$ 5-brane is extended along the line with the slope, $\tan^{-1}(q_1/p_1)$.   More precisely, 5- and 7-brane configurations are listed in Table \ref{tab:TypeIIB brane cofniguration}.
\begin{table}[H]
    \centering
    \begin{tabular}{c|ccccc|cc|ccc}
    &0&1&2&3&4&5&6&7&8&9\\
    \hline
    NS5 $(0,1)$& $\times$&$\times$&$\times$&$\times$&$\times$&$\times$ &&&&\\
    D5 $(1,0)$& $\times$&$\times$&$\times$&$\times$&$\times$& &$\times$&&&\\
    $(p,q)$& $\times$&$\times$&$\times$&$\times$&$\times$&$\vartheta $&$\vartheta$&&&\\
    7-brane& $\times$&$\times$&$\times$&$\times$&$\times$&&&$\times$&$\times$&$\times$\\
\end{tabular}
\caption{Type IIB 5-brane worldvolume configuration. The $(x^6,x^5)$-plane is referred to as the $(p,q)$-plane. The directions that each brane is extended are marked with $\times$ or $\vartheta$. } 
\label{tab:TypeIIB brane cofniguration}
\end{table}
\begin{figure}[t]
     \centering
     \includegraphics[scale=.19]{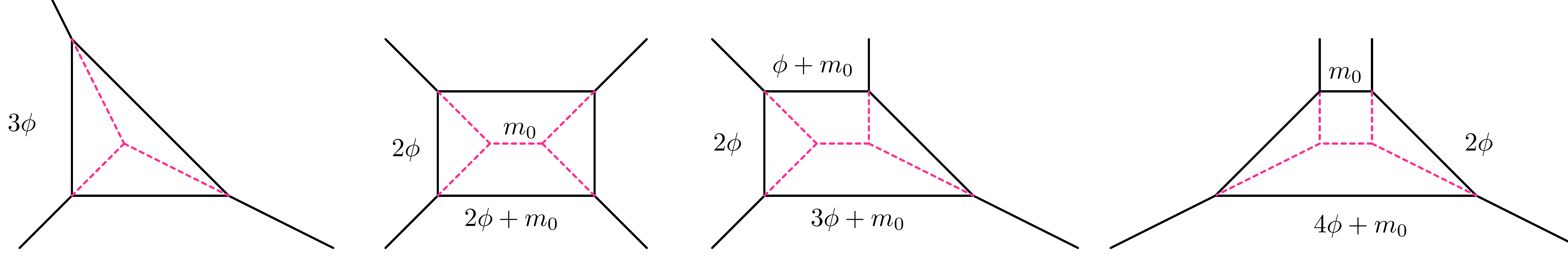}
     \caption{5-brane webs for $E_0$-theory, $SU(2)_{0}$ ,  $SU(2)_{\pi}$, and $SU(2)_{2\pi}$ theories. In geometry, they correspond to local $\mathbb{P}^2$, $\mathbb{F}_0$, $\mathbb{F}_1$, and $\mathbb{F}_2$ respectively.}
     \label{fig:web-SU2-localP2}
 \end{figure}

Such charge conserving 5-brane configurations form a web diagram of 5-branes on the $(p,q)$-plane, which we refer to as a 5-brane web. The $\tau$ is defined up to an $\mathrm{SL}(2, \mathbb{Z})$ action and does not affect the low energy theory. %Typically 5-brane web is drawn with $\tau=i$. 
With the choice $\tau=i$, a few simple 5-brane webs are given in Figure  \ref{fig:web-SU2-localP2}. 
The first 5-brane is for the simplest non-Lagrangian theory in five dimensions called the $E_0$-theory, which is associated with a local $\mathbb{P}^2$. The other webs are for the 5d   $SU(2)_\theta$ gauge theory with no flavor and the discrete $\theta$ parameter $0$, $\pi$, and $2\pi$, respectively. Note that as the discrete $\theta$ allows only two parameters, $SU(2)_{2\pi}$ is equivalent to $SU(2)_{0}$ up to decoupled states.  Enclosed faces (compact surfaces in geometry) correspond to the Coulomb branch of the theory and hence the number of enclosed faces is the same as the dimension of the Coulomb branch of the theory. As a 5-brane web is drawn in a Coulomb phase, one can associate the edges of 5-brane webs with physical parameters like the bare couplings or W-boson masses as shown in Figure \ref{fig:web-SU2-localP2}. 
It follows that the areas of enclosed faces correspond to the monopole string tensions. It is hence straightforward to extract the prepotential $\mathcal{F}$ from a 5-brane web for a given theory, by computing the monopole string tensions and integrating them with respect to the K\"ahler parameters. For instance, the tension of monopole string for the $E_0$ theory and the $SU(2)_{0,\pi}$ theories can be computed from the 5-brane webs given in Figure \ref{fig:web-SU2-localP2}: $    T_{E_0}= \frac{9}{2}\phi^2$ and $T_{SU(2)_{0,\pi}} = 2\phi(2\phi+m_0)$, 
and one easily gets the prepotential by integrating them with respect to $\phi$.

\begin{figure}[t]
     \centering
     \includegraphics[scale=.2]{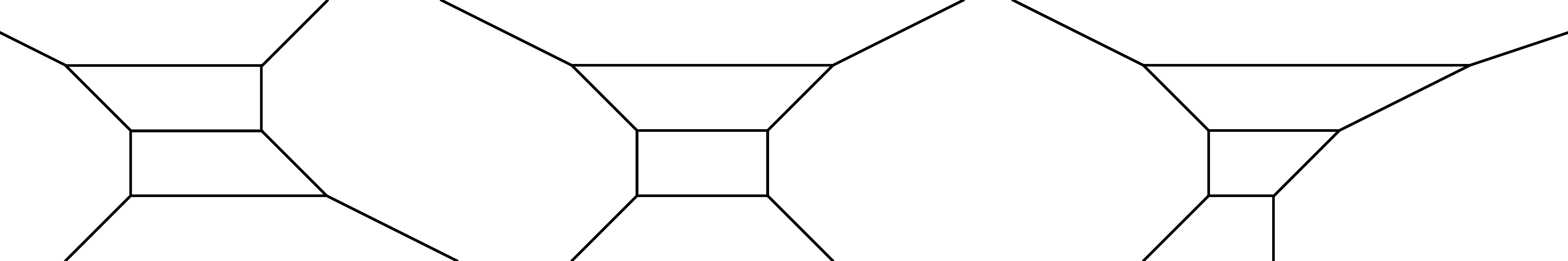}
     \caption{5-brane webs for the $SU(3)_{\kappa=0,1,2}$ theories without matter.}
     \label{fig:web-SU3}
 \end{figure}

In this way, a 5-brane web for the $SU(N)_\kappa$ gauge theory at the Chern-Simons level $\kappa$ can be readily generated, as $SU(N)$ theories with different $\kappa$ have different monopole string tensions. For instance, 5-brane webs for the $SU(3)_\kappa$ theories without matter  are listed in Figure~\ref{fig:web-SU3}. One can easily see that the prepotential for the $SU(3)_0$ theory from the 5-brane web agrees with that from the associated geometry \eqref{eq:F1F1} in the infinite coupling limit $m_0\to0$. 
Hypermultiplets in the fundamental representation (flavors) are introduced by external D5-branes or D7-branes. We note here that though there is a large class of theories described by 5-brane webs, there are still some theories whose 5-brane configurations are not known, which are theories of higher Chern-Simons level or hypermultiplets in some higher dimensional representations. See for such examples \cite{Bergman:2013aca,Bergman:2014kza, Hayashi:2018bkd,Hayashi:2018lyv, Hayashi:2019yxj,Bergman:2020myx, Kim:2020hhh}.

Brane web realizations for gauge groups other than $SU(N)$ are also possible with the help of orientifold 5- or 7-branes. An O7-plane produces a point-wise reflected image around its location and, on the other hand, an O5-plane produces a plane reflected image on the $(p,q)$-plane. This implies that the $SO(N)$ and $Sp(N)$ gauge theories can be realized by introducing respectively an O7$^+$- and O7$^-$-plane, or instead,  an O5-plane. 
It is worthy of mentioning that even a 5-brane web for the theories of exceptional gauge group $G_2$ are recently obtained through the Higgsing of the $SO(7)$ gauge theory with a spinor~\cite{Zafrir:2015ftn,Kim:2018gjo,Hayashi:2018bkd}.

%-----------------
\subsection{7-branes and discrete symmetries}

5-brane webs are often depicted with 7-branes which are attached to the ends of external 5-branes of the same charge. Through the Hanany-Witten moves, one sees various aspects of the theories from 5-brane webs. For instance, the equivalence between the $SU(2)_0$ gauge theory and the $SU(2)_{2\pi}$ gauge theory can be shown via Hanany-Witten moves followed by an $\mathrm{SL}(2,\mathbb{Z})$ transformation in Figure \ref{fig:web-SU2-localP2}. Flavor symmetry would be another good example that shows the usefulness of 7-branes. To analyze the global symmetry of a theory from the perspective of 7-brane configuration on a 5-brane web, let us first set up some notations for 7-branes. 
For frequently used 7-branes, we name them as follows: 
\begin{align}
    \mathbf{A}=[1,0],\qquad \mathbf{B}=[1,-1],\qquad  \mathbf{C}=[1,1],\qquad 
    \mathbf{N}=[0,1]\ ,
\end{align}
up to the overall sign. For other 7-branes, we denote them by their charge $\mathbf{X}_{[p,q]}$.

7-branes create monodromy cuts on the $(p,q)$-plane. When a 7-brane with charge $[p,q]$ crosses the cut of another 7-brane of different 7-brane charge, the charge $[p,q]$ of the 7-brane across the cut  change. How the charge is modified can be read off from the monodromy matrix. For a $[p,q]$ 7-brane, its monodromy matrix is
 \begin{equation}
 K_{[p,q]}= \mathbbm{1}_{2\times 2}+ (p,q)^t(p,q) S = \begin{pmatrix} 
 1+pq &- p^2 \\
  q^2  &  1-pq
 \end{pmatrix},
 \label{eq:monodromy matrix}
 \end{equation}
where $\mathbbm{1}_{2\times 2}$ is the $2\times 2$ identity matrix and $S=\big(\substack{\;0 \ -1 \\  1 \ \  0}\,\big) $. Under an $\mathrm{SL}(2,\mathbb{Z})$ transformation $g$ acting on a charge vector $(p,q)^t$, the monodromy matrix transforms as $gK_{[p,q]}g^{-1}$. The monodromies for $\mathbf{A}, \mathbf{B}, \mathbf{C}$ 7-branes are  $K_{\bf A}=T^{-1}, K_{\bf B}=ST^2, K_{\bf C}= T^2S$ respectively with 
 $T=\big(\substack{\;1 \ \ 1 \\ \, \,0 \ \ 1}\,\big) $.  

Here, $T$ and $S$ are two generators of $\mathrm{SL}(2,\mathbb{Z})$ whose presentation is given by $S^2=-\mathbbm{1}_2 $ and $(ST)^3=-\mathbbm{1}_2=(TS)^3$. An $[r,s]$ 7-brane crossing counterclockwise the cut of 7-brane of charge $[p,q]$ becomes a 7-brane of charge $[r_K,s_K]=K_{[p,q]}\cdot(r,s)^t$.
An $[r,s]$ 7-brane crossing clockwise the cut of 7-brane of charge $[p,q]$ becomes a 7-brane of charge
$[r_M,s_M]=M_{[p,q]}\cdot(r,s)^t$ where $M_{[p,q]}=K_{[p,q]}^{-1}$.
We write such a 7-brane shuffling by listing 7-branes from the left to right counterclockwise, $\mathbf{X}_{[r,s]}\mathbf{X}_{[p,q]} =\mathbf{X}_{[p,q]}\mathbf{X}_{[r_K,s_K]}$, or $\mathbf{X}_{[p,q]}\mathbf{X}_{[r,s]} =\mathbf{X}_{[r_M,s_M]}\mathbf{X}_{[p,q]}$. For example, $\bf AB=NA$ means a 7-brane $\mathbf{B}$ going across the cut of $\mathbf{A}$ clockwise becomes a 7-brane $\mathbf{N}$. 
We also have $\bf AB = BN$, which means $\mathbf{A}$ goes across the cut of $\mathbf B$ counterclockwise and becomes $\mathbf{N}$. Together, one finds $\bf AB=NA=BN$. The followings are some of useful relations of 7-branes:
 \begin{align}
     \bf CA=AN=NC, \qquad 
     \bf ABC=BCA,  \qquad BC = X_{[3,-1]}B\ .
    \label{eq:useful 7-branes} 
 \end{align}

\begin{figure}[t]
     \centering
     \includegraphics[scale=.2]{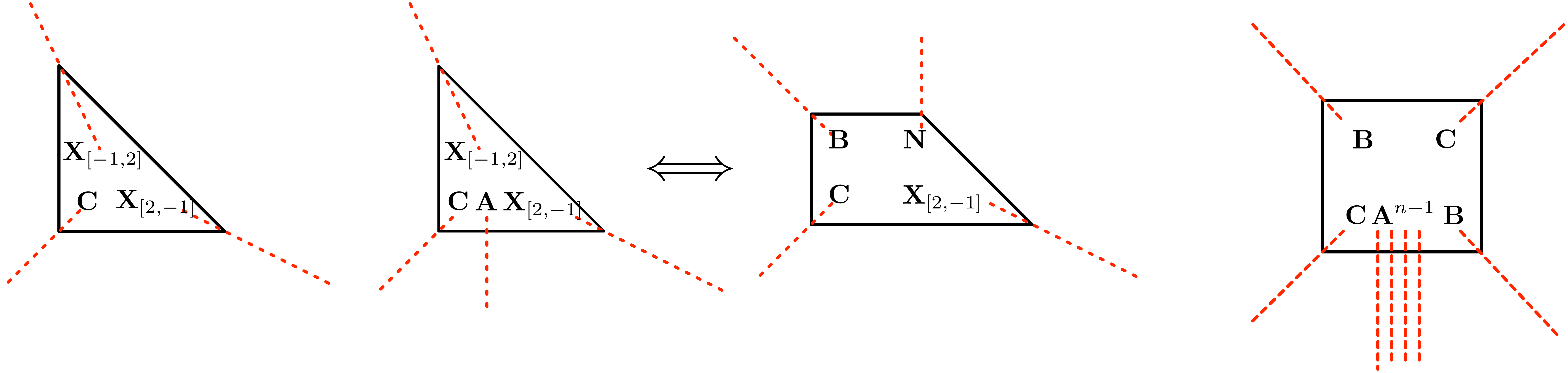}
     \caption{7-brane configurations for 5d rank-1 theories. The first web is the $E_0$ theory of $\widehat{\mathbf{E}}_0$ 7-brane configuration. The second and the third ones are equivalent and describe the $SU(2)_\pi$ theory of $\widehat{\widetilde{\mathbf{E}}}_1$. The last one is generic 7-brane configuration of $SU(2)+(n-1)\mathbf{F}$ of  $\widehat{\mathbf{E}}_n$ associated with $E_n$ global symmetry.}
     \label{fig:rank-1 webs}
 \end{figure}

The enhanced global symmetry of a 5d SCFT from the perspective of F-theory background with $ADE$-type singularities was extensively studied in 
\cite{Sen:1996vd,Dasgupta:1996ij,Gaberdiel:1997ud,Gaberdiel:1998mv,DeWolfe:1998eu, DeWolfe:1999hj}. To read off the flavor symmetry of a 5d SCFT, one puts all the 7-branes inside compact faces without any 5-brane attached, and then $ADE$ global symmetries of a 5d SCFT are given as 7-brane configuration of $ADE$-type singularity.  For $SU(N)$ symmetry, the corresponding 7-brane configuration contains $\mathbf{A}^{N}$, and for $SO(2N)$ symmetry, it contains $\mathbf{A}^{N}\mathbf{B}\mathbf{C}$. In particular, $E_n$ symmetry has various configurations, $\widehat{\mathbf{E}}_n$,  given as follows \cite{Gaberdiel:1997ud,Gaberdiel:1998mv,DeWolfe:1998eu,DeWolfe:1999hj}:
\begin{align}
  & \widehat{\bf E}_0={\bf X}_{[2,-1]}{\bf X}_{[-1,2]}{\bf C} \nonumber \\
  & \widehat{\widetilde{\bf E}}_1 = {\bf A} {\bf X}_{[2,-1]}{\bf X}_{[-1,2]}{\bf C} \nonumber \\
  & \widehat{\bf E}_n 
    = {\bf A}^{n-1}{\bf BC}{\bf BC} \qquad (n\ge1)\ , 
    \label{eq:En-7branes}
 \end{align}
 where $E_1=SU(2)$, $E_2=SU(2)\times U(1)$, $E_3=SU(3)\times SU(2)$, $E_4=SU(5)$, and  $E_5=SO(10)$. Generic 7-brane configurations for the rank-1 theories are depicted in Figure \ref{fig:rank-1 webs}. 

\begin{figure}[t]
     \centering
     \includegraphics[scale=.2]{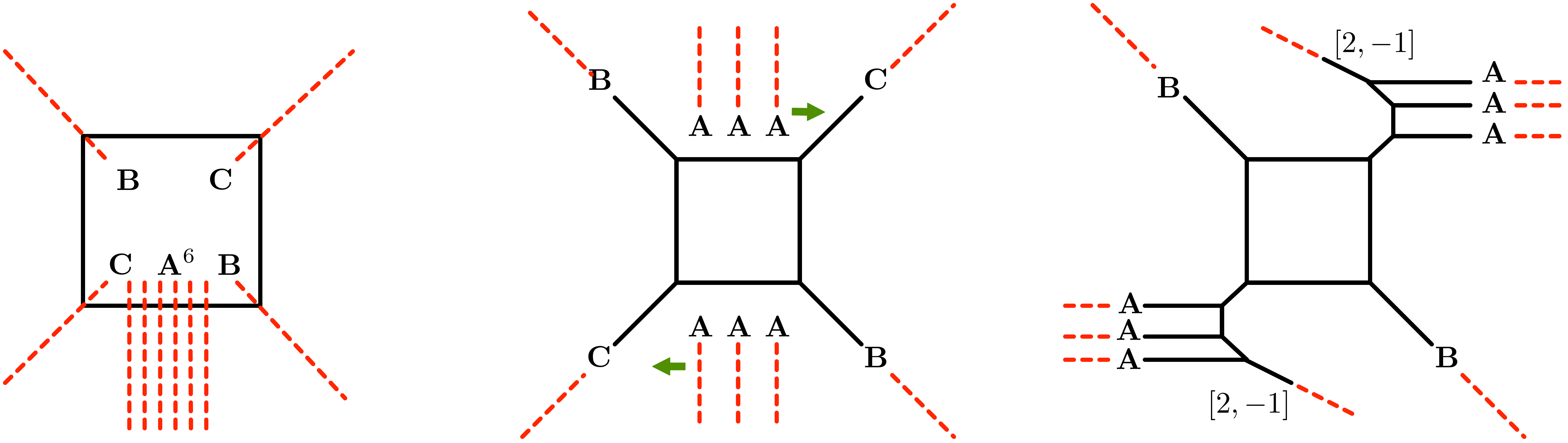}
     \caption{Various 7-brane configurations for 5d $SU(2)+6\mathbf{F}$ and a 5-brane web obtained via Hanany-Witten moves. The branch cuts of 7-branes are denoted by dotted lines.}
     \label{fig:su2+6F-7-brane}
 \end{figure}

As a concrete example, consider 7-brane configurations for the 5d $SU(2)$ gauge theory with 6 fundamental hypermultiplets ($SU(2)+6\mathbf{F}$) given in Figure \ref{fig:su2+6F-7-brane}. In the first configuration, the 7-branes $\bf A^6BCBC$ are all inside a compact face  where the cut of each 7-brane is denoted by a red dotted line. Note that 5-brane surrounding the face bends around whenever it crosses the cut of each 7-brane. This 7-brane configuration is precisely that of the $E_6$ singularity, which coincides with the enhanced global symmetry of the 5d $SU(2)+6\mathbf{F}$ theory as expected. 
In the second web in Figure \ref{fig:su2+6F-7-brane}, after implementing \eqref{eq:useful 7-branes} or $\bf A^3BC=BCA^3$, one can pull out all the 7-branes from the compact face via the Hanany-Witten (HW) transitions,
where additional 5-branes are created when 7-branes $\bf B$ and $\bf C$ are pulled out, while no 5-brane is created for $\bf A$'s. In the last configuration, the cuts of 7-branes $\bf A$s are aligned along the direction of their charge and moved away, which creates a typical 5-brane configuration for 5d $SU(2)+6\mathbf{F}$.

\begin{figure}[t]
     \centering
     \includegraphics[scale=.18]{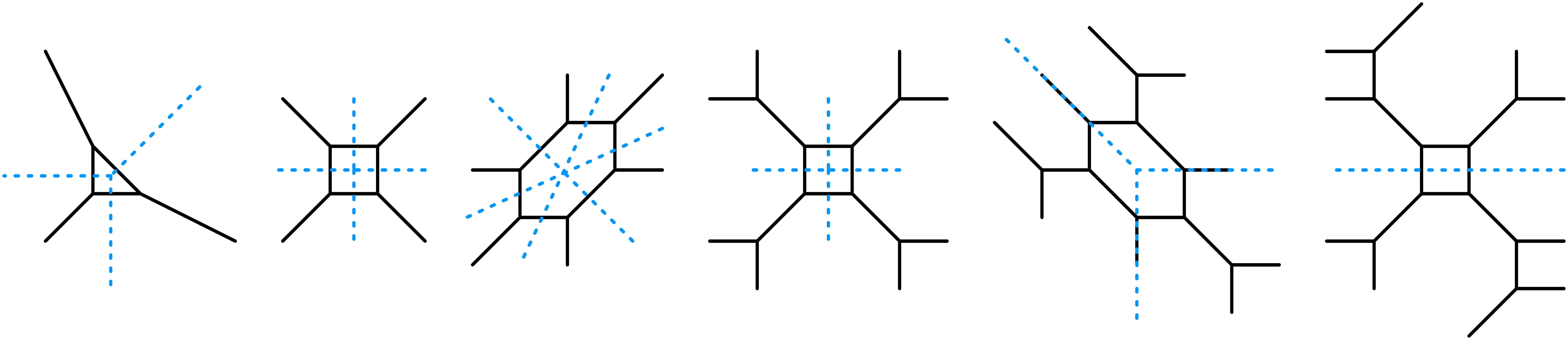}
     \caption{5-brane webs for 5d rank-1 theories showing $\mathbb{Z}_N$ discrete symmetries, where the blue dotted lines denote the action of discrete symmetries.   From the left, local $\mathbb{P}^2$, $SU(2)_0$, $SU(2)+2\mathbf{F}$, $SU(2)+4\mathbf{F}$, $SU(2)+5\mathbf{F}$, and $SU(2)+6\mathbf{F}$.}
     \label{fig:rank1-Zn}
 \end{figure}

Notice that the last 5-brane web in Figure \ref{fig:su2+6F-7-brane} reveals a discrete $\mathbb{Z}_2$ symmetry, which is a 180 degree rotation of  the 5-brane web. In fact, many 5-brane web diagrams enjoy similar discrete symmetries. To see this, consider 5-brane webs for rank-1 theories. As depicted in Figure \ref{fig:rank1-Zn}, 
the $E_0$-theory or the geometry of a local $\mathbb{P}^2$ has a $\mathbb{Z}_3$ symmetry;
  $SU(2)_0$  without flavor  has $\mathbb{Z}_2,\mathbb{Z}_4$ symmetry;
$SU(2)+2\mathbf{F}$ has $\mathbb{Z}_2,\mathbb{Z}_3, \mathbb{Z}_6$ symmetry; $SU(2)+4\mathbf{F}$ has $\mathbb{Z}_2,\mathbb{Z}_4$ symmetry; $SU(2)+5\mathbf{F}$ has a $\mathbb{Z}_3$ symmetry;
$SU(2)+6\mathbf{F}$ has a $\mathbb{Z}_2$ symmetry.

We remark here that the Type IIB coupling $\tau$ is not fixed  for a $\mathbb{Z}_2$ symmetric web, while it should be tuned to see other $\mathbb{Z}_n$ symmetries. However, the 5-brane webs in this paper are drawn at $\tau=i$ as these are more familiar to most readers.
The discrete symmetries can be seen in a more manifest way if we choose $\tau=e^{2\pi i/3}, i,  e^{\pi i/3}$ respectively for $n=3,4,6$ when constructing a web diagram, where $\mathbb{Z}_n$ discrete symmetry is realized as a $2\pi/n$ discrete rotation.

\begin{figure}[t]
     \centering
     \includegraphics[scale=.23]{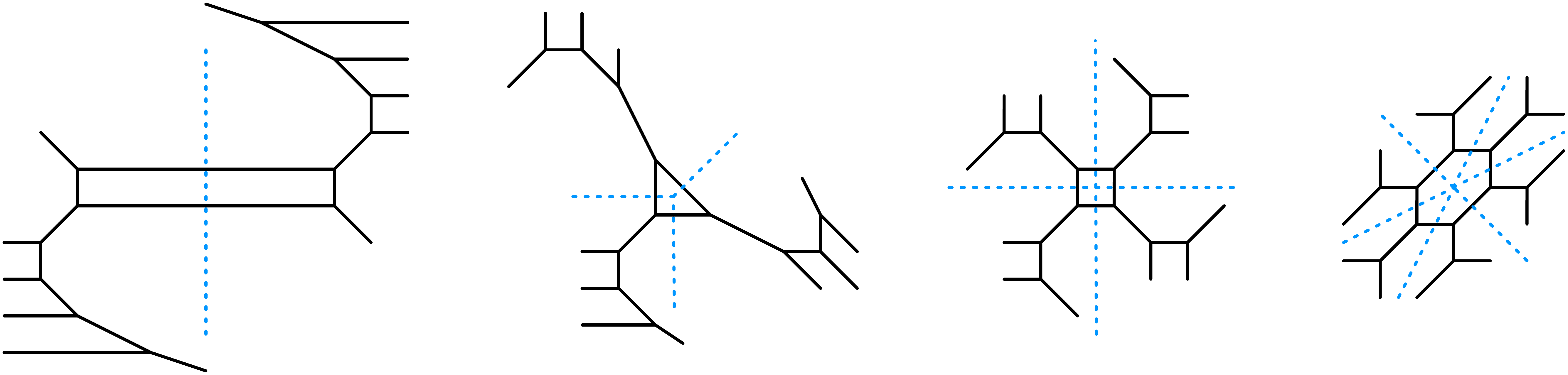}
     \caption{5-brane webs for $SU(2)+8\mathbf{F}$ explicitly revealing  $\mathbb{Z}_{2},\mathbb{Z}_{3},\mathbb{Z}_{4},\mathbb{Z}_{6}$ discrete symmetries (from the left to the right). 
     }
     \label{fig:E8-Z2346}
 \end{figure}

The 5-brane web for the 5d $SU(2)+8\mathbf{F}$ theory, which is a KK theory for the 6d $E$-string on a circle, is particularly interesting because it enjoys all the  $\mathbb{Z}_{2,3,4,6}$ discrete symmetries. See the following deformations of 5-brane Tao web \cite{Kim:2015jba} with various HW moves, as depicted in Figure \ref{fig:E8-Z2346}. Also, brane webs for all other rank-1 SCFTs in Table \ref{tab:SU2 discrete symmetry} having discrete symmetries can be obtained from the 5-brane web for this KK theory by integrating out a number of hypermultiplets.

\begin{table}[H]
    \centering
\begin{tabular}{ c | c c c c }
% \hline
% Symmetry &  & & &  \\
\hline
local $\mathbb{P}^2$ && $\mathbb{Z}_3$ & &  \\
 zero flavor $SU(2)_0$ & $\mathbb{Z}_2$ &&$\mathbb{Z}_4$ &  \\  
 $SU(2)+2\mathbf{F}$ & $\mathbb{Z}_2$&$\mathbb{Z}_3$&&$ \mathbb{Z}_6$ \\
 $SU(2)+4\mathbf{F}$ &  $\mathbb{Z}_2$&&$\mathbb{Z}_4$ & \\
 $SU(2)+5\mathbf{F}$ && $\mathbb{Z}_3$ &&\\
$SU(2)+6\mathbf{F}$ &  $\mathbb{Z}_2$ &&&\\
$SU(2)+8\mathbf{F}$ & $\mathbb{Z}_2$& $\mathbb{Z}_3$&$\mathbb{Z}_4$&$\mathbb{Z}_6 $\\
\hline
\end{tabular}
\caption{Discrete $\mathbb{Z}_{2,3,4,6}$ symmetries of 5d rank-1 theories.}
\label{tab:SU2 discrete symmetry}
\end{table}

%%%%%%%%%%%%%%%%%%%%%%%%%%%%%%%%%%%%%%
\section{S-fold on Face  }\label{sec:3}

We now consider a class of discrete global symmetries of 5d SCFTs and some 6d SCFTs,   in particular, in terms of  their 5-brane web realizations, and study how to  gauge them. Specifically, we will discuss the quotients or S-foldings of discrete symmetries that are generated by rotations of the brane webs and also $\mathrm{SL}(2,\mathbb{Z})$ transformations. These symmetries are 0-form symmetries which act on the particle spectrum in the Coulomb phase. There are two distinct classes of such discrete symmetries in 5-brane webs: one class involves a fixed point in the middle of a compact face in $(p,q)$ 5-branes, while the other involves a fixed point at the center of a line or at a vertex of intersecting $(p,q)$ 5-branes. In this section, we discuss how to gauge the former class of $\mathbb{Z}_n$ symmetries. The S-fold of discrete symmetries in the latter class will be discussed in the next section.

Along the way, we propose a systematic algorithm for identifying  the theories resulting from the $\mathbb{Z}_n$ S-folds. Our proposal is based on the consistency of 7-brane monodromies in the 5-brane webs and also on cubic prepotentials on the Coulomb branch of the moduli space. We then present non-trivial examples of the $\mathbb{Z}_n$ S-folds of various 5-brane webs.

\subsection{Prescription}\label{sec:prescription-face}

As discussed in the previous section, we are interested in the $\mathbb{Z}_n$ symmetries that rotate the $(p,q)$-plane by certain discrete angles together with $\mathrm{SL}(2,\mathbb{Z})$ actions. We first review the $\mathbb{Z}_2$ quotient of $(p,q)$ 5-brane webs using an orientifold 7-plane. We will then generalize this procedure to the S-folds of $\mathbb{Z}_3,\mathbb{Z}_4,\mathbb{Z}_6$ symmetries later.

\begin{figure}[t]
     \centering
     \includegraphics[scale=.15]{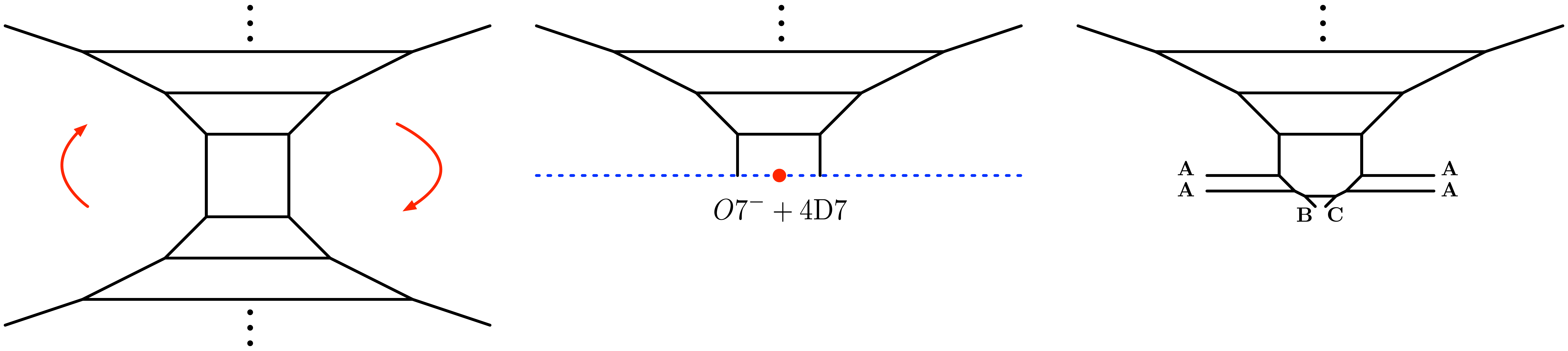}
     \caption{The $\mathbb{Z}_2$ quotient and the insertion of an O7$^-$-plane and four D7-branes at the center of a $180$ degree rotation. Equivalently one can insert $\mathbf{A}^4 {\mathbf{B}} {\mathbf{C}}$ 7-branes. 
     }
     \label{fig:Z2gauging-O7}
 \end{figure}

The $\mathbb{Z}_2$ symmetry in a 5-brane web is generated by a $180$ degree rotation of the $(p,q)$-plane. The quotient with this $\mathbb{Z}_2$ symmetry can be implemented in Type IIB theory by locating an O7$^-$-plane at the fixed point of the rotation, which is the center of the $(p,q)$-plane, as depicted in Figure \ref{fig:Z2gauging-O7}. This is, however, not sufficient, since we expect that the asymptotic 5-brane configuration on the remaining half-plane must be maintained after the $\mathbb{Z}_2$ quotient. 
We need additional 7-branes in order to make the final deficit angle be $\pi$ so that the identification by the $\mathbb{Z}_2$ rotation can be achieved. The exact 7-brane configuration at the singularity for the $\mathbb{Z}_2$ folding is %
\begin{align}
     \mathcal{S}_{D_4}=4\mathrm{D}7+\mathrm{O}7^-.
\end{align} 
This configuration has a deficit angle $\pi$, a constant axio-dilaton field of any value, and a monodromy   $-\mathbbm{1}$.\footnote{ There is a subtlety for the monodromy $-\mathbbm{1}$. This changes $(p,q)$ 5-brane to the $(-p,-q)$ 5-brane which is an anti-brane. The additional $\pi$ rotation on the $(p,q)$ plane brings it back to the original brane. This can be seen more easily by drawing an arrow on a 5-brane  to indicate the direction.}  The $\mathrm{O}7^-$-plane can be split to a pair of ${\bf B}$ and ${\bf C}$ 7-branes. 
Thus, one finds an equivalent 7-brane configuration
\begin{align}
     \mathcal{S}_{D_4}=\mathbf{A}^4\mathbf{BC} .
\end{align}  

Therefore, the $\mathbb{Z}_2$ quotient, which has a fixed point in the middle of a face, can be performed by two steps. First cut off half of the 5-brane webs and then insert  an O7$^-$-plane and four D7-branes at the $\mathbb{Z}_2$ singularity. The 7-brane configuration at the fixed point makes there resulting 5-brane web smoothly merged along the cut.   
 We illustrate this in Figure \ref{fig:Z2gauging-O7}.

\begin{figure}[t]
     \centering
     \includegraphics[scale=.22]{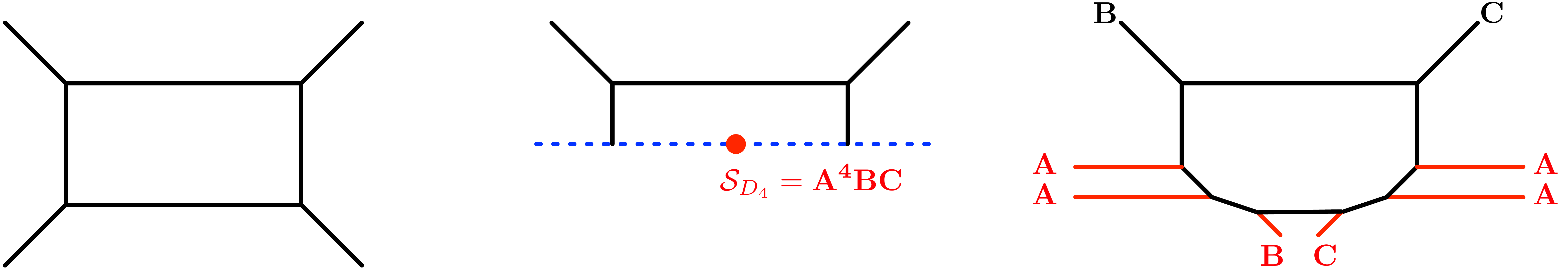}
     \caption{The $\mathbb{Z}_2$ fold of the 5d $SU(2)_0$ theory. It is to insert the 7-branes of $D_4$ singularity,  $\mathcal{S}_{D_4}=\mathbf{A^4BC}$,  at the fixed point of the $\mathbb{Z}_2$ action (180 degree rotation) of the $\mathbb{F}_0$ surface. The resulting 5-brane configuration is that of the $SU(2)$ gauge theory with 4 flavors.
     }
     \label{fig:Z2-pureSU2}
 \end{figure}

As a simple example, consider the 5d $SU(2)_0$ gauge theory without matter which is realized by the brane web of a Hirzebruch surface $\mathbb{F}_0$  shown in Figure \ref{fig:Z2-pureSU2}. The brane web has a $\mathbb{Z}_2$ symmetry rotating it by 180 degrees. The $\mathbb{Z}_2$ symmetry has a fixed point in the middle of the $\mathbb{F}_0$ which therefore belongs to the first class of discrete symmetries. After the $\mathbb{Z}_2$ quotient, we get the third brane web in Figure \ref{fig:Z2-pureSU2} and this gives rise to the 5d $SU(2)$ gauge theory with four fundamental hypermultiplets.

The generalization of the $\mathbb{Z}_2$ fold on a face to other $\mathbb{Z}_{3,4,6}$ symmetries is straightforward. First, divide the 5-brane webs to $n$ equal slices at the fixed point, keep only one of them. Second, insert an appropriate 7-brane configuration at the tip. The symmetry of the 5-brane web requires a special value of the axio-dilaton field $\tau$. The 7-brane inserted should carry the same axio-dilaton field, the deficit angle $2\pi(n-1)/n$, the right monodromy matrix $K_{\mathbb{Z}_n}$, so that the 5-brane webs get smoothly merged along two boundaries of a single slice of angle $2\pi/n$. The relevant 7-brane configurations for $\mathbb{Z}_n$ S-folds are summarized  in Table~\ref{tab:7-branes and Zn S-folding}.

\begin{table}[h]
    \centering
    \begin{tabular}{c|c|c|c|c}
    \hline
      S-fold &  7-brane   $\mathcal{S}$ & deficit angle   &  axio-dilaton $\tau$   & monodromy $K_{\mathbb{Z}_n}$ \\ \hline
       $\mathbb{Z}_2$  
        & $\mathcal{S}_{D_4}= {\bf A}^{\! 4}{\bf B}{\bf C}~~\,$
        %${\rm D}7^4{\rm O}7^-={\bf A}^{\! 4}\widetilde{\bf B}\widetilde{\bf C}$  
         & $\pi$ &  $\mathrm{any}$  &  $-\mathbbm{1}_2$ \\
         $\mathbb{Z}_3$ & $\mathcal{S}_{E_6}={\bf A}^{\! 5}{\bf B}{\bf C}{\bf B}$
        % ${\bf A}^{\! 5}\widetilde{\bf B}\widetilde{\bf C}\widetilde{\bf C}$ 
        & $\frac{_4}{^3}\pi $ & 
          $ e^{\frac{2\pi i }{3}}$   
        & $(ST)^4=({0 \ \ \ 1 \atop - 1 \ -1 })$\\
        $\mathbb{Z}_4$   &
        $\mathcal{S}_{E_7}={\bf A}^{\! 6}{\bf B}{\bf C}{\bf B}$
        %${\bf A}^{\! 6}\widetilde{\bf B}\widetilde{\bf C}\widetilde{\bf C}$
        &  $\frac{_3}{^2}\pi $ & 
         $i$   
        & $S^3=({0 \ \  1 \atop -1 \ 0 })$\\
        $\mathbb{Z}_6$ & $\mathcal{S}_{E_8}={\bf A}^{\! 7}{\bf B}{\bf C}{\bf B}$
        % ${\bf A}^{\! 7}\widetilde{\bf B}\widetilde{\bf C}\widetilde{\bf C}$ 
        & $\frac{_5}{^3}\pi $
        &     $ e^{\frac{\pi i}{3}} $ 
        & $(TS)^5=({0 \ \  1 \atop -1 \ 1})$ \\ \hline
    \end{tabular}
    \caption{ The characteristics of the 7-branes for the $\mathbb{Z}_n$ S-foldings. %For $D_4$, $\mathcal{S}_{D_4}=4\mathrm{D}7+\mathrm{O}7^{-}.$  
    } 
    \label{tab:7-branes and Zn S-folding}
\end{table}

\noindent In the table, the deficit angles are just for the $\mathbb{Z}_{2,3,4,6}$ foldings on the $(p,q)$-plane. The monodromy matrix $K_{\mathbb{Z}_n}$ for each case leaves the axio-dilaton field $\t$ invariant and satisfies also the identity $K_{\mathbb{Z}_n}^n=\mathbbm{1}$. 
The 7-brane configuration for $\mathbb{Z}_n$ has the exactly right monodromy.\footnote{In the convention used in \cite{Gaberdiel:1997ud},   the 7-branes $\mathcal{S}$ for $D_4$ and $E_{n}$ singularities are given as 
$\mathbf{A}^4\widetilde{\mathbf{B}}\widetilde{\mathbf{C}}$ and $\mathbf{A}^{n-1}\widetilde{\mathbf{B}}\widetilde{\mathbf{C}}\widetilde{\mathbf{C}}$, respectively. Here,  $\widetilde{\mathbf{B}}=[3,-1]$ and $\widetilde{\mathbf{C}}=[1,-1]=\mathbf{B}$ are related to the original $\mathbf{B}=[1,-1]$ and $\mathbf{C}=[-1,-1]$ by the $\SL(2, \mathbb{Z}_2)$ transformation $T^2$, respectively. % transformation, respectively. 
As one can show easily that  $\widetilde{\mathbf{B}}\widetilde{\mathbf{C}}=\mathbf{B}\mathbf{C}$ and  $\widetilde{\mathbf{B}}\widetilde{\mathbf{C}}\widetilde{\mathbf{C}}=\mathbf{B}\mathbf{C}\mathbf{B}$, we express the 7-branes $\mathcal{S}$ in terms of $\mathbf{B}$ and $\mathbf{C}$ branes rather than $\widetilde{\mathbf{B}}, \widetilde{\mathbf{C}}$.}

\begin{figure}[t]
     \centering
     \includegraphics[scale=.40]{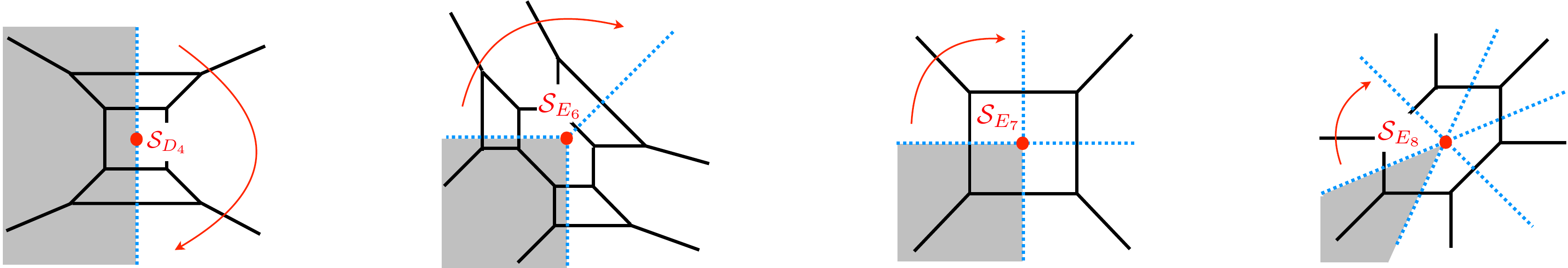}
     \caption{The $\mathbb{Z}_2,\mathbb{Z}_3,\mathbb{Z}_4,\mathbb{Z}_6$ S-folding prescriptions (from left to right). We choose the gray region as the fundamental domain of $(p,q)$-plane after the $\mathbb{Z}_n$ quotient. The $\mathbb{Z}_n$ singularity is labelled by $\mathcal{S}$
      with a subscript denoting the global symmetry assigned at the singularity.}
     \label{fig:rule-face}
 \end{figure}

Here, we have fixed our convention for the $\mathrm{SL}(2,\mathbb{Z})$ frame of the 5-brane webs as follows. First we orient the 5-branes so that D5 or (1,0) brane to be horizontal and NS5 or (0,1) brane to be vertical.  
For $\mathbb{Z}_{3,6}$, we may also need a reflection, for example, along the vertical axis which transforms   $(1,1)$-brane to  $(-1,1)$ brane, and vice versa.   
From the 5-brane tension formula \eqref{eq:bpstension}, we note that, for $\mathbb{Z}_3$, $(1,0),(0,1),(1,-1)$ 5-branes have the same tension at $\tau=e^{2\pi i /3}$, and for $\mathbb{Z}_6$, $(1,0),(0,1),(1,1)$ have the same tension at $\tau=e^{\pi i /3}$. 
See Figure \ref{fig:rule-face} for some examples for this adjustment of symmetric 5-brane webs.   
Second,  for such a 5-brane web,  we choose the fundamental domain for the $\mathbb{Z}_n$ quotient to contain the bottom-left corner of the original brane web.
Third, we organize the brane web using an $\mathrm{SL}(2,\mathbb{Z})$ transformation such that the bottom 5-brane of the face involving the $\mathbb{Z}_n$ fixed point becomes a D5-brane ending at the boundary of the fundamental domain. 
It follows then that D7-branes (or $\mathbf{A}$'s)
inserted at the singularity can move freely downward across D5-branes without any brane creations by the Hanany-Witten effect. We illustrate this convention in Figure \ref{fig:rule-face}.
In this case, the rest of the 5-brane webs in clockwise succession are obtained by  multiple applications of the monodromy matrix $K_{\mathbb{Z}_n}$ in Table~\ref{tab:7-branes and Zn S-folding} on the 5-brane web in the fundamental region. One can check the above prescription to web diagrams in Figure~\ref{fig:rule-face}.

The rank of the gauge group after the $\mathbb{Z}_n$ S-folding can be easily read off from the rank of the original theory. The $n$ compact faces in a $\mathbb{Z}_n$ orbit in the original 5-brane webs are identified under the $\mathbb{Z}_n$ quotient. This implies that the rank of the original 5d theory $\mathcal{T}$ associated to the 5-brane web is reduced, after the $\mathbb{Z}_n$ S-folding, by
\begin{align}\label{eq:rank-F-gauging}
  r(\mathcal{T}) \ \ \rightarrow \ \ r(\mathcal{T}/\mathbb{Z}_n) = \frac{r(\mathcal{T})-1}{n} + 1 \ ,
\end{align}
where $r(\mathcal{T})$ denotes the rank of the original theory $\mathcal{T}$ and $r(\mathcal{T}/\mathbb{Z}_n)$ is the rank of the theory after the $\mathbb{Z}_n$ quotient.

Likewise, the flavor symmetry of the original 5-brane web is reduced under the $\mathbb{Z}_n$ S-folding. The rank of the flavor symmetry in a brane web without orientifold planes is given by the number of external 7-branes minus ``3'' where ``3'' comes from the overall movement of the branes on the $(p,q)$-plane and the 5-brane charge conservation constraint. The positions of the asymptotic 7-branes on the $(p,q)$-plane are described in the 5d field theory by mass parameters for the flavor symmetry.  As the  external 7-branes in a $\mathbb{Z}_n$ orbit are identified with each other, the rank of the  flavor symmetry in the original web decreases after the S-folding.
 We note however that this reduced symmetry is not the full symmetry of the resulting 5-brane web. Since the $\mathbb{Z}_n$ quotient leads to a singularity hosting additional 7-branes, we should also take into account the flavor symmetries coming from these 7-branes. The full  flavor symmetry of the 5d theory after the $\mathbb{Z}_n$ quotient is thus the product of the flavor symmetry from the original 5-brane web which survives under the $\mathbb{Z}_n$ folding and the flavor symmetry from the 7-branes placed at the singularity. It is not straightforward to identify the precise flavor symmetry  because the flavor symmetry can be enhanced at the UV CFT fixed point.
Instead, we propose a formula for the rank of the flavor symmetry after the S-folding as follows\footnote{This formula can fail to capture the correct rank of the flavor symmetry when the brane web involves decoupled free matters.} :
\begin{align}\label{eq:flavor-F-gauging}
  r_F(\mathcal{T}/\mathbb{Z}_n) = \frac{r_F(\mathcal{T})+3}{n}-1+r_F(\mathcal{S}) \ ,
\end{align}
where $r_F(\mathcal{T})$ denotes the rank of the flavor symmetry in the original theory $\mathcal{T}$, $r_F(\mathcal{T}/\mathbb{Z}_n)$ is the rank of the flavor symmetry after the $\mathbb{Z}_n$ quotient.
The first term on the right-hand side of \eqref{eq:flavor-F-gauging} counts the number of asymptotic 7-branes remaining after $\mathbb{Z}_n$ folding. However, the position of one of the external 7-branes is fixed by $(p,q)$ 5-brane charge conservation, which results in the factor $-1$ on the right-hand side. The last term $r_F(\mathcal{S})$ is the rank of the flavor symmetry from the extra 7-branes at the singularity.

Now we discuss the S-fold in terms of the prepotentials $\mathcal{F}$ of 5d SCFTs on the Coulomb branch of the moduli space. The $\mathbb{Z}_n$ symmetry we are interested in can be spontaneously broken at a generic point on the Coulomb branch. The $\mathbb{Z}_n$ symmetry in general emerges only along special loci on the Coulomb branch with particular K\"ahler parameters.  The prepotential at this special point on the Coulomb branch is indeed invariant under the $\mathbb{Z}_n$ symmetry.
We will take the prepotential at the special point and discuss the $\mathbb{Z}_n$ quotient of it. Also, unless explicitly stated, we will assume that the mass parameters are all switched off since generic mass parameters can explicitly break the $\mathbb{Z}_n$ symmetry.

Let us explain how the prepotential is modified under the $\mathbb{Z}_n$ quotient. We start with the prepotential written in terms of the K\"ahler parameters in the original 5d theory.
Since the $\mathbb{Z}_n$ action permutes compact faces in an orbit of its action, it also permutes the K\"ahler parameters for the compact faces. It then follows that the $\mathbb{Z}_n$ quotient identifies all the K\"ahler parameters in the orbit. Accordingly, the K\"ahler parameters of the original prepotential in the orbit should be identified with one another. %each other.
For example, if a $\mathbb{Z}_3$ action permutes three compact faces as $(\phi_1, \phi_2 ,\phi_3)\rightarrow (\phi_2, \phi_3 ,\phi_1)$ where $\phi_i$ are the K\"ahler parameters for the faces, then all these 3 parameters in the prepotential will be identified, i.e. $\phi_1=\phi_2=\phi_3$, after the $\mathbb{Z}_3$ quotient.

For a 5-brane web, the prepotential is computed by integrating the areas of compact faces in terms of the K\"ahler parameters $\phi_i$. Note that the physical domain after the $\mathbb{Z}_n$ quotient is the $1/n$ sub-region of the original $(p,q)$ web as drawn in Figure \ref{fig:rule-face}. Thus, the $\mathbb{Z}_n$ S-fold affects the areas of the compact faces in the following way. First, the areas of faces in a $\mathbb{Z}_n$ orbit that are interchanged with other faces do not change, though their K\"ahler parameters are identified according to the rule explained above. Second, on the other hand, the face containing the $\mathbb{Z}_n$ fixed point is 
cut into its $1/n$ fraction, and thus its area reduces to $1/n$ of the area in the original brane web. The prepotential after the $\mathbb{Z}_n$ quotient must be consistent with these results.

We find, after integrating the areas of the faces in the resulting 5-brane web, the prepotential of the 5d theory $\mathcal{T}/\mathbb{Z}_n$ becomes
\begin{align}\label{eq:gauged-F}
  \mathcal{F}(\mathcal{T}/\mathbb{Z}_n) = \frac{1}{n}\mathcal{F}(\mathcal{T})\big|_{\phi_{S(i)} = \phi_i} \ ,
\end{align}
where  $\mathcal{F}(\mathcal{T})\big|_{\phi_{S(i)} = \phi_i}$ stands for the prepotential of the original theory written in terms of K\"ahler parameters with the identifications $\phi_{S(i)} = \phi_i$ for all $i$ permuted as $i\rightarrow S(i)$ under the $\mathbb{Z}_n$ action. This prepotential formula is also consistent with the fact that the total volume of the Calabi-Yau 3-fold related to the 5-brane web is reduced by a factor of $1/n$ under the $\mathbb{Z}_n$ quotient. We remark that the same prepotential formula works for the $\mathbb{Z}_n$ S-folds in the second class which we will discuss in the next section.

%An instructive example would be the $SU(2)$ theory with $N_f$ fundamental hypermultiplets ($SU(2)+N_f\mathbf{F}$) as the corresponding prepotential is very simple,
%\begin{align}
%    6\mathcal{F}(\mathcal{T}_{SU(2)+N_f\mathbf{F}}) = (8-N_f) \phi^3\ .
%    \label{eq:FFSU(2)+Nf}
%\end{align} 
%Let us consider the $SU(2)_0$ theory without flavor. 
An instructive example would be the $SU(2)_0$ theory without flavor. 
The corresponding brane web has a $\mathbb{Z}_4$ symmetry where the $\mathbb{Z}_4$ fixed point is located in the middle of the  $\mathbb{F}_0$ surface.
The S-fold with this symmetry can be discussed in terms of the prepotential as well as the 7-branes at the singularity. The prepotential of the $SU(2)_0$ theory is given by
\begin{align}
  6\mathcal{F}(\mathcal{T}_{SU(2)_0}) = 8\phi^3 \ .
\end{align}
Our prescription says that the 5d theory after the $\mathbb{Z}_4$ quotient will have the prepotential
\begin{align}
  6\mathcal{F}(\mathcal{T}_{SU(2)_0}/\mathbb{Z}_4) =\frac{1}{4}\times6 \mathcal{F}(\mathcal{T}_{SU(2)_0}) =  2\phi^3 \ .
\end{align}
One may notice that this coincides with the prepotential of the $SU(2)$ gauge theory with 6 fundamental hypers. This leads us to assert that the $\mathbb{Z}_4$ S-fold of the $SU(2)_0$ theory gives rise to the  $SU(2)$ gauge theory with 6 fundamentals.

\begin{figure}[t]
     \centering
     \includegraphics[scale=.23]{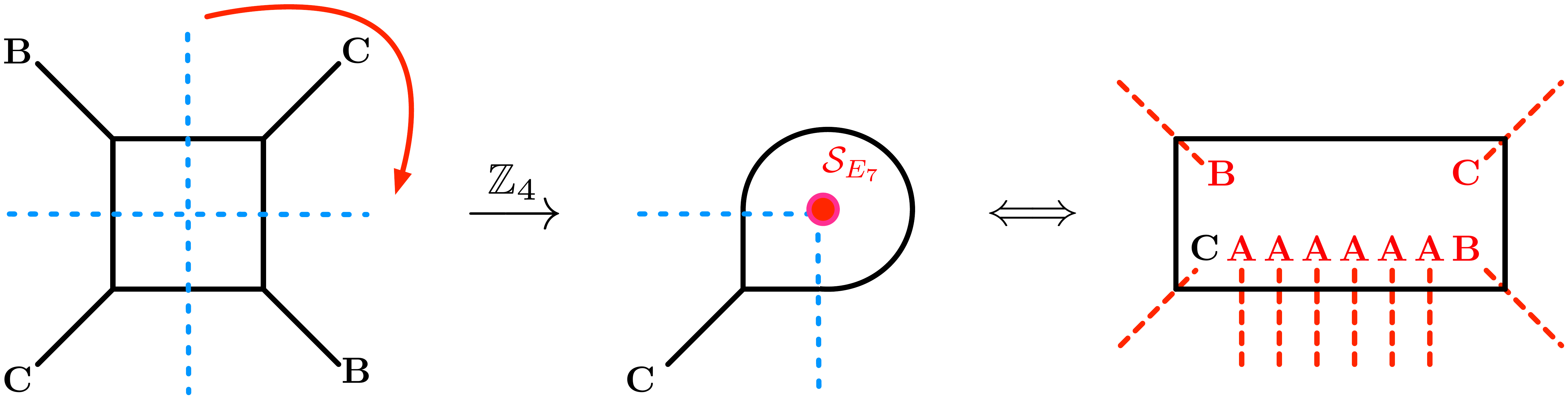}
     \caption{ The $\mathbb{Z}_4$ S-fold of the $SU(2)_0$ theory. Insertion of $\mathcal{S}_{E_7}=\bf A^6BCB$ at the center of the face of the Coulomb branch leads to the 7-brane configuration for $SU(2)+6\mathbf{F}$ depicted in Figure \ref{fig:su2+6F-7-brane}, which has an $E_7$ flavor symmetry.}
     \label{fig:pureSU2-Z4}
 \end{figure}

The same conclusion can be drawn by using the 7-brane singularity. The $\mathbb{Z}_4$ quotient in the brane web leaves only a quarter of the plane which has a single external 7-brane and also hosts a $\mathbb{Z}_4$ singularity with 7-branes of $E_7$ flavor symmetry. The equations (\ref{eq:rank-F-gauging}) and (\ref{eq:flavor-F-gauging}) tell us that the 5d theory after the $\mathbb{Z}_4$ quotient is a rank-1 theory with $E_7$ flavor symmetry. See Figure \ref{fig:pureSU2-Z4}. 
The resulting theory turns out to be the $SU(2)$ gauge theory with 6 fundamentals that precisely agrees with the prepotential calculation.
Note that the $SU(2)$ flavor symmetry of the original $SU(2)_0$ theory is broken by the $\mathbb{Z}_4$ quotient, but the extra 7-branes inserted at the $\mathbb{Z}_4$ singularity provides the $E_7$ flavor symmetry.

We will now present various examples of the $\mathbb{Z}_n$ S-fold having a singularity on a face.

%---------------------
\subsection{Rank-1 theories}\label{sec:rank-1}
The S-folds of the rank-1 theories are summarized in Table \ref{tab:symmetries for rank1}. 

\begin{table}[H]
    \centering
\begin{tabular}{  m{5.3em} | c c c c }
 \hline
 & $\mathbb{Z}_2$ &$\mathbb{Z}_3$ &$\mathbb{Z}_4$ & $\mathbb{Z}_6$ \\
\hline
local $\mathbb{P}^2$ && $E_6$ & &  \\
 $SU(2)_0$ & $E_5$ &&$E_7$ &  \\  
 $SU(2)+2\mathbf{F}$ & $E_6$&$E_7$&&$E_8$ \\
 $SU(2)+4\mathbf{F}$ &  $E_7$&&$E_8$ & \\
 $SU(2)+5\mathbf{F}$ && $E_8$ &&\\
$SU(2)+6\mathbf{F}$ & $E_8$ &&&\\
$SU(2)+8\mathbf{F}$ & $\hat{E}_8$& $\hat{E}_8$&$\hat{E}_8$&$\hat{E}_8 $\\
\hline
\end{tabular}
\caption{Flavor symmetries after the $\mathbb{Z}_{2,3,4,6}$ quotient of 5d rank-1 theories. Here, $E_5=D_5$.}
\label{tab:symmetries for rank1}
\end{table}

\begin{figure}[t]
     \centering
     \includegraphics[scale=.18]{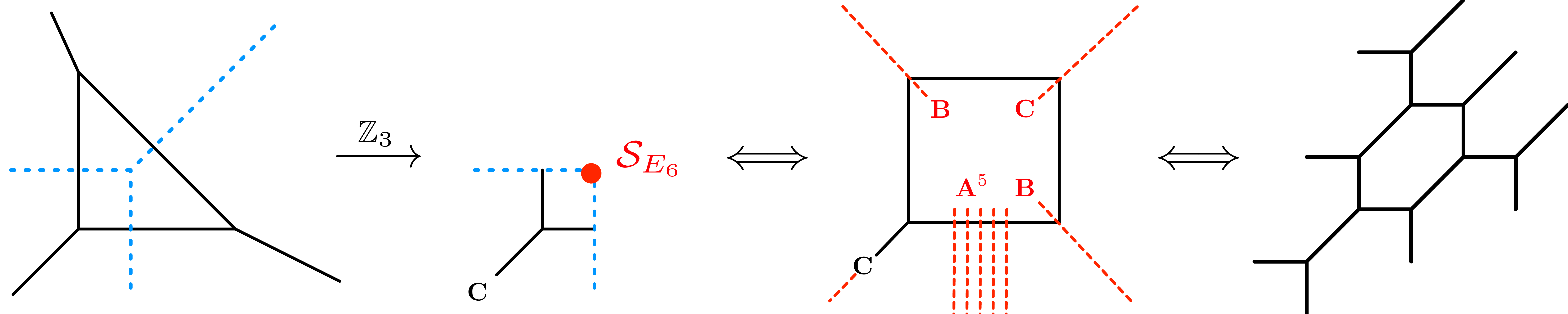}
     \caption{The $\mathbb{Z}_3$ quotient of the 5d SCFT of a local $\mathbb{P}^2$ (the $E_0$ theory), leading to $SU(2)+5\mathbf{F}$ or $T_3$ theory.}
     \label{fig:localP2-Z3}
 \end{figure}

\paragraph{Local $\mathbb{P}^2$.}
The simplest 5d SCFT can be engineered by a local $\mathbb{P}^2$ embedded in a Calabi-Yau 3-fold. This theory has a one-dimensional Coulomb branch without flavor symmetry. As shown in the 5-brane web in Figure \ref{fig:localP2-Z3}, 
this theory has a $\mathbb{Z}_3$ symmetry rotating the brane diagram by 120 degrees.

For S-folding the $\mathbb{Z}_3$ discrete symmetry, one cuts the 5-brane webs into 3 slices and keeps a slice with   a single external 7-brane, for example, the 
$[1,1]$ 7-brane, and then inserts  the 7-brane of type $\mathcal{S}_{E_6}={\bf A}^5{\bf B}{\bf C}{\bf B}$ with $E_6$ flavor symmetry at the tip of the slice as in the second diagram in Figure~\ref{fig:localP2-Z3}.  The resulting diagram is equivalent to the 5-brane web for 
  a rank-1 theory with $E_6$ flavor symmetry. We thus find that the $\mathbb{Z}_3$ quotient of the 5d theory on a local $\mathbb{P}^2$ gives rise to the $SU(2)+5\mathbf{F}$ theory or $T_3$ theory as shown in Figure \ref{fig:localP2-Z3}. 

We can understand this result in terms of the cubic prepotential. The $\mathbb{Z}_3$ quotient reduces the volume of the local $\mathbb{P}^2$ by $1/3$ and thus one computes the prepotential of the $\mathbb{Z}_3$ gauged theory as
\begin{align}
    6\mathcal{F}_{\mathbb{P}^2} = 9\phi^3  %\ \ \rightarrow \ \ 
    \quad \overset{\mathbb{Z}_3}{\longrightarrow} \quad 
    6\mathcal{F}_{\mathbb{P}^2/\mathbb{Z}_3} = \frac{1}{3}\times 6\mathcal{F}_{\mathbb{P}^2} = 3\phi^3 \ .
\end{align}
The result agrees with the prepotential for the $SU(2)+5\mathbf{F}$ theory which has an $E_6$ flavor symmetry, as expected.

There is a more geometric understanding of the $\mathbb{Z}_3$-folding of the local $\mathbb{P}^2$.
Let us consider the GLSM description of the local $\mathbb{P}^2$ Calabi-Yau 3-fold. It can be described by the    4-supersymmetric $U(1)$ GLSM of three fields $X_a, a=1,2,3$ of unit charge and a single field $P$ of charge $-3$ with FI term $\xi$, which is proportional to the Coulomb parameter $\phi$. The $\mathbb{Z}_3$ symmetry is the cyclic permutation of $X_1,X_2,X_3$. %
 \begin{table}[h]
     \centering
     \begin{tabular}{c|ccc|c}
          &  $X_1$ & $X_2$ & $X_3$ & $P$  \\ \hline
       $U(1)$    & 1 & 1 & 1 & -3
     \end{tabular}
     \caption{GLSM for local $\mathbb{P}^2$}
     \label{tab:localP2-U(1)}
 \end{table}%
\noindent We introduce new chiral fields $Y_1=(X_1+X_2+X_3) P^{-1/3}$, $Y_2=(X_1+e^{2\pi i/3}X_2+ e^{-2\pi i /3} X_3)P^{-1/3}$, and $Y_3=(X_1+e^{-2\pi i /3} X_2+e^{2\pi i /3} X_3)P^{-1/3}$ which are gauge invariant up to $\mathbb{Z}_3$. In the new variables $Y_1,Y_2, Y_3$ with $\xi=0$, the geometry is basically   $\mathbb{C}^3/\mathbb{Z}_3$. 
Additional  quotient with the  $\mathbb{Z}_3$ cyclic permutation of  $Y_1,Y_2,Y_3$  leads to the geometry $\mathbb{C}^3/\mathbb{Z}_3\times\mathbb{Z}_3$ which is exactly the geometry for the $T_3$ theory as we expect.

\paragraph{$SU(2)_0$ theory.}

Let us consider the 5-brane web for the 5d $\mathcal{N}=1$  $SU(2)_0$ theory without any flavors. The brane web has a $\mathbb{Z}_4$ symmetry rotating the diagram by 90 degrees. The quotient with this $\mathbb{Z}_4$ symmetry gives rise to the $SU(2)+6\mathbf{F}$ theory which was discussed in section \ref{sec:prescription-face} using both the 7-branes at the $\mathbb{Z}_4$ singularity and the prepotential change under the $\mathbb{Z}_4$ folding.

We can also gauge the $\mathbb{Z}_2$ symmetry rotating the diagram by 180 degrees. As depicted in Figure \ref{fig:Z2-pureSU2}, the $\mathbb{Z}_2$ quotient can be thought of as the $\mathbb{Z}_2$ orbifold action on the web which is realized by the introduction of an O7$^-$-plane and four D7-branes at the $\mathbb{Z}_2$ singularity. Notice that the resulting web diagram is that of the 5d $SU(2)+4\mathbf{F}$ theory. The monodromy of the combination of an O7$^-$-plane and four D7's is the same as that of the ${\bf A^4 {B}{C}}$ 7-brane configuration. The global symmetry of the $\mathbb{Z}_2$ gauged theory is enhanced in the UV fixed point to $E_5=D_5$ symmetry that can be explicitly seen by bringing an external 7-brane inside the face. Namely, one gets the 7-brane configuration for $E_5$, $\bf (A^4BCBC)$ which is equivalent to a 7-brane configuration for $D_5$ ${\bf (A^5BCX_{[5,2]})}$. See also  \cite{DeWolfe:1998eu} for detailed computation of this equivalence.

To check consistency, we can compare the cubic prepotential of the gauged theory. Under the $\mathbb{Z}_2$ quotient, the prepotential is halved compared to the original prepotential, 
\begin{align}
  6\mathcal{F}_{SU(2)_0} = 8\phi^3  
  %\rightarrow
  \quad \overset{\mathbb{Z}_2}{\longrightarrow} \quad 
   6\mathcal{F}_{SU(2)_0/\mathbb{Z}_2} = \frac{1}{2}\times 6\mathcal{F}_{SU(2)_0} = 4\phi^3 \ ,
\end{align}
which is indeed the prepotential of the $SU(2)+4\mathbf{F}$ theory.

\paragraph{$SU(2)+2\mathbf{F}$ theory.}
\begin{figure}[t]
     \centering
     \includegraphics[scale=.30]{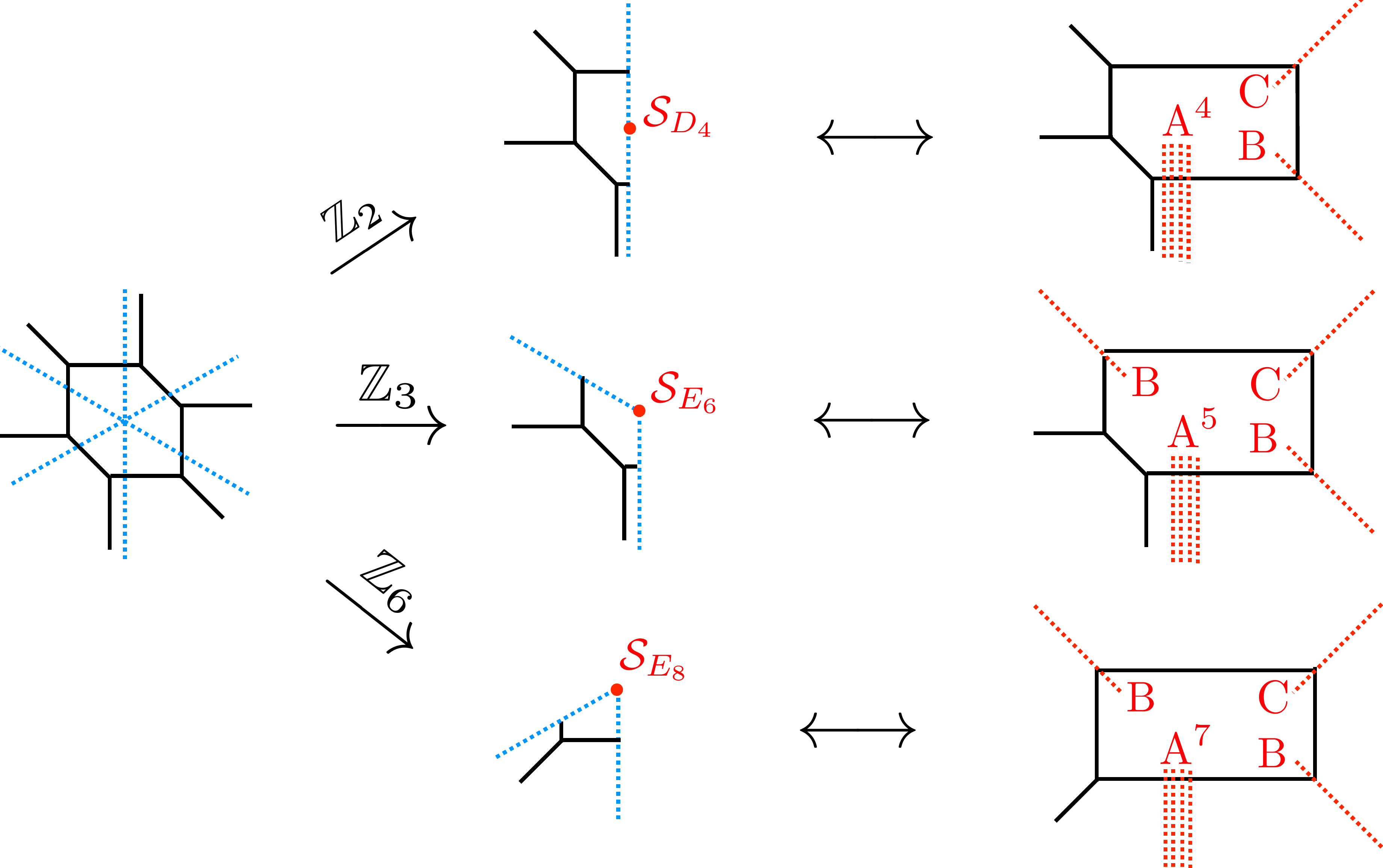}
     \caption{
     Brane web for the $SU(2)+2\mathbf{F}$ theory and its $\mathbb{Z}_{2,3,6}$ quotients. For $\mathbb{Z}_6$, the first diagram is reflected along the vertical line and then its lower left corner slice becomes the second bottom diagram. }
     \label{fig:su2-2F}
 \end{figure}
The 5-brane web for the $SU(2)+2\mathbf{F}$ theory is depicted in Figure \ref{fig:su2-2F}. 
The brane web for the theory has a $\mathbb{Z}_6$ symmetry at $\tau=e^{\pi i/3}$ which rotates the diagram by 60 degrees together with an ${\rm SL}(2,\mathbb{Z}_2)$ action on $(1,0),(0,1),(1,1)$ branes of the same tension (more concretely after a vertical reflection of the first diagram in Figure \ref{fig:su2-2F}).  At $\tau=e^{2\pi i /3}$, one gets a $\mathbb{Z}_3$ symmetry which rotates the diagram by 120 degrees and relates $(1,0),(0,1),(1,-1)$ 5-branes of the same tension. At $\tau=i$, this theory has a $\mathbb{Z}_2$ symmetry which is a 180 degree rotation of the diagram.

Let us first perform the $\mathbb{Z}_2$ quotient of the 5-brane web. The $\mathbb{Z}_2$ quotient then leaves 3 external 7-branes together with ${\bf A^4 BC}$ 7-branes at the $\mathbb{Z}_2$ fixed point. The resulting theory is the $SU(2)+5\mathbf{F}$ theory having an $E_6$ global symmetry in the UV limit. The $D_4$ flavor symmetry at the $\mathbb{Z}_2$ singularity combines with the $U(1)^2$ flavor symmetry of the external 7-branes and enhances to the $E_6$ symmetry as drawn in Figure \ref{fig:su2-2F}. The same theory can be obtained by looking at the cubic prepotential after the $\mathbb{Z}_2$ quotient as
\begin{align}
  6\mathcal{F}_{SU(2)+2\mathbf{F}} = 6\phi^3  \quad \overset{\mathbb{Z}_2}{\longrightarrow} \quad   \frac{1}{2}\times 6\mathcal{F}_{SU(2)+2\mathbf{F}}   = 3\phi^3 =6\mathcal{F}_{SU(2)+5\mathbf{F}}  \ .
\end{align}

The $\mathbb{Z}_3$ S-folding can be performed in a similar manner. After the $\mathbb{Z}_3$ quotient, we %will 
have a brane diagram with 2 external 7-branes and ${\bf A^5 {B}{C}{B}}$ 7-branes at the $\mathbb{Z}_3$ singularity. This gives rise to the $SU(2)+6\mathbf{F}$ theory with an $E_7$ global symmetry in the UV limit. It also follows from the prepotential computation 
\begin{align}
  6\mathcal{F}_{SU(2)+2\mathbf{F}}  \quad \overset{\mathbb{Z}_3}{\longrightarrow} \quad   \frac{1}{3}\times 6\mathcal{F}_{SU(2)+2\mathbf{F}}   = 2\phi^3 =6\mathcal{F}_{SU(2)+6\mathbf{F}}  \ ,
\end{align}
that the resulting theory is the $SU(2)+6\mathbf{F}$ theory.

Lastly, consider the $\mathbb{Z}_6$ quotient of the $SU(2)+2\mathbf{F}$ theory. One easily finds that the resulting web diagram will have a single external 7-brane with additional ${\bf A^7 {B}{C}{B}}$ 7-branes at the $\mathbb{Z}_6$ singularity. This brane web engineers the $SU(2)+7\mathbf{F}$ theory which arises from the UV SCFT with an $E_8$ global symmetry. We computes the prepotential as
\begin{align}
  6\mathcal{F}_{SU(2)+2\mathbf{F}}  \quad \overset{\mathbb{Z}_6}{\longrightarrow} \quad   \frac{1}{6}\times 6\mathcal{F}_{SU(2)+2\mathbf{F}}   = \phi^3=6\mathcal{F}_{SU(2)+7\mathbf{F}}   \ ,
\end{align}
which agrees with the prepotential of the $SU(2)+7\mathbf{F}$ theory.

\begin{figure}[t]
     \centering
     \includegraphics[scale=.25]{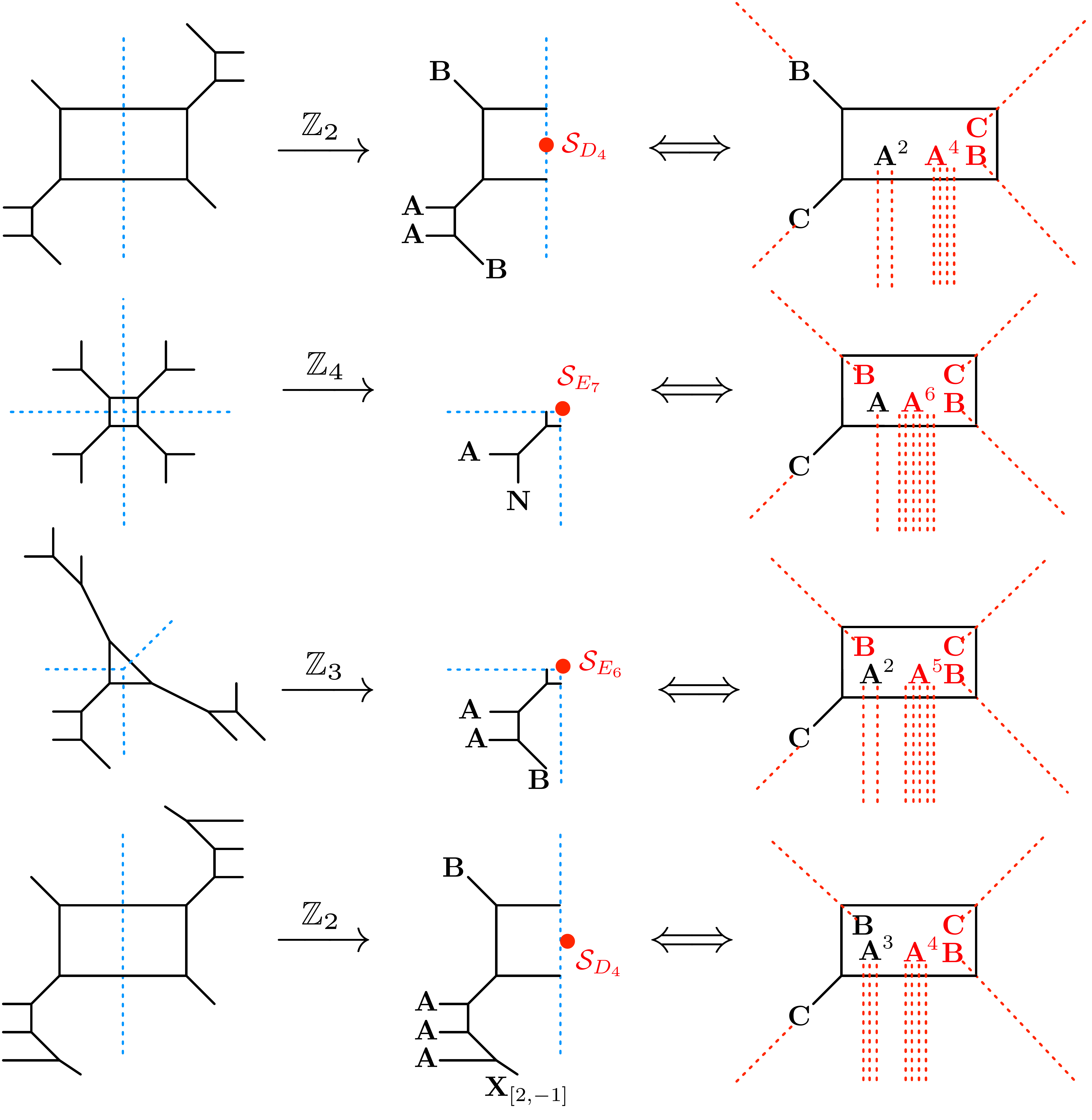}
     \caption{Other S-foldings of rank-1 theories and the resulting brane webs of $\widehat{\mathbf{E}}_n$ configurations, given in \eqref{eq:En-7branes} and also in Figure \ref{fig:rank-1 webs}. }
     \label{fig:rank1-Z234}
 \end{figure}
\paragraph{Other rank-1 theories.}

By employing the S-folding prescription for other rank-1 5d theories which have SCFT fixed points, we can easily read off the 5d theories resulting from the $\mathbb{Z}_n$ S-foldings. The results is summarized as follows (also see Figure \ref{fig:rank1-Z234}):

\begin{itemize}

  \item $SU(2)+4\mathbf{F} \quad \overset{\mathbb{Z}_2}{\longrightarrow} \quad SU(2)+6\mathbf{F}$ 

  \item $SU(2)+4\mathbf{F} \quad \overset{\mathbb{Z}_4}{\longrightarrow} \quad SU(2)+7\mathbf{F}$

  \item $SU(2)+5\mathbf{F} \quad \overset{\mathbb{Z}_3}{\longrightarrow} \quad SU(2)+7\mathbf{F}$

  \item $SU(2)+6\mathbf{F} \quad \overset{\mathbb{Z}_2}{\longrightarrow} \quad SU(2)+7\mathbf{F}$
\end{itemize}

%------------------------
\subsection{More examples}\label{sec:3.3}
In this subsection, we list some of representative or intriguing theories of the $\mathbb{Z}_n$ S-folds on face.

\paragraph{$SU(2N)_0$ theory.}
The 5d $SU(2N)_0$ theory at the CS level zero has a $\mathbb{Z}_2$ symmetry and the $\mathbb{Z}_2$ fixed point lies at the center of the face in the middle of the corresponding brane web. According to our prescription, the $\mathbb{Z}_2$ quotient of this theory leads to the $Sp(N)+4\mathbf{F}$ theory or the $SU(N+1)_{N+1}+4\mathbf{F}$ theory which is dual to the former theory. This quotient of the brane web is illustrated in Figure  \ref{fig:Z2gauging-O7}. The brane web result is consistent with the prepotential calculation, 
\begin{align}
    6\mathcal{F}_{SU(2N)_0/\mathbb{Z}_2} = \frac12 \times 6\mathcal{F}_{SU(2N)_0}\Big|_{\phi_{2N-1}=\phi_1,~ \phi_{2N-2}=\phi_2, \cdots,~ \phi_{N+1}=\phi_{N-1}} =6\mathcal{F}_{Sp(N)+4\mathbf{F}}.
\end{align}

As a concrete example, consider the  $SU(6)_0$ theory without flavor. We calculate the prepotential for the theory after the $\mathbb{Z}_2$ S-folding as 
\begin{align}
    6\mathcal{F}_{SU(6)_0}
    \quad \overset{\mathbb{Z}_2}{\longrightarrow} &\quad  \frac12 \times 6\mathcal{F}_{SU(6)_0}(\phi_1,\cdots,\phi_5)\Big|_{\phi_{5}=\phi_1,~ \phi_{4}=\phi_2}\cr
    &=8\phi_1^3+6\phi_1^2\phi_2-12\phi_1\phi_2^2+8\phi_2^3-6\phi_2\phi_3^2+4\phi_3^3\cr
    &=6\mathcal{F}_{Sp(3)+4\mathbf{F}} (\phi_1,\phi_2,\phi_3)\ .
\end{align}
The result agrees with the prepotential of the $Sp(3)+4{\bf F}$ theory having $SO(8)\times U(1)_I$ global symmetry which we proposed for the $\mathbb{Z}_2$ gauged theory.

It is straightforward to generalize this $\mathbb{Z}_2$ quotient to the brane webs containing $2n$ fundamental flavors. One readily finds that the $\mathbb{Z}_2$ quotient of the $SU(2N)_0+2n{\bf F}$ theory gives rise to the $Sp(N)+(n+4){\bf F}$ theory and therefore
\begin{align}
    6\mathcal{F}_{SU(2N)_0+2n\mathbf{F}}
    \quad \overset{\mathbb{Z}_2}{\longrightarrow} \quad  6\mathcal{F}_{Sp(N)+(n+4)\mathbf{F}}\ .
\end{align}

\begin{figure}
\centering
\begin{minipage}{.3\textwidth}
  \centering
     \includegraphics[scale=.35]{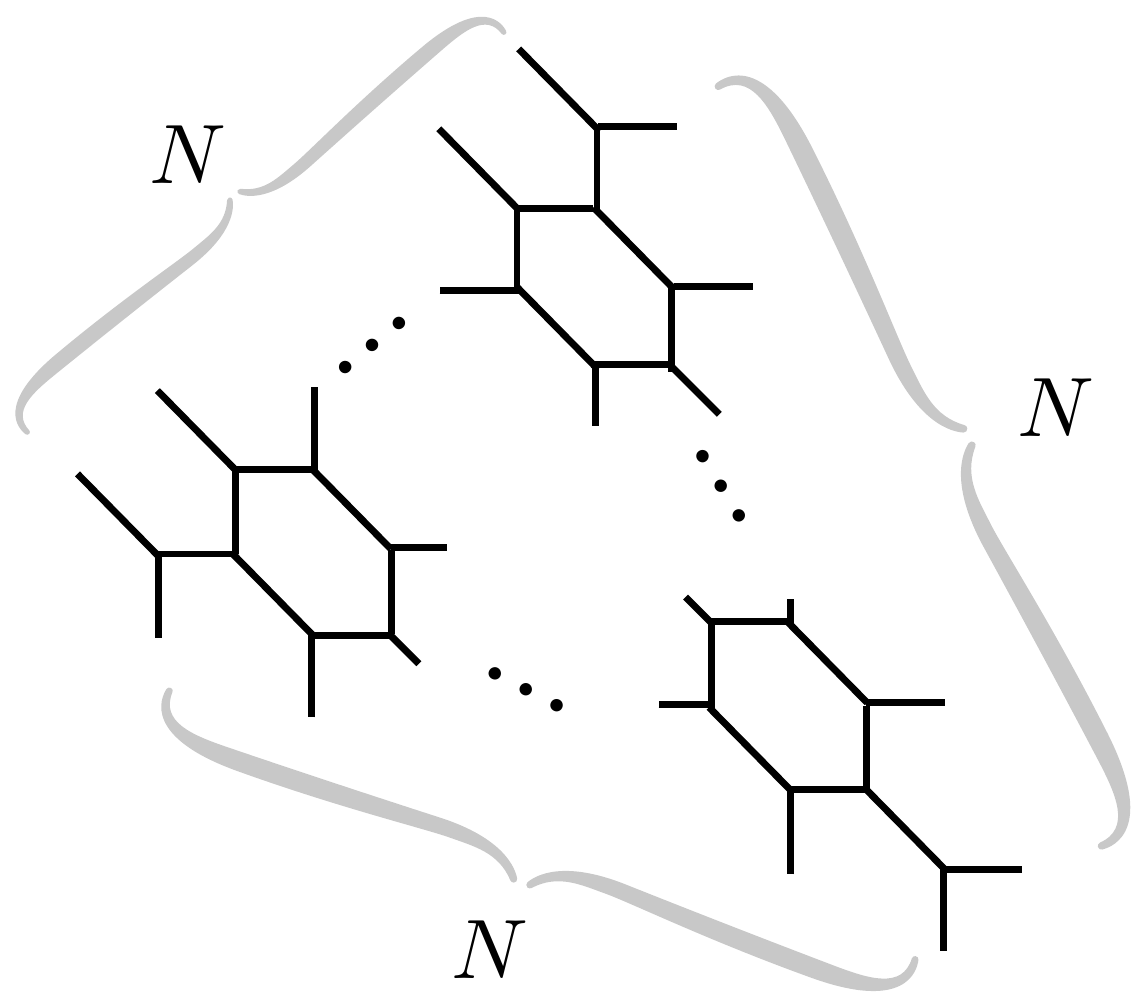}
     %\caption{5-brane webs for $T_N$-theories.}
     %\label{fig:Tn-web}
\end{minipage}%
\begin{minipage}{.7\textwidth}
  \centering
  \includegraphics[scale=.35]{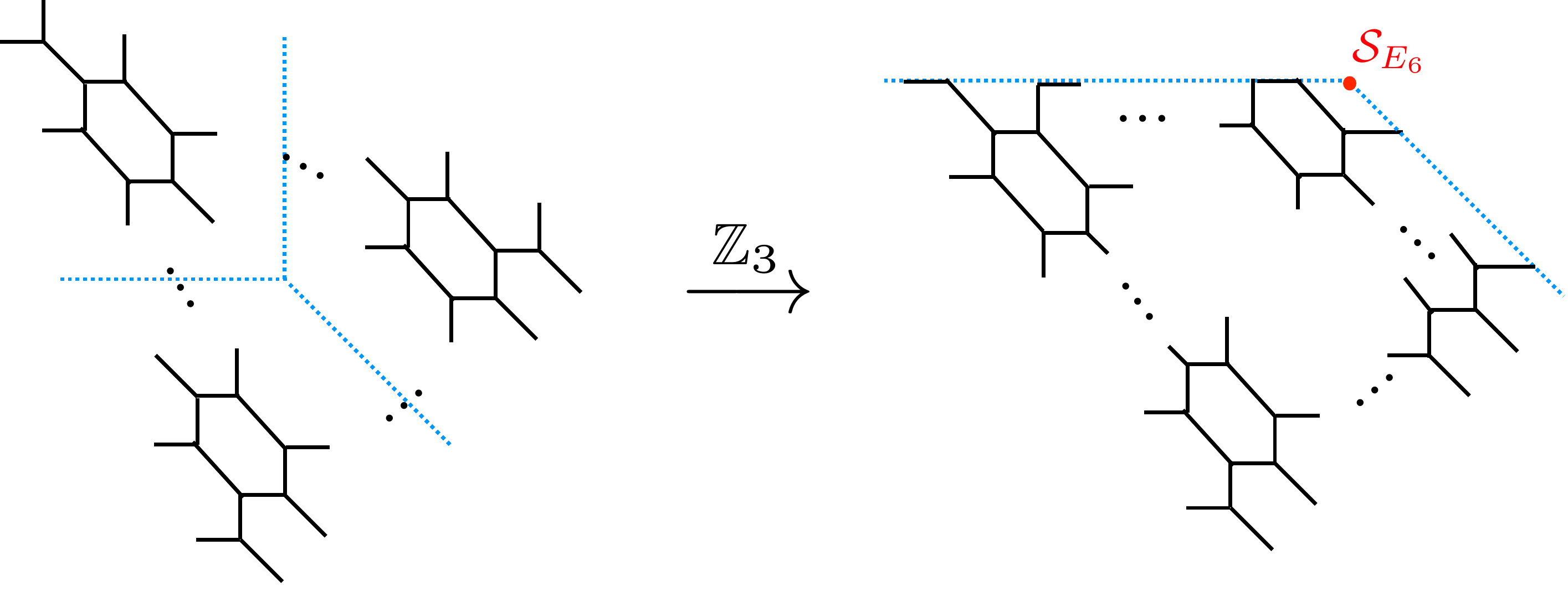}
     %\caption{$\mathbb{Z}_3$ gauging of the $T_{3k}$ theory.}
     %\label{fig:T3N-fold}
\end{minipage}
\caption{A 5-brane web of $T_{N}$ theory and $\mathbb{Z}_3$ quotient of the $T_{3k}$ theory.}
     \label{fig:T3N-fold}
\end{figure}

\paragraph{$T_{N}$ theory with $N=0$ mod $3$.}
An interesting family of 5d SCFTs having a discrete rotational symmetry of brane webs are $T_N$ theories  \cite{Gaiotto:2009we,Benini:2009gi} as depicted in Figure \ref{fig:T3N-fold}. It has a quiver gauge theory description given as
 $T_N=[N]-SU(N-1)-\cdots-SU(2)-[2]$ \cite{Hayashi:2013qwa,Hayashi:2014hfa}, where $[n]$ denotes $n$ flavors. The $T_N$ theory is engineered in M-theory compactified on an orbifold 3-fold $\mathbb{C}^3/\mathbb{Z}_N\times \mathbb{Z}_N$ where two $\mathbb{Z}_N$ factors are generated by phase rotations of three complex coordinates by $(\omega,\omega^{-1},1)$ and $(1,\omega,\omega^{-1})$ with $\omega^N=1$. The orbifold construction tells us that this theory has an $(N-1)(N-2)/2$ dimensional Coulomb branch of the moduli space and possesses a large global symmetry $SU(N)\times SU(N)\times SU(N)$ at the CFT fixed point. The orbifold singularity can be resolved by turning on the Coulomb branch parameters and mass parameters for the global symmetry. At a particular point of the Coulomb branch, this theory can be represented by a 5-brane web, 
containing $(N-1)(N-2)/2$ honeycomb-shaped faces as well as $3N$ external 5-branes \cite{Benini:2009gi} as Figure \ref{fig:T3N-fold}.

The $T_N$ theory, at $\tau = e^{2\pi i/3}$, has a $\mathbb{Z}_3$ symmetry rotating the 5-brane web by 120 degrees. The three $SU(N)$ global symmetries are permuted under the $\mathbb{Z}_3$ action. One finds that the $\mathbb{Z}_3$ action has a fixed point in the middle of a compact face when $N=0$ mod 3, whereas when $N\neq0$ mod 3, the $\mathbb{Z}_3$ fixed point is placed at a junction of 5-branes. In this section, we shall discuss the S-folding of the first cases for $N=0$ mod 3. The S-folding of the other cases will be discussed in the next section.

The $\mathbb{Z}_3$ folding of the $T_{3k}$ theories has been studied previously in \cite{Acharya:2021jsp}. In that work, the $\mathbb{Z}_3$ S-fold is realized by a $\mathbb{Z}_3$ quotient of the 5-brane web together with a stack of 7-branes carrying an $E_6$ algebra introduced at the fixed point. This is illustrated in Figure \ref{fig:T3N-fold}.
Once we replace the stack of 7-branes by $\bf A^5BCB$ 7-branes with monodromy cuts, their prescription is in accordance with our prescription for these cases.
The 5d theory after the $\mathbb{Z}_3$ quotient becomes a 5d SCFT with rank $(3k^2-3k+2)/2$ and global symmetry $SU(3k)\times E_6$ \cite{Acharya:2021jsp}.

For example, the $T_3$ theory when $k=1$ is described by the 5d $SU(2)$ gauge theory with 5 fundamentals, and the $\mathbb{Z}_3$ quotient of this theory has already been discussed in the rank-1 examples above. See the 3rd example of Figure \ref{fig:rank1-Z234}. The resulting theory is the $SU(2)$ gauge theory with 7 fundamental hypers. The prepotentials of two theories before and after the $\mathbb{Z}_3$ quotient are related as
\begin{align}
  6\mathcal{F}_{SU(2)+5\mathbf{F}}= 3\phi^3  \quad \overset{\mathbb{Z}_3}{\longrightarrow} \quad  %6\mathcal{F}_{SU(2)+7\mathbf{F}} =
  \frac{1}{3}\times 6\mathcal{F}_{SU(2)+5\mathbf{F}}   = \phi^3 = 6\mathcal{F}_{SU(2)+7\mathbf{F}} \ .
\end{align}
In the CFT limit, the $SU(3)\times E_6$  global symmetry is enhanced to $E_8$.

\begin{figure}[t]
     \centering
     \includegraphics[scale=.25]{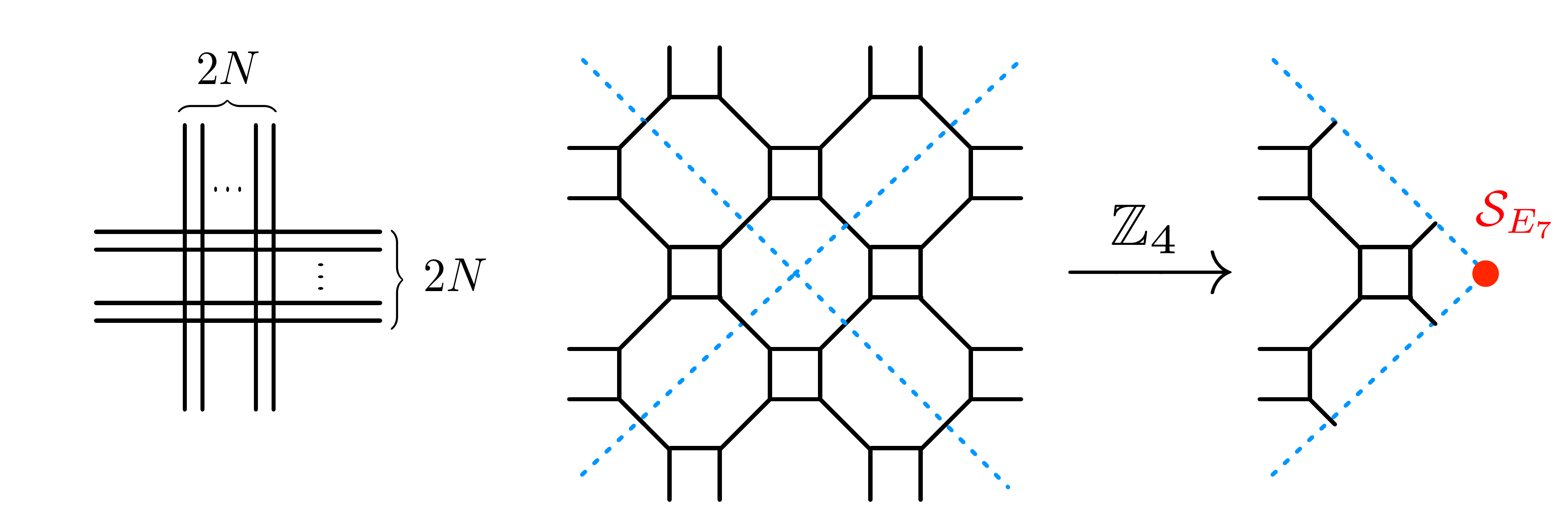}
     \caption{5d $+_{2N, 2N}$-theory and the $\mathbb{Z}_4$ quotient of the $+_{4,4}$-theory}
     \label{fig:higher-+_2N-Z4}
 \end{figure}

\begin{figure}[t]
     \centering
     \includegraphics[scale=.25]{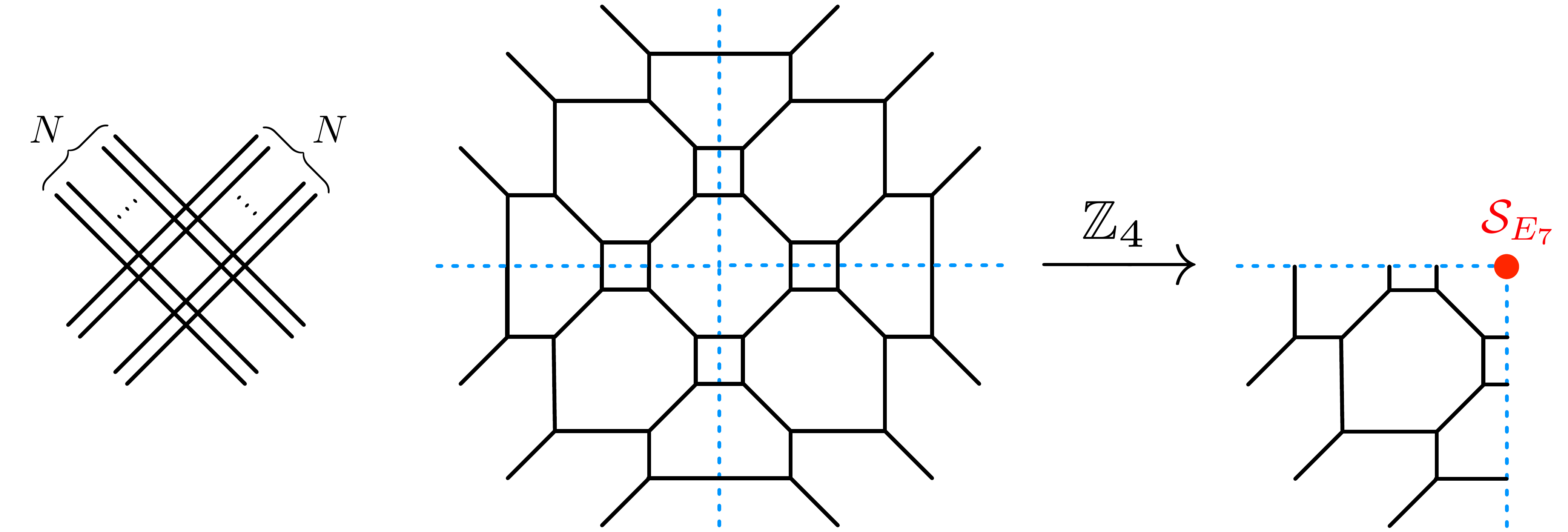}
     \caption{5d $X_{N,N}$-theory and the $\mathbb{Z}_4$ quotient of the $X_{3,3}$-theory}
     \label{fig:higher-X_N-Z4}
 \end{figure}

\paragraph{$+_{N,N}$ and $X_{N, N}$ theories.}
Typical quiver theories of  $\mathbb{Z}_4$ symmetry are $+_{N,N}$-theory \cite{Aharony:1997bh} and $X_{N, N}$-theory \cite{Bergman:2018hin}, which are named after the shape of asymptotic 5-brane configurations, as depicted in Figure 
\ref{fig:higher-+_2N-Z4} and in Figure  \ref{fig:higher-X_N-Z4}, respectively.
The $+_{N, N}$-theory is composed of 
$(N-1)$ copies of $SU(N)$ theories of the $N$ flavors on the first and last gauge nodes, and the corresponding web has  $N$ D5-branes and $N$ NS5-branes. The theory has an $(N-1)^2$ dimensional Coulomb branch and $SU(N)^4\times U(1)$ global symmetry. Depending on whether $N$ is even or odd, one can apply the S-folding on a face or on a line, respectively. Here, we only consider the even case, $+_{2N, 2N}$-theory, 
\begin{align}
    [2N]-\underbrace{SU(2N) - SU(2N) - \cdots - SU(2N)}_\text{$2N-1$ nodes}-[2N], \label{eq:_2N2N}
\end{align}
which has $SU(2N)^4\times U(1)$ global symmetry.

One can now implement the $\mathbb{Z}_4$ folding on a face by inserting the 7-brane configuration associated with $\mathbb{Z}_4$, i.e. $\mathcal{S}_{E_7}= {\bf A^6BCB}$, at the center of the web, as shown in Figure \ref{fig:higher-+_2N-Z4}. 
After the $\mathbb{Z}_4$ quotient, we obtain a 5d theory with a Coulomb branch of dimension $N^2-N+1$. This theory has $SU(2N)\times E_7$ symmetry which is consistent with the flavor rank prediction in \eqref{eq:flavor-F-gauging}. For example, the $+_{2,2}$-theory is actually the $SU(2)+4\mathbf{F}$ theory and its $\mathbb{Z}_4$ S-fold leads to the $SU(2)+7{\bf F}$ showing $E_8\supset E_7\times SU(2)$ symmetry enhancement as discussed in section \ref{sec:rank-1}. Higher rank generalization of the $\mathbb{Z}_4$ S-fold is straightforward as one can see from Figure \ref{fig:higher-+_2N-Z4} for $+_{4,4}$-theory, $[4]-SU(4)-SU(4)-SU(4)-[4]$.

The $X_{N,N}$ theory is another interesting quiver theory with a $\mathbb{Z}_4$ symmetry, 
\begin{align}
    SU(2)\!-\!SU(4)\!-\!\cdots\!-\!SU(2N\!-\!2)\!-\!SU(2N)\!-\!SU(2N\!-\!2)\!-\!\cdots\!-\!SU(2)\, .
\end{align}
This theory has $SU(N)^4\times U(1)$ global symmetry. As shown in  Figure  \ref{fig:higher-X_N-Z4}, one can implement the $\mathbb{Z}_4$ folding on a face by inserting the 7-branes of type $\mathcal{S}_{E_7}= {\bf A^6BCB}$ in the middle of the Coulomb branch of the  $SU(2N)$ center node. After the $\mathbb{Z}_4$ quotient, we obtain a 5d theory having a $(2N^2-2N+1)$ dimensional Coulomb branch and $SU(N)\times E_7$ global symmetry, which is also consistent with \eqref{eq:rank-F-gauging} and \eqref{eq:flavor-F-gauging}. 
As a concrete example, consider $X_{3,3}$-theory, whose 5-brane configuration is depicted in Figure \ref{fig:higher-X_N-Z4}. For instance, after the $\mathbb{Z}_4$ quotient of the $X_{3,3}$-theory, we obtain a 5d theory with a 4-dimensional Coulomb branch and $SU(3)\times E_7$ global symmetry.

\begin{figure}[t]
     \centering
     \includegraphics[scale=.23]{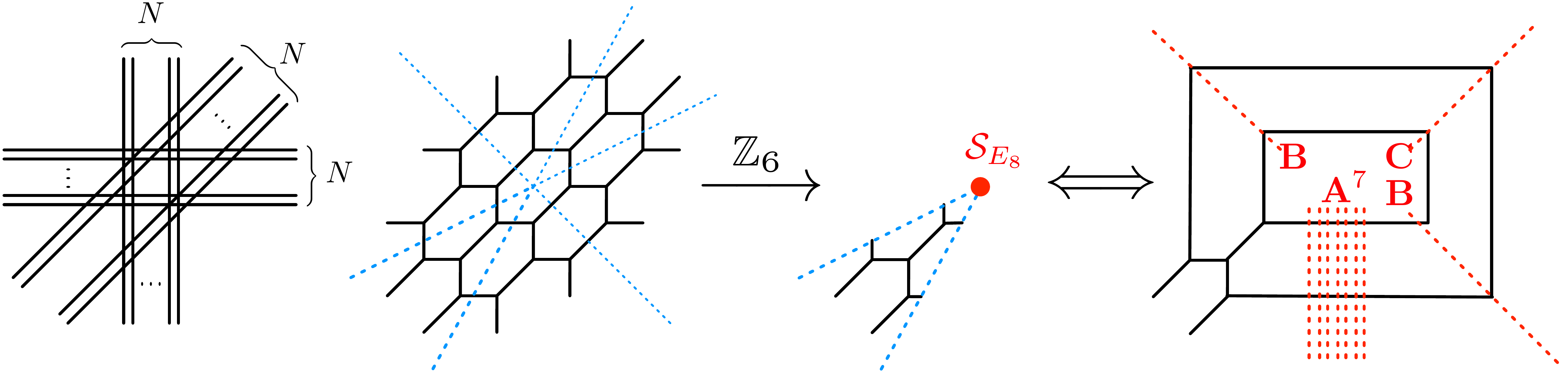}
     \caption{
    $\pluslash_N$-theory, which is of $\mathbb{Z}_6$ symmetry (which also has $\mathbb{Z}_2, \mathbb{Z}_3$). The $\mathbb{Z}_6$ quotient of the $\pluslash_2$-theory,  $[2]\!-\!SU(3)\!-\!SU(4)\!-\!SU(3)\!-\![2]$, becomes 5d $Sp(2)+1\mathbf{AS}+7\mathbf{F}$.}
     \label{fig:higher-rank-Z6}
 \end{figure}

\paragraph{$\pluslash_N$-theory.}
A representative of 5d SCFTs with a $\mathbb{Z}_6$ discrete symmetry having a fixed point in the middle of the center face of the brane web is the $\pluslash_N$-theory described by the quiver
\begin{align}\label{eq:Z6quiver}
    [N]\!-\!SU(N\!+\!1)\!-\!\cdots\!-\!SU(2N\!-\!1)\!-\!SU(2N)\!-\!SU(2N\!-\!1)\!-\!\cdots\!-\!SU(N\!+\!1)\!-\![N],
\end{align}
which has $SU(N)^6\times U(1)^3$ global symmetry  \cite{Bergman:2018hin}. 
The 5-brane webs for the $\pluslash_N$-theory are given in Figure \ref{fig:higher-rank-Z6}, which clearly reveal a $\mathbb{Z}_6$ symmetry. 

The $\mathbb{Z}_6$ quotient of this theory gives rise to a 5d SCFT with a Coulomb branch of dimension $(N^2-N+2)/2$ and $SU(N)\times E_8$ global symmetry which is consistent with \eqref{eq:rank-F-gauging} and \eqref{eq:flavor-F-gauging}. In particular,  the $\pluslash_2$ theory when $N=2$, which is a quiver theory $[2]\!-\!SU(3)\!-\!SU(4)\!-\!SU(3)\!-\![2]$, after the $\mathbb{Z}_6$ quotient leads to 5d $Sp(2)+1\mathbf{AS}+7\mathbf{F}$ theory, as shown in Figure \ref{fig:higher-rank-Z6}. 
This theory also has $\mathbb{Z}_2$ and $\mathbb{Z}_3$ symmetries. We can gauge them by inserting the 7-branes of $D_4,E_6$ singularities, respectively.

%------------------------

\section{S-fold on Line or Vertex}\label{sec:4}
In this section, we discuss the S-folding of discrete symmetries in the second class. The $\mathbb{Z}_n$ symmetries of the second class have a fixed point in the middle of an edge or at a vertex of intersecting 5-branes. The quotient with these symmetries is more involved since we need to take into account the 5-branes touching the fixed point.
We will propose a simple prescription to perform the S-fold in the second class and present some interesting examples which will support our prescription.

%--------------------
\subsection{Prescription}

The idea of S-folds in the second class is almost the same as that in the first class. First of all, after the $\mathbb{Z}_n$ quotient,  the fundamental domain of the 5-brane web is reduced to a $1/n$ slice of the original $(p,q)$-plane. 

It follows that extra 7-branes should be inserted at the $\mathbb{Z}_n$ fixed point in such a way that their monodromies are consistent with the brane configuration in the fundamental domain. We claim that the additional 7-branes inserted at the fixed point should be as follows:
\begin{itemize}
  \item $\mathbb{Z}_2$ quotient : $\mathcal{S}_{D_4}={\bf {A}^4 BC}$ 7-branes 

  \item $\mathbb{Z}_3$ quotient : $\mathcal{S}_{E_6}={\bf {A}^5 BCB}$ 7-branes 

  \item $\mathbb{Z}_4$ quotient : $\mathcal{S}_{E_7}={\bf {A}^6 BCB}$ 7-branes 

\end{itemize}
These 7-branes are the same as those for the $\mathbb{Z}_n$ symmetries of the first class studied in the previous section. We remark that the 5-brane web configuration having $\Z_6$ symmetry of the second class always turns to the first class by turning on Coulomb parameters.

\begin{figure}[t]
     \centering
     \includegraphics[scale=.50]{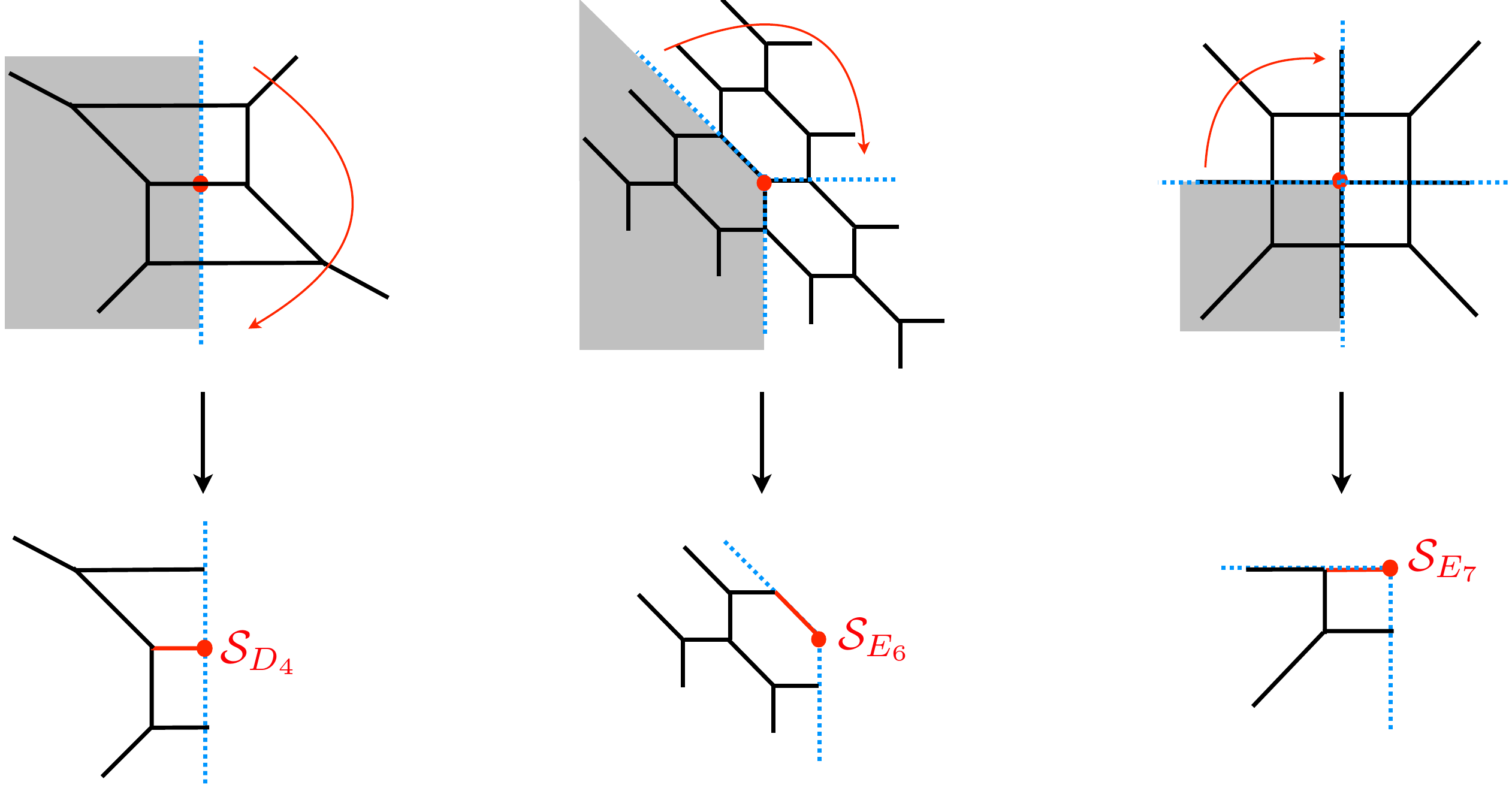}
     \caption{The prescriptions for the $\mathbb{Z}_2,\mathbb{Z}_3,\mathbb{Z}_4$ S-folds (from left to right). After the $\mathbb{Z}_n$ quotients, the brane webs reduce to the bottom three diagrams. The red line in the bottom diagrams denotes the 5-brane ending on one of extra 7-branes at the singularity.}
     \label{fig:fig-rule-vertex}
 \end{figure}

We still need to take care of the 5-branes intersecting at the $\mathbb{Z}_n$ fixed point in the brane web. When the $\mathbb{Z}_n$ quotient acts, the $n$ 5-branes meeting at the fixed point fold up and become a single  5-brane suspended between the fixed point and the other part of the 5-brane web. Suppose the folded 5-brane carries a 5-brane charge $(p,q)$. Then the  5-brane charge conservation demands that this 5-brane should end on one of the  7-branes at the singularity with charge $[p,q]$. Consequently, the final brane configuration, after the $\mathbb{Z}_n$ quotient on a line or at a vertex, is the $1/n$ subregion of the original brane web together with extra 7-branes inserted at the $\mathbb{Z}_n$ singularity and a (full) 5-brane ending on one of the extra 7-branes. This is illustrated in Figure \ref{fig:fig-rule-vertex}.   

The rank of the gauge group is reduced by a factor of $1/n$ after the $\mathbb{Z}_n$ quotient since the quotient simply folds $n$ faces in a $\mathbb{Z}_n$ orbit into a single face,
\begin{align}\label{eq:gauge-line}
  r(\mathcal{T}/\mathbb{Z}_n) = \frac{r(\mathcal{T})}{n}  \ .
\end{align}
For the same reason, the total area of the compact faces is reduced by a factor of $1/n$. This forces the cubic prepotential of the $\mathbb{Z}_n$ gauged theory $\mathcal{T}/\mathbb{Z}_n$ to also be reduced to $1/n$ of the original prepotential with K\"ahler parameters within $\mathbb{Z}_n$ orbits identified with each other as shown in (\ref{eq:gauged-F}).

The rank of the flavor group of the $\mathbb{Z}_n$ gauged theory is 
\begin{align}\label{eq:flavor-line}
  &r_F(\mathcal{T}/\mathbb{Z}_n) =  \frac{r_F(\mathcal{T})+3}{n}-1+r_F(\mathcal{S}) -\kappa\ , 
\end{align}
where $r_F(\mathcal{S})=4,6,7$ for $\mathbb{Z}_{2,3,4}$, respectively. Here, $\kappa=1$ if the 5-brane attached to a 7-brane at the singularity can be pulled out of the 5-brane web by the Hanany-Witten move, or $\kappa=0$ otherwise.

The $\mathbb{Z}_2$ quotient can also be performed by placing an O7$^-$ plane at the fixed point in the middle of a D5-brane. In this case, the singularity must accommodate four additional D7-branes for the twisted sectors and one of these D7-branes should be attached to the 5-brane folded in half. Resolving the O7$^-$-plane by a pair of ${\bf B}$ and ${\bf C}$ 7-branes will then lead to our prescription above for the $\mathbb{Z}_2$ quotient.

\begin{figure}[t]
     \centering
     \includegraphics[scale=.18]{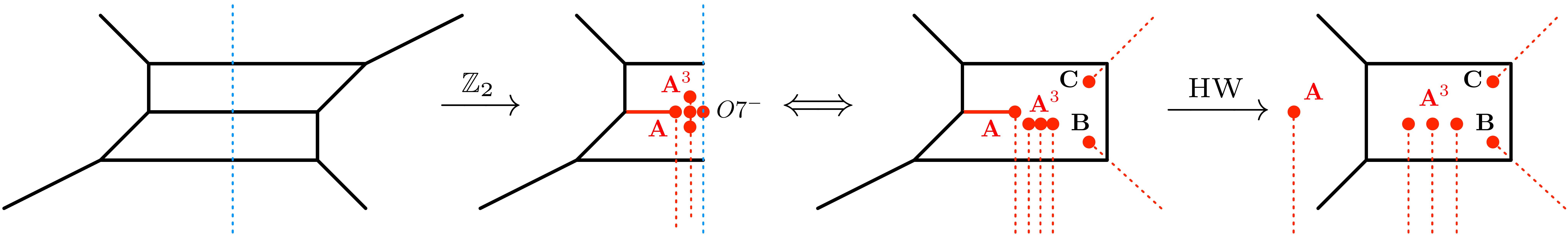}
     \caption{The $\mathbb{Z}_2$ quotient of 5d $SU(3)_0$ theory without flavor leading to  $SU(2)+3\mathbf{F}$. Here a free hypermultiplet associated with a 7-brane $\mathbf{A}$ outside of the 5-brane web is decoupled by Hanany-Witten (HW) moves. }
     \label{fig:pureSU3-Z2}
 \end{figure}
A simple example of the S-fold on a line is the $\mathbb{Z}_2$ quotient of the 5d $SU(3)_0$ gauge theory with no flavor. From the 5-brane perspective in Figure \ref{fig:pureSU3-Z2}, the $\mathbb{Z}_2$ symmetry is the rotation of the brane web by 180 degrees and we can gauge it as follows.
We insert four ${\bf A}$-branes on top of an O7$^-$-plane at the singularity and the halved 5-brane of the middle edge is attached to one of the $\bf A$-branes. The O7$^-$-plane is resolved into the ${\bf B}$ and ${\bf C}$ 7-branes in the third diagram of the Figure \ref{fig:pureSU3-Z2}. One can also readily see that an $\bf A$ 7-brane attached to the middle edge can be pulled out of the 5-brane loop via the Hanany-Witten move. Taking into account the charge change of 7-branes going through the monodromy cut of the $\bf A$ 7-brane attached to the middle, one finds that the remaining 7-brane configuration inside the 5-brane loop becomes that of $SU(2)+3\mathbf{F}$, as depicted in the last diagram in Figure 
\ref{fig:pureSU3-Z2}. The $SU(2)+3\mathbf{F}$ theory  in the UV CFT limit has an $E_4=SU(5)$ enhanced global symmetry which agrees with our rank formulae in (\ref{eq:gauge-line}) and in (\ref{eq:flavor-line}). 

The same conclusion can be drawn using the cubic prepotential. The original   $SU(3)_0$ theory without flavor has the cubic prepotential in (\ref{eq:F1F1}). The $\mathbb{Z}_2$ folding on the  $SU(3)_0$ theory with zero flavor then results in the prepotential change $\mathcal{F}\to\mathcal{F}/2$ with the identification of the K\"ahler parameters as  $\phi_1=\phi_2=\phi$. This leads to the prepotential of the $\mathbb{Z}_2$ gauged theory given by
\begin{align}
     6\mathcal{F}_{SU(3)_0/\mathbb{Z}_2} = \frac12\times 6\mathcal{F}_{SU(3)_0}(\phi_1,\phi_2)\big|_{\phi_1=\phi_2=\phi} = 5\phi^3  \ .
\end{align}
Indeed, the resulting prepotential is precisely that of the $SU(2)+3\mathbf{F}$ theory.

Let us now exhibit more examples of the $\mathbb{Z}_n$ quotients on a line or a vertex.

\subsection{Examples} 
\begin{figure}[t]
     \centering
     \includegraphics[scale=.154]{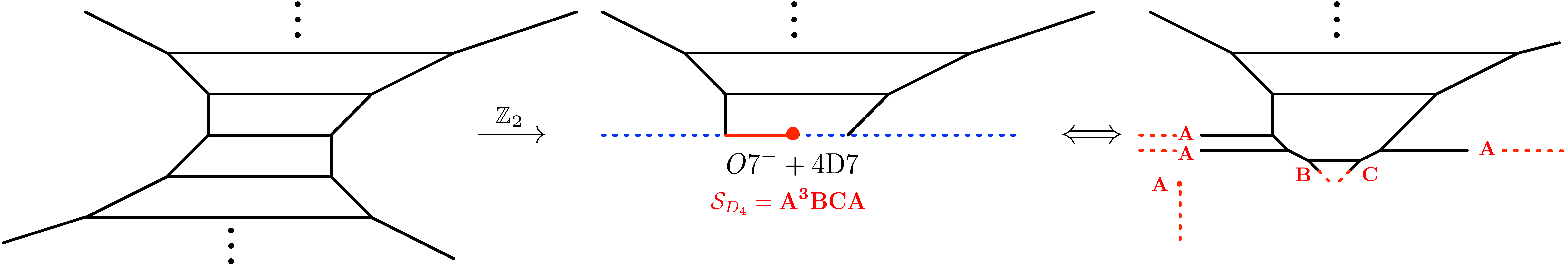}
     \caption{The $\mathbb{Z}_2$ quotient of   $SU(2N+1)_0$ theory with no flavor. The resulting theory is $Sp(N)+3\mathbf{F}$, after the Hanany-Witten transition associated with a D7-brane (or $\mathbf{A}$).}
     \label{fig:SOodd-Z2}
 \end{figure}
\paragraph{$SU(2N+1)_0$ theory.}
We first consider the $\mathbb{Z}_2$ quotient of the $SU(2N+1)_0$ theory which is a simple generalization of the $\mathbb{Z}_2$ quotient of the $SU(3)_0$ theory above. The resulting theory after the $\Z_2$ folding is given by
$Sp(N)+3\mathbf{F}$, as depicted in Figure \ref{fig:SOodd-Z2}.   
One can check this explicitly from the prepotential calculation
\begin{align}
    6\mathcal{F}_{SU(2N+1)_0/\mathbb{Z}_2} = \frac12 \times 6\mathcal{F}_{SU(2N+1)_0}\Big|_{\phi_{2N}=\phi_1,~ \phi_{2N-1}=\phi_2, ~\cdots,~ \phi_{N+1}=\phi_{N}} =6\mathcal{F}_{Sp(N)+3\mathbf{F}}.
\end{align}
The global symmetry after the quotient is $SO(6)\times U(1)$ which is consistent with \eqref{eq:flavor-line}.

As a concrete example, consider the  $SU(5)_0$ theory without flavor. 
We calculate the prepotential for the theory after the $\mathbb{Z}_2$ folding as 
\begin{align}
    6\mathcal{F}_{SU(5)_0}
    \quad \overset{\mathbb{Z}_2}{\longrightarrow} &\quad  \frac12 \times 6\mathcal{F}_{SU(5)_0}(\phi_1,\cdots,\phi_4)\Big|_{\phi_{4}=\phi_1,~ \phi_{3}=\phi_2}\cr
    &=8\phi_1^3+3\phi_1^2\phi_2-9\phi_1\phi_2^2+5\phi_2^3\cr
    &=6\mathcal{F}_{Sp(2)+3\mathbf{F}} (\phi_1,\phi_2).
\end{align}
This result agrees with the 5-brane web description given in Figure \ref{fig:SOodd-Z2}.

It is also straightforward to add more flavors to this theory while maintaining the $\mathbb{Z}_2$ symmetry. We then easily find that the $\mathbb{Z}_2$ quotient of the $SU(2N+1)_0+2n{\bf F}$ theory gives rise to the $Sp(N)+(n+3){\bf F}$ theory and therefore
\begin{align}
    6\mathcal{F}_{SU(2N+1)_0+2n\mathbf{F}}
    \quad \overset{\mathbb{Z}_2}{\longrightarrow} \quad  6\mathcal{F}_{Sp(N)+(n+3)\mathbf{F}}\ .
\end{align}

\paragraph{$T_N$ theory with $N\neq 0$ mod $3$.}
\begin{figure}[t]
     \centering
     \includegraphics[scale=.44]{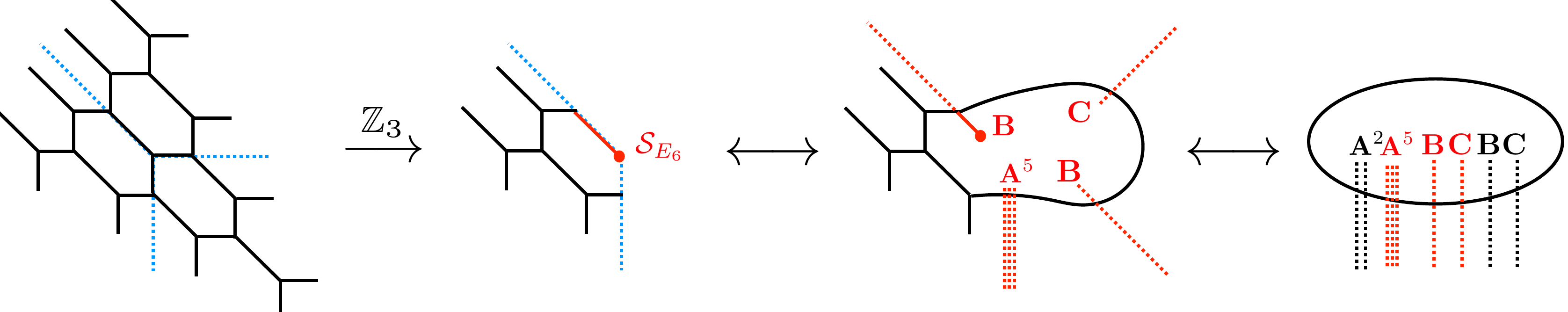}
     \caption{The $\mathbb{Z}_3$ quotient of $T_4$ theory.
     We insert $\bf A^5{B}{C}{B}$ at the vertex connected to the $(1,-1)$ 5-brane line. This leads to a 7-brane configuration for $SU(2)+7{\bf F}$ theory which has an $E_8$ flavor symmetry in CFT limit.}
     \label{fig:T4}
 \end{figure}

The 5-brane configuration for the $T_N$ theory with $N\neq 0$ mod $3$ has a $\mathbb{Z}_3$ symmetry with a fixed point at a vertex of three 5-branes. We can gauge this $\mathbb{Z}_3$ symmetry using our S-folding prescription. The $\mathbb{Z}_3$ folding then introduces extra ${ \bf A^5BCB}$ type 7-branes carrying $E_6$ global symmetry to the fixed point. In addition, the three 5-branes joining at the vertex are folded into a single 5-brane attached to one of the extra 7-branes. In the end, we obtain a 5d SCFT of Coulomb branch dimension $(N-1)(N-2)/6$ and of global symmetry $SU(N)\times SO(10)\times U(1)$ for $N\neq 4$ or $E_8$ for $N=4$.  The global symmetry can be further enhanced in the UV limit.

We remark that the $\mathbb{Z}_3$ quotient associated with $2\pi/3$ rotation of the brane web for the $T_N$ theories with $N\neq 0$ mod $3$ has been studied in \cite{Acharya:2021jsp}. However, unlike for the $T_{3k}$ cases we discussed in Section \ref{sec:3.3}, our S-folding prescription for the $\mathbb{Z}_3$ symmetry differs from the proposal in that literature. The proposal in \cite{Acharya:2021jsp} is that the $\mathbb{Z}_3$ quotient introduces additional singularity with $SU(3)$ global symmetry and thus the gauged theory will have a flavor symmetry group of rank $N+1$. This is different from our prescription which predicts after the $\mathbb{Z}_3$ quotient a flavor symmetry of rank $N+5$ for $N>4$ or an $E_8$ flavor symmetry for $N=4$. We will show with concrete examples below that our prescription for the $\mathbb{Z}_3$ S-fold is consistent with the prepotential computation, which may provide  evidence in favor of our S-folding prescription, 
while the results of \cite{Acharya:2021jsp} do not correspond to the prepotentials of the $\mathbb{Z}_3$ gauged theories.

For example, our S-folding prescription for the $\mathbb{Z}_3$ symmetry in the $T_4$ brane web is shown in Figure \ref{fig:T4}. The $\mathbb{Z}_3$ quotient introduces the $E_6$ type 7-branes at the singularity.
Then, as depicted by a red line in the second and third diagrams, the folded 5-brane ends on a 7-brane of charge $[-1,1]$ inserted at the $\mathbb{Z}_3$ fixed point. Since this 5-brane can be removed by the Hanany-Witten transition, the $E_6$ flavor symmetry of the additional 7-branes at the singularity is reduced to $SO(10)$. Combined with the rank-3 flavor symmetry from the external 5-branes, the full flavor symmetry of the gauged theory enhances to $E_8$ flavor symmetry, which is shown in the last diagram of Figure \ref{fig:T4}. Therefore, we predict that the $\mathbb{Z}_3$ quotient of the $T_4$ theory gives rise to a 5d rank-1 SCFT with $E_8$ global symmetry which is the UV theory of the $SU(2)$ gauge theory with 7 fundamental hypermultiplets. 

This can be also checked by the $\mathbb{Z}_3$ quotient of the prepotential of the $T_4$ theory. We compute
\begin{align}
  6\mathcal{F}_{T_4}\! &= 5\phi_1^3+5\phi_2^3+5\phi_3^3\! -\!3\phi_1^2\phi_2\!-\!3\phi_1\phi_2^2\!-\!3\phi_1^2\phi_3\!-\!3\phi_1\phi_3^2\!-\!3\phi_3^2\phi_2\!-\!3\phi_3\phi_2^2+6\phi_1\phi_2\phi_3 \nonumber \\
  \overset{\mathbb{Z}_3}{\longrightarrow}& \qquad 6\mathcal{F}_{T_4/\mathbb{Z}_3} = \frac{1}{3}\times 6\mathcal{F}_{T_4}|_{\phi_1=\phi_2=\phi_3=\phi} = \phi^3 =6\mathcal{F}_{SU(2)+7\mathbf{F}} \ .
\end{align}
Therefore, the $\mathbb{Z}_3$ quotient acting on the prepotential of the $T_4$ theory agrees with the prepotential of the $SU(2)+7{\bf F}$ theory as expected.

\begin{figure}[t]
     \centering
     \includegraphics[scale=.14]{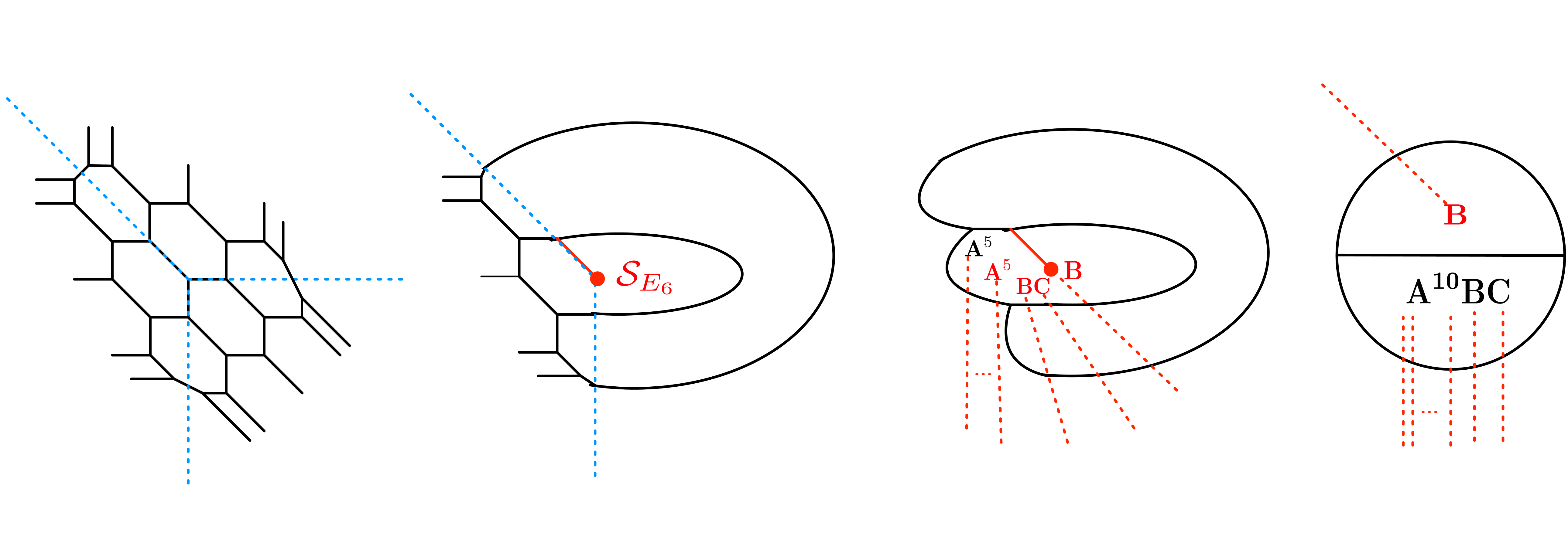}
     \caption{The $\mathbb{Z}_3$ quotient of $T_5$ theory.
     We insert $\bf A^5{B}{C}{B}$ at the vertex connected to the $(1,-1)$ 5-brane line. This leads to a 7-brane configuration for $SU(3)_\frac12+9\mathbf{F}$ \cite{Hayashi:2018lyv} or $Sp(2)+9\mathbf{F}$  \cite{Hayashi:2019jvx}, 
      which has an $SO(20)$ flavor symmetry.}
     \label{fig:T5-Z3-SO20}
 \end{figure}
Similarly, we can perform the $\mathbb{Z}_3$ S-folding in the $T_5$ brane web. As shown in Figure \ref{fig:T5-Z3-SO20}, we obtain a 5d rank-2 SCFT with $SO(20)$ flavor symmetry. This theory is the UV theory of the $SU(3)$ gauge theory with 9 fundamentals at Chern-Simons level $\frac{1}{2}$. In this case, the 5-brane ending on a 7-brane at the singularity cannot be removed by the Hanany-Witten move which is contrary to the $T_4$ example above. We can draw the same conclusion using the prepotential calculation. The prepotential of the $T_5$ brane web in the first diagram of Figure \ref{fig:T5-Z3-SO20} is
\begin{align}
  6\mathcal{F}_{T_5} =&~ 5(\phi_1^3+\phi_2^3+\phi_3^3)+6(\phi_4^3+\phi_5^3+\phi_6^3)-3\phi_1\phi_4(\phi_1+\phi_4)-3\phi_2\phi_4(\phi_2+\phi_4) \nonumber \\
  &-3\phi_1\phi_6(\phi_1+\phi_6)-3\phi_3\phi_6(\phi_3+\phi_6) -3\phi_2\phi_5(\phi_2+\phi_5)-3\phi_3\phi_5(\phi_3+\phi_5) \nonumber \\
  &-3\phi_4\phi_5(\phi_4+\phi_5)-3\phi_5\phi_6(\phi_5+\phi_6)-3\phi_4\phi_6(\phi_4+\phi_6) + 6\phi_1\phi_4\phi_6+6\phi_2\phi_4\phi_5 \nonumber \\
  &+6\phi_3\phi_5\phi_6+6\phi_4\phi_5\phi_6 \ .
\end{align}
After the $\mathbb{Z}_3$ quotient, we obtain
\begin{align}
  6\mathcal{F}_{T_5/\mathbb{Z}_3} &= \frac{1}{3}\times 6\mathcal{F}_{T_5}|_{\phi_1=\phi_2=\phi_3,\phi_4=\phi_5=\phi_6} \nonumber \\
  &=5\phi_1^3-6\phi_1^2\phi_4 + 2\phi_4^3 \ ,
\end{align}
which indeed agrees with the prepotential of the $SU(3)_{\frac{1}{2}}+9{\bf F}$ theory in the chamber realized by the Calabi-Yau 3-fold of ${\rm Bl}_3\mathbb{F}_0 \cup \rm{Bl}_6\mathbb{F}_2$.

\paragraph{Non-Lagrangian $B_N$ theory with $N\neq0$ mod $3$.}
\begin{figure}[t]
     \centering
     \includegraphics[scale=.18]{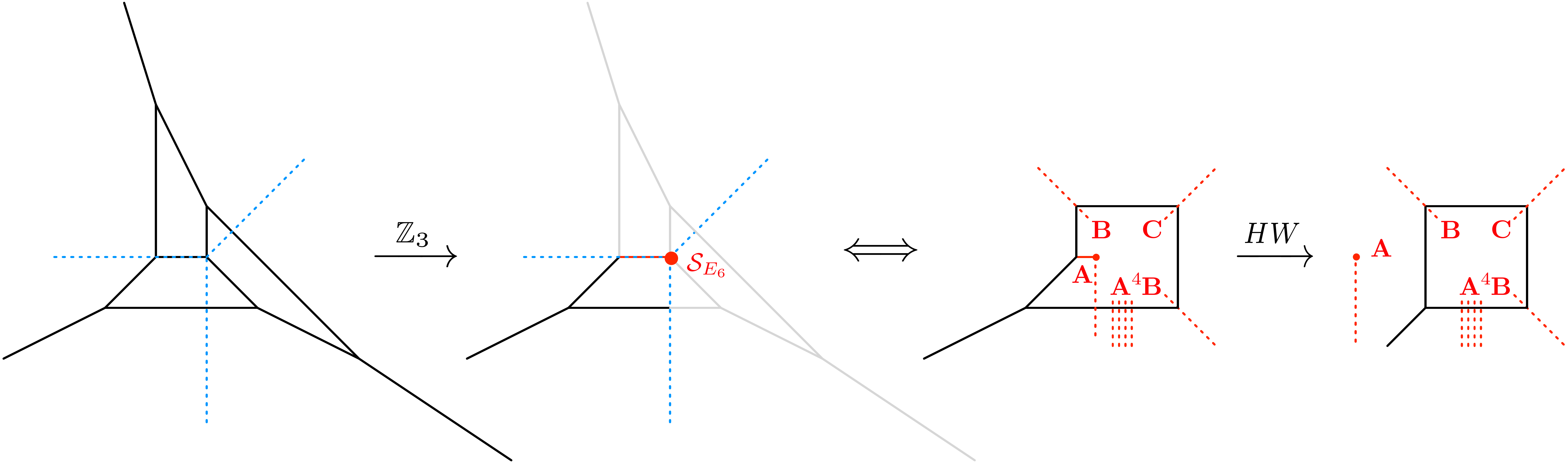}
     \caption{The $\mathbb{Z}_3$ quotient of a rank-3 non-Lagrangian theory, $B_4$, leading to $SU(2)+4\mathbf{F}$.}
     \label{fig:B4}
 \end{figure}
Some non-Lagrangian theories also have  $\mathbb{Z}_3$ symmetries. For example, the $B_N$ theory \cite{Morrison:2020ool} which can be obtained from the $T_N$ theory by decoupling hypermultiplets. As their $\Z_3$ folding can be done in the same way as done for the $T_N$ theory, we here introduce two simple examples of them, the $B_4$ and $B_5$ theories. They can be obtained from $T_4$ and $T_5$ theories by successive decoupling hypermultiplets while keeping the $\mathbb{Z}_3$ symmetry. 

One obtains the $B_4$ theory by decoupling 9 hypermultiplets from the $T_4$ theory. The corresponding 5-brane web is given in Figure \ref{fig:B4}, and the $\mathbb{Z}_3$ quotient leads to $SU(2)+4\mathbf{F}$ theory. This can be also confirmed from the prepotential as follows:
\begin{align}
  6\mathcal{F}_{B_4}\! &= 8\phi_1^3+8\phi_2^3+8\phi_3^3 \!-\!6\phi_1^2\phi_2\!-\!6\phi_2^2\phi_3\!-\!6\phi_3^2\phi_1\!+6\phi_1\phi_2\phi_3 \nonumber \\
  \overset{\mathbb{Z}_3}{\longrightarrow}& \qquad 6\mathcal{F}_{B_4/\mathbb{Z}_3} = \frac{1}{3}\times 6\mathcal{F}_{B_4}|_{\phi_1=\phi_2=\phi_3=\phi} = 4\phi^3 = 6\mathcal{F}_{SU(2)+4\mathbf{F}}  \ .
\end{align}
\begin{figure}[t]
     \centering
     \includegraphics[scale=.16]{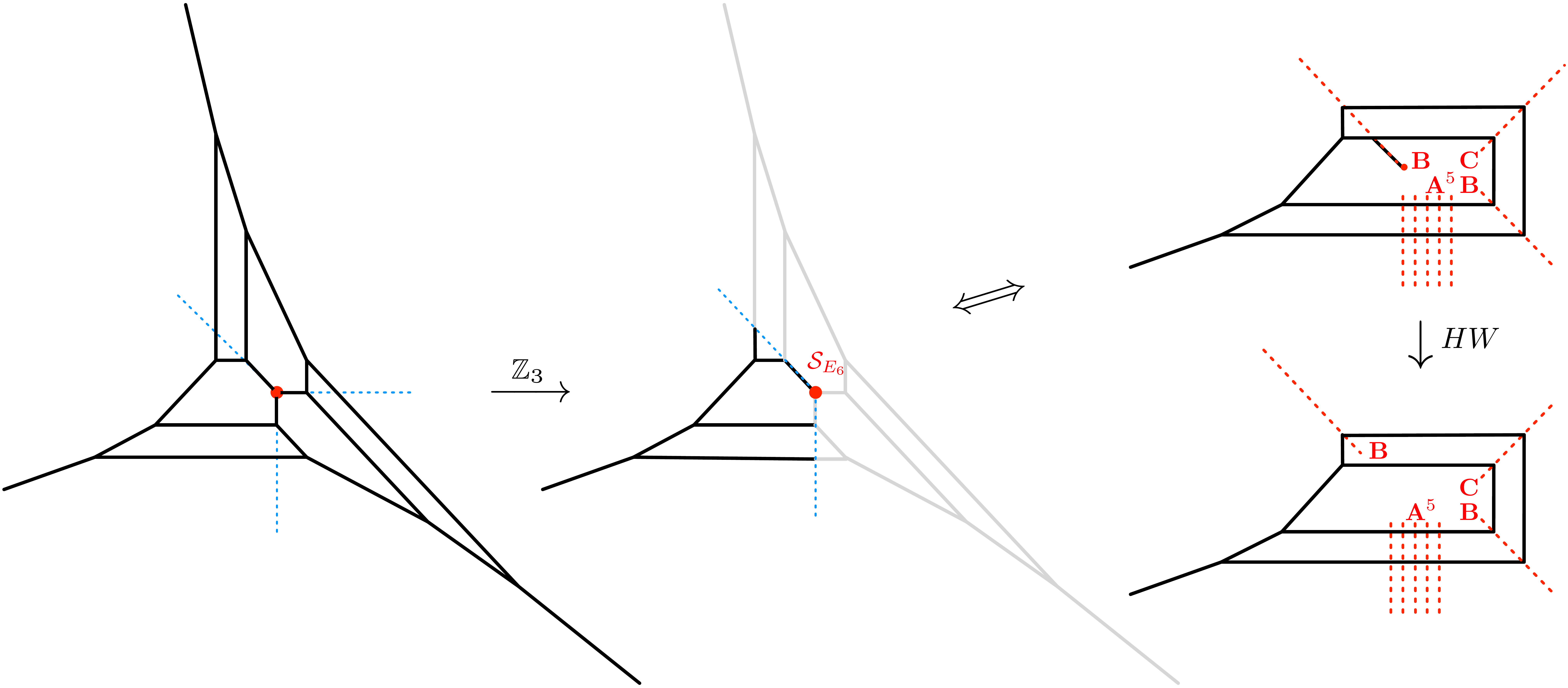}
     \caption{The $\mathbb{Z}_3$ quotient of a rank-6 non-Lagrangian theory, $B_5$, leading to 5d $Sp(2)+5\mathbf{F}$. }
     \label{fig:B5}
 \end{figure}

In a similar way, one obtains the $B_5$ theory by decoupling 12 hypermultiplets from the $T_5$ theory. The corresponding 5-brane web is given in Figure \ref{fig:B5}, and the $\mathbb{Z}_3$ quotient leads to $Sp(2)+5\mathbf{F}$ theory. This can be also also confirmed form the prepotential as follows: 
\begin{align}
  6\mathcal{F}_{B_5}\! =&~ 8(\phi_1^3+\!\phi_2^3+\!\phi_3^3)\!+\!7(\phi_4^3+\!\phi_5^3+\!\phi_6^3)\!-6\phi_1^2\phi_4\!-6\phi_2^2\phi_5\!-6\phi_3^2\phi_6\!-3\phi_4\phi_5(\phi_4+\!\phi_5) \nonumber \\
  &-3\phi_5\phi_6(\phi_5+\phi_6) -3\phi_6\phi_4(\phi_6+\phi_4) -9\phi_2\phi_4^2+3\phi_2^2\phi_4-9\phi_3\phi_5^2+3\phi_3^2\phi_5 \nonumber \\
  &-9\phi_1\phi_6^2+3\phi_1^2\phi_6+ 6\phi_1\phi_4\phi_6+6\phi_2\phi_4\phi_5+6\phi_3\phi_5\phi_6+6\phi_4\phi_5\phi_6 \ .
\end{align}
After the $\mathbb{Z}_3$ quotient, we obtain
\begin{align}
  6\mathcal{F}_{B_5/\mathbb{Z}_3} &= \frac{1}{3}\times 6\mathcal{F}_{B_5}|_{\phi_1=\phi_2=\phi_3,\phi_4=\phi_5=\phi_6} \nonumber \\
  &=8\phi_1^3-3\phi_1^2\phi_4-3\phi_1\phi_4^2 + 3\phi_4^3 \ ,
\end{align}
which agrees with the prepotential of the $Sp(2)+5\mathbf{F}$ theory (or $SU(3)_\frac52+5\mathbf{F}$) in the chamber $\phi_4>\phi_1$ corresponding to the geometry of $\mathbb{F}_1 \cup {\rm dP}_6$ glued along the $e$ curve with $e^2=-1$ in $\mathbb{F}_1$.

\paragraph{Pinwheel theory.} 

\begin{figure}[t]
     \centering
     \includegraphics[scale=.22]{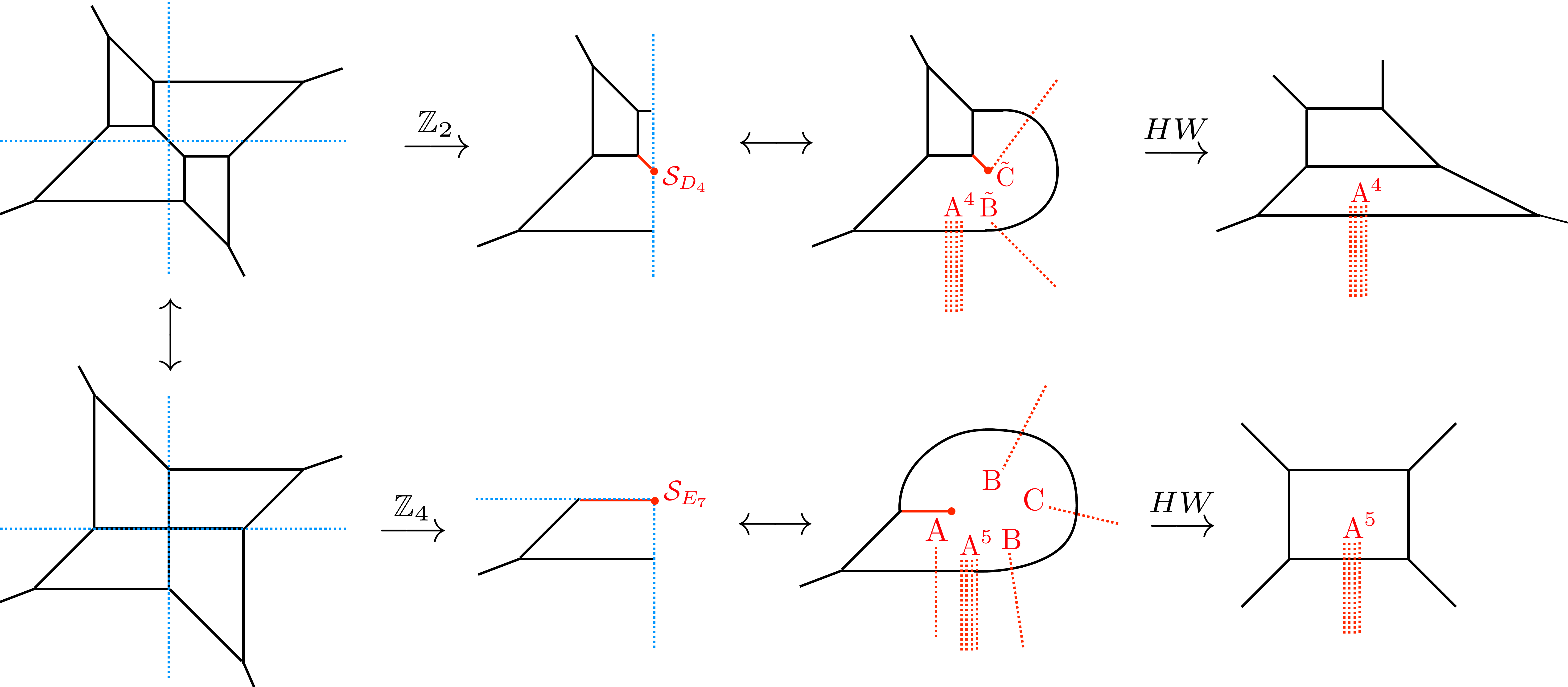}
     \caption{
    S-folds of a rank-4 non-Lagrangian theory called ``Pinwheel theory''. The $\mathbb{Z}_2$ quotient (top) gives rise to the $SU(3)_2+4{\bf F}$ theory, and the $\mathbb{Z}_4$ quotient (bottom) leads to the $SU(2)+5{\bf F}$ theory. Here, we used a different resolution of an $O7^-$ as $ \widetilde{\mathbf{B}}=[3, -1]$ and $ \widetilde{\mathbf{C}}=[1, -1]=\mathbf{B}$ so that the $\mathcal{S}_{D_4}$ is attached to the internal $(1,-1)$ 5-brane.}
     \label{fig:pinwheel}
\end{figure}

The simplest 5d SCFT having a $\mathbb{Z}_4$ symmetry of the second class is a rank-4 non-Lagrangian theory with $U(1)$ global symmetry so-called ``Pinwheel theory'' which is depicted in Figure \ref{fig:pinwheel}. The prepotential on generic points of the Coulomb branch of this theory is given by
\begin{align}
  6\mathcal{F}_{PW} &= 8\phi_1^3+7\phi_2^3+7\phi_3^3+8\phi_4^3-3\phi_1^2\phi_2-3\phi_1\phi_2^2-3\phi_3^2\phi_4-3\phi_3\phi_4^2-3\phi_2^2\phi_3-3\phi_2\phi_3^2 \nonumber \\
  &\quad -6\phi_1^2\phi_3-6\phi_4^2\phi_2+6\phi_1\phi_2\phi_3 + 6\phi_2\phi_3\phi_4 \ .
\end{align}

When $\phi_1=\phi_4$ and $\phi_2=\phi_3$, this theory has a $\mathbb{Z}_2$ symmetry generated by $\pi$ rotation of the brane web. The $\mathbb{Z}_2$ quotient can be illustrated by the diagrams at the top of Figure \ref{fig:pinwheel}. So the $\mathbb{Z}_2$ quotient of the Pinwheel theory yields the 5d $SU(3)_2$ gauge theory with 4 fundamental hypers. The prepotential computation for the $\mathbb{Z}_2$ gauge theory given by
\begin{align}
  6\mathcal{F}_{PW/\mathbb{Z}_2} = \frac{1}{2}\times 6\mathcal{F}_{PW}|_{\phi_1=\phi_4,\phi_2=\phi_3} = 8\phi_1^3+4\phi_2^3-9\phi_1^2\phi_2+3\phi_1\phi_2^2\ ,
\end{align}
agrees with the prepotential of the $SU(3)_2+4{\bf F}$ theory. 

The theory has a $\mathbb{Z}_4$ symmetry permuting 4 compact faces in the brane web when all the K\"ahler parameters are the same, i.e. $\phi_1=\phi_2=\phi_3=\phi_4$. 
The $\mathbb{Z}_4$ quotient can be illustrated by the diagrams at the bottom of Figure \ref{fig:pinwheel}. In this case, as drawn in the 3rd diagram at the bottom, the folded 5-brane ends on a [1,0] 7-brane and we can remove this 5-brane by the Hanany-Witten move. Consequently, the theory after the $\mathbb{Z}_4$ quotient becomes a rank-1 5d SCFT with $E_6$ global symmetry which can be described by the $SU(2)$ gauge theory with 5 fundamentals at low energy. This can be checked by computing the prepotential of the $\mathbb{Z}_4$ gauged theory as
\begin{align}
  6\mathcal{F}_{PW/\mathbb{Z}_4} = \frac{1}{4}\times 6\mathcal{F}_{PW}|_{\phi_1=\phi_2=\phi_3=\phi_4} = 3\phi_1^3 \ .
\end{align}
The result is precisely the prepotential of the $SU(2)+5{\bf F}$ theory.

%----------------------------------
%------------------------
\section{6d Cases}\label{sec:5}
A large class of 6d SCFTs has 5-brane descriptions as 5d KK theories which are 6d theories on a circle with/without a twist.  Many of these 5-brane webs show some of $\mathbb{Z}_n$ symmetries. Our prescriptions discussed for 5d SCFTs work well for S-folding discrete symmetries in these 6d theories. We list some of them along with gauge group types.

\begin{figure}[t]
     \centering
     \includegraphics[scale=.23]{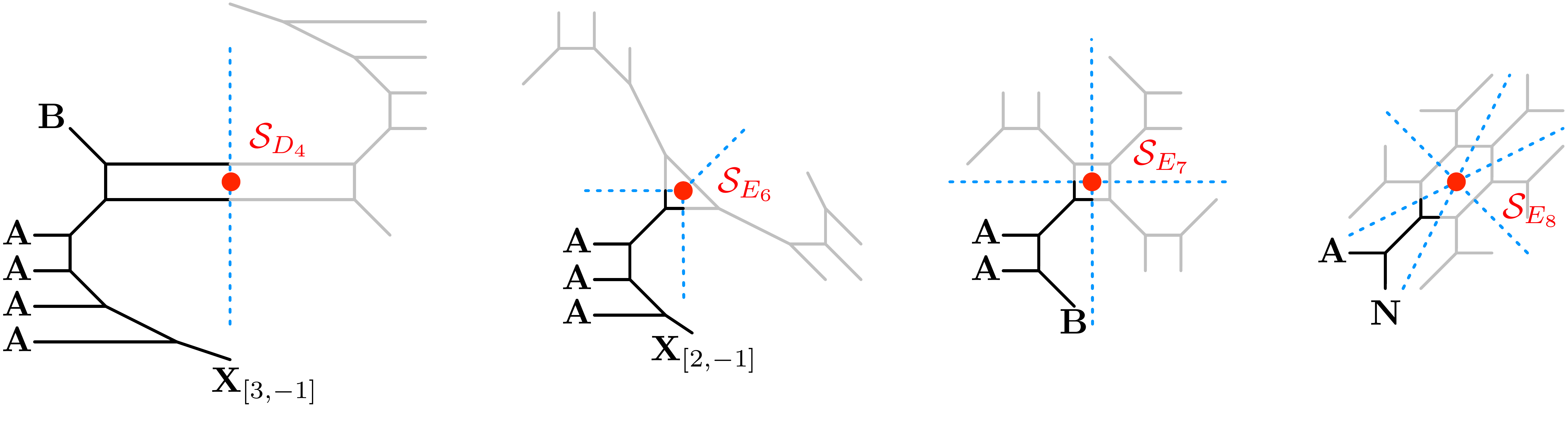}
     \caption{The $\mathbb{Z}_{2,3,4,6}$ quotients of 6d $E$-string on a circle (or 5d  $SU(2)+8\mathbf{F}$).}
     \label{fig:E8-gauging}
 \end{figure}

\paragraph{SU-type.}
As a simple example, consider the 6d $E$-string theory on a circle, which corresponds to the 5d $SU(2)+8\mathbf{F}$ theory \cite{Witten:1995gx,Ganor:1996mu,Seiberg:1996vs}.  The 5-brane web for $SU(2)+8\mathbf{F}$ can be deformed to reveal $\mathbb{Z}_{2,3,4,6}$ symmetries as shown in Figure \ref{fig:E8-Z2346}. The $E$-string theory is in particular interesting as it allows all the $\mathbb{Z}_n$ quotients. For each discrete symmetry, one can apply the $\mathbb{Z}_{n}$ S-folding on a face by inserting the corresponding 7-branes at the center of the Coulomb branch which is the fixed point of $\mathbb{Z}_n$ symmetric rotation, as depicted in Figure \ref{fig:E8-gauging}. %
The flavor symmetry of the $\mathbb{Z}_n$ gauge theory is all affine $E_8$ symmetry, which can be seen from rearrangements of 7-brane configurations for $SU(2)+8\mathbf{F}$ as \begin{align}
    {\bf
    (A^8BCBC)}
&\Longleftrightarrow 
     {\bf (BCA^4{[A^4BC]})}= ({\bf BA^4X_{[3,-1]}}[\mathcal{S}_{D_4}])
     &&\rightarrow\quad \mathbb{Z}_2
\nonumber\\
&\Longleftrightarrow 
     {\bf (CA^3{[A^5BCB]})}= ({\bf A^3X_{[2,-1]}}[\mathcal{S}_{E_6}])&&\rightarrow\quad \mathbb{Z}_3\nonumber\\
&\Longleftrightarrow 
    {\bf (CA^2{[A^6BCB]})}= ({\bf A^2B}[\mathcal{S}_{E_7}]) &&\rightarrow\quad\mathbb{Z}_4\nonumber\\
    &\Longleftrightarrow  {\bf (CA{[A^7BCB]}) }= ({\bf AN}[\mathcal{S}_{E_8}]) &&\rightarrow \quad \mathbb{Z}_6
          \ ,
\end{align}
where $\mathcal{S}_{D_4,E_6,E_7,E_8}$ are inserted at the center of the compact face. Therefore, the theory after the $\mathbb{Z}_{2,3,4,6}$ quotient is a rank-1 theory with an affine $E_8$ flavor symmetry, which is again the $E$-string theory on a circle:
\begin{align}
    5d~SU(2)+8 \mathbf{F} ~
    \xrightarrow[]{~\mathbb{Z}_{2,3,4,6} \text{~}} ~ 5d~SU(2)+8 \mathbf{F}
\ .
\end{align}

We note that, as listed in Table \ref{tab:symmetries for rank1}, one can reproduce various $\mathbb{Z}_n$ S-folds of other 5d rank-1 SCFTs by decoupling flavors in a way that $\mathbb{Z}_n$ symmetry is kept unbroken. This is a generic feature of other KK theories. 
In other words, RG-flows from a $\mathbb{Z}_n$ gauged KK theory will lead to 5d SCFTs arising from the $\mathbb{Z}_n$ S-folds of the 5d SCFTs which we obtain by $\mathbb{Z}_n$ preserving RG-flows from the original KK theory.

\begin{figure}[t]
     \centering
     \includegraphics[scale=.18]{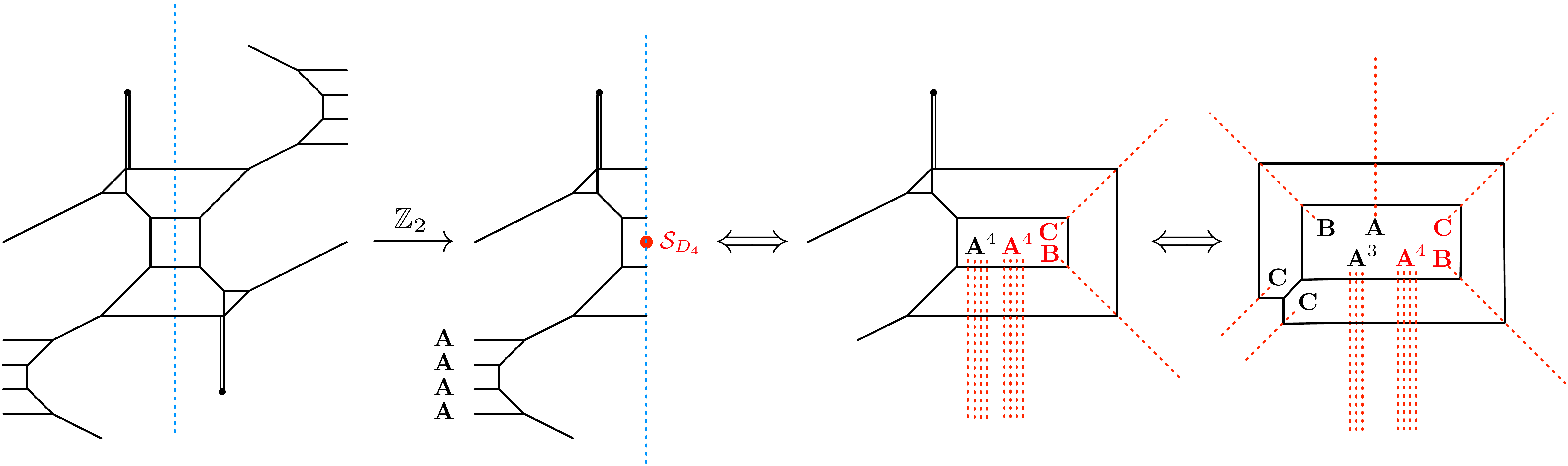}
     \caption{The $\mathbb{Z}_2$ quotient of 5d $SU(4)_0+2\mathbf{AS}+8\mathbf{F}$ giving rise to a 5-brane configuration for rank-2 $E$-string on a circle, 5d $Sp(2)+1\mathbf{AS}+8\mathbf{F}$. The inner 5-brane loop contains a 7-brane configuration $\bf (A^7BCAB)=(A^8BCB)$ showing an $E_8$ global symmetry.}
\label{fig:SU4+2AS+8F}
 \end{figure}

For higher ranks, there are many KK theories of $SU$-type gauge group with various matter contents which enjoy a $\mathbb{Z}_2$ discrete symmetry. 
Here, we list some interesting examples.
First, consider the 5d  $SU(4)$ theory with two antisymmetric hypermultiplets ($\mathbf{AS}$) and 8 fundamental hypermultiplets, $SU(4)_0+2 \mathbf{AS}+8 \mathbf{F}$ \cite{Bergman:2015dpa, Zafrir:2015rga}. The corresponding 5-brane web is given in Figure \ref{fig:SU4+2AS+8F}, which has a $\mathbb{Z}_2$ discrete symmetry.
The $\mathbb{Z}_2$ quotient of this theory gives a rank-2 theory naively with a $SO(16)\times SU(2)$ symmetry. By a little manipulation of the 7-brane configuration, one can show that the global symmetry is enhanced to $E_8\times SU(2)$ and its 7-brane configuration coincides with that of the rank-2 $E$-string on a circle which is dual to the 5d  $Sp(2)+1\mathbf{AS}+8\mathbf{F}$ theory, as explicitly shown in Figure \ref{fig:SU4+2AS+8F}: 
\begin{align}
    5d~SU(4)_0+2 \mathbf{AS}+8 \mathbf{F} ~
    \xrightarrow[]{~\mathbb{Z}_2 \text{~}} ~ 5d~Sp(2)+1\mathbf{AS}+8 \mathbf{F}
\ . 
\end{align}

\begin{figure}[t]
     \centering
     \includegraphics[scale=.218]{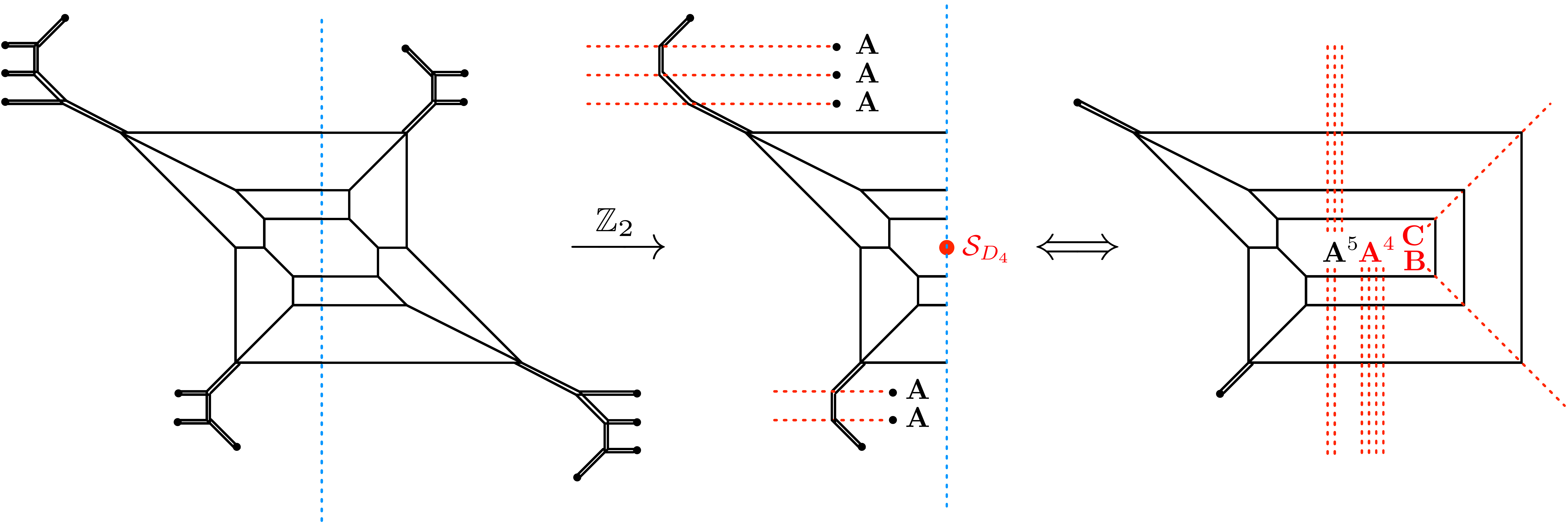}
     \caption{The $\mathbb{Z}_2$ quotient of 5d $SU(6)_0+1\mathbf{TAS}+10\mathbf{F}$ leads to a 5-brane configuration for 5d $Sp(3)+\frac12\mathbf{TAS}+\frac{19}{2}\mathbf{F}$.}
     \label{fig:SU6+TAS}
 \end{figure}

Another interesting KK theory is the 5d $SU(6)$ gauge theory with one hypermultiplet in the rank-3 antisymmetric representation ($\mathbf{TAS}$) and 10 fundamental hypermultiplets,  $SU(6)_0+1 \mathbf{TAS}+10 \mathbf{F}$ \cite{Jefferson:2017ahm}. The corresponding 5-brane web is discussed in \cite{Hayashi:2019yxj}. One can see from the first brane web in Figure \ref{fig:SU6+TAS} that it has a $\mathbb{Z}_2$ symmetry, and one readily finds that the $\mathbb{Z}_2$ quotient of this theory 
%$SU(6)_0+1 \mathbf{TAS}+10 \mathbf{F}$ 
yields a 5-brane configuration for a rank-3 KK theory as depicted in Figure \ref{fig:SU6+TAS}. This brane configuration is, in fact, a brane web for the 5d $Sp(3)$ gauge theory with one half-hypermultiplet in the rank-3 antisymmetric representation and $19$ half-hypermultiplets in the fundamental representation,  $Sp(3)+\frac12\mathbf{TAS}+\frac{19}{2} \mathbf{F}$, which is another KK theory  \cite{Jefferson:2017ahm} whose brane web is obtained in \cite{Hayashi:2019yxj},
\begin{align}
    5d~SU(6)_0+1 \mathbf{TAS}+10 \mathbf{F} ~
    \xrightarrow[]{~\mathbb{Z}_2 \text{~}} ~ 5d~Sp(3)+\frac12\mathbf{TAS}+\frac{19}{2} \mathbf{F}
\ .
\end{align}

\begin{figure}[t]
     \centering
     \includegraphics[scale=.17]{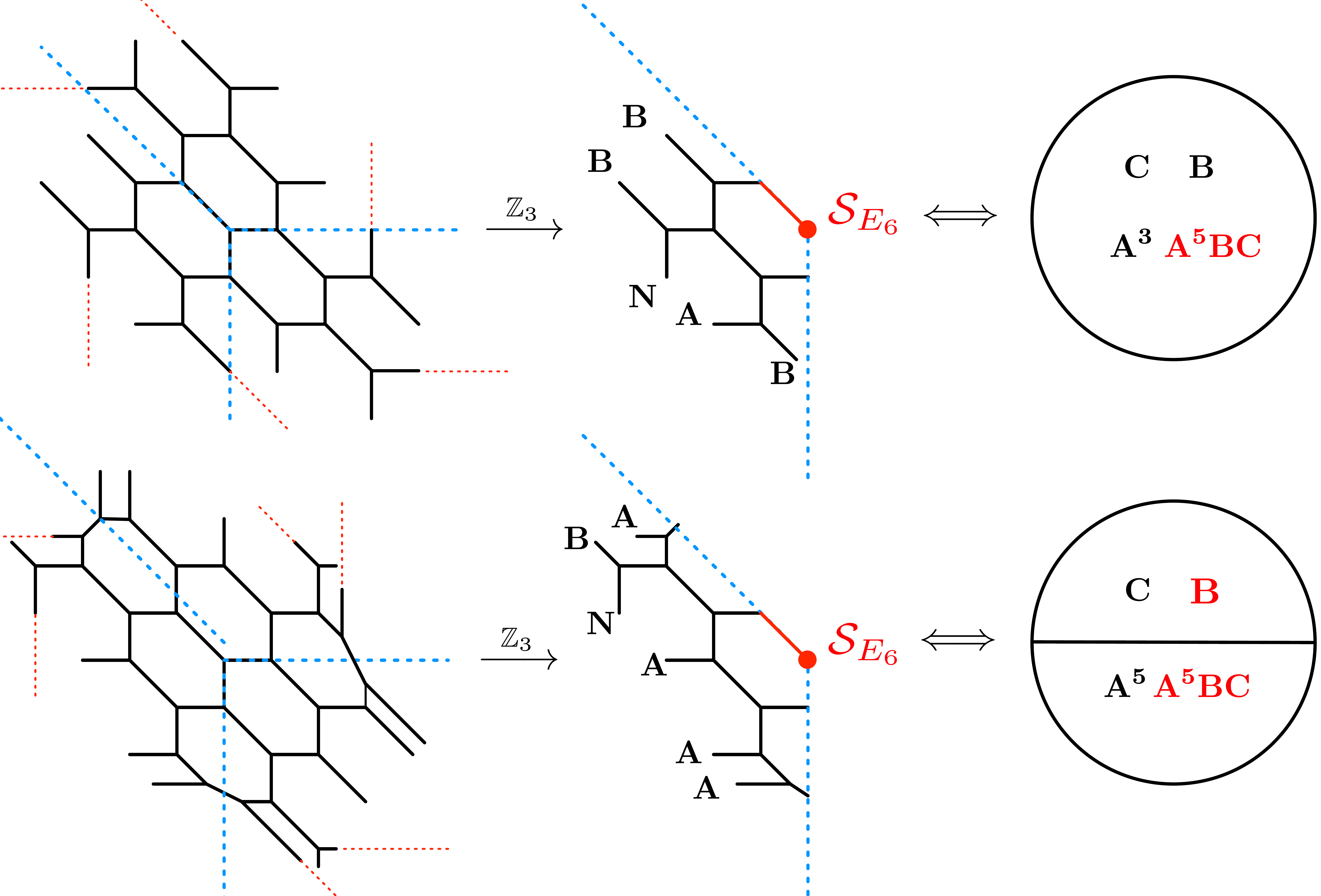}
     \caption{The $\mathbb{Z}_3$ quotient of $\widehat{T}_3$ and $\widehat{T}_4$ theories, giving rise to $SU(2)+8\mathbf{F}$ and $SU(3)_0+10\mathbf{F}$ respectively.}
     \label{fig:TNtao-Z3}
 \end{figure}

As discussed in \cite{Zafrir:2015rga, Hayashi:2015zka}, one can add more flavors to a $T_N$ theory to turn it to a KK theory described by a Tao web diagram in \cite{Kim:2015jba}. We denote such a KK theory by $\widehat{T}_N$. This theory has a quiver gauge theory description given by
\begin{align}
    \widehat{T}_N=[N+2]-SU(N-1)-SU(N-2)-\cdots-SU(3)-SU(2)-[3]\ .
\end{align}
Note that $\widehat{T}_3$ is the 5d $SU(2)+8\mathbf{F}$ theory or the circle compactification of the 6d $E$-string theory; $\widehat{T}_4$ is the 6d $SU(3)+11\mathbf{F}+1\mathbf{AS}$ theory on a circle; $\widehat{T}_5$ is the 6d $SU(6)+15\mathbf{F}+1/2\mathbf{TAS}$ theory on a circle \cite{Zafrir:2015rga,Hayashi:2015zka,Eckhard:2020jyr}.
The $\mathbb{Z}_{3}$ quotient of $\widehat{T}_N$ can be done in the same way as done for ${T}_N$. One finds that, as illustrated in Figure \ref{fig:TNtao-Z3}, 
\begin{align}
    \widehat{T}_3 ~&
    \xrightarrow[]{~\mathbb{Z}_3 \text{~}} ~ \widehat{T}_3\ ,\cr
    \widehat{T}_4 ~&
    \xrightarrow[]{~\mathbb{Z}_3 \text{~}} ~ \widehat{T}_3\ ,\cr
    \widehat{T}_5 ~&
    \xrightarrow[]{~\mathbb{Z}_3 \text{~}} ~ (D_5,D_5)\ ,
\end{align}
where $(D_5,D_5)$ refers to the 6d $(D_5,D_5)$ conformal matter in \cite{DelZotto:2014hpa} on a circle which is dual to the 5d $SU(3)_0+ 10\mathbf{F}$ theory \cite{Hayashi:2015fsa}. 

One can also add more flavors to the $+_{N,N}$, $X_{N,N}$, and $
\pluslash_N$ theories  of $\mathbb{Z}_4$ and $\mathbb{Z}_6$ symmetries discussed in the previous sections in order to promote them to KK theories. The resulting theories, which we call  $\widehat{+}_{N,N}$, $\widehat{X}_{N,N}$, and $
\widehat{\pluslash}_N$, can be described by Tao web diagrams. As done for $\widehat{T_N}$, implementing the $\mathbb{Z}_n$ S-folding is straightforward. We just present a simple example that leads to the rank-2 $E$-string, which is the $\mathbb{Z}_6$ S-fold of the $\widehat{\pluslash}_2$ theory:
\begin{align}
  \widehat{\pluslash}_2=  [4]\!-\!SU(3)\!-\!SU(4)\!-\!SU(3)\!-\![4]~
    \xrightarrow[]{~\mathbb{Z}_6 \text{~}} ~ 5d~Sp(2)+1\mathbf{AS}+8 \mathbf{F}.
\end{align}
%

%---------------------------------
\paragraph{Sp-type.}
In the previous section, we discussed that one can perform the $\mathbb{Z}_2$ quotient of the $SU(3)_0$ gauge theory with $2n \mathbf{F}$ to obtain 
\begin{align}
    SU(3)_0+2n \mathbf{F} \quad 
    \xrightarrow[]{~\mathbb{Z}_2 \text{~}} \quad SU(2)+(3+n)\mathbf{F},
\end{align}
where $n\le 5$.
The maximal case when $n=5$ corresponds to a KK theory $SU(3)_0+10\mathbf{F}$. This theory is dual to 5d $Sp(2)+10\mathbf{F}$ and also the 6d $(D_5, D_5)$ conformal matter theory on a circle \cite{Gaiotto:2015una,Hayashi:2015fsa}. This implies that the $\mathbb{Z}_2$ folding of the 6d $(D_5, D_5)$ conformal matter theory reduces to the 6d $E$-string theory or the $(D_4, D_4)$ conformal matter theory.
\begin{figure}[t]
     \centering
     \includegraphics[scale=.27]{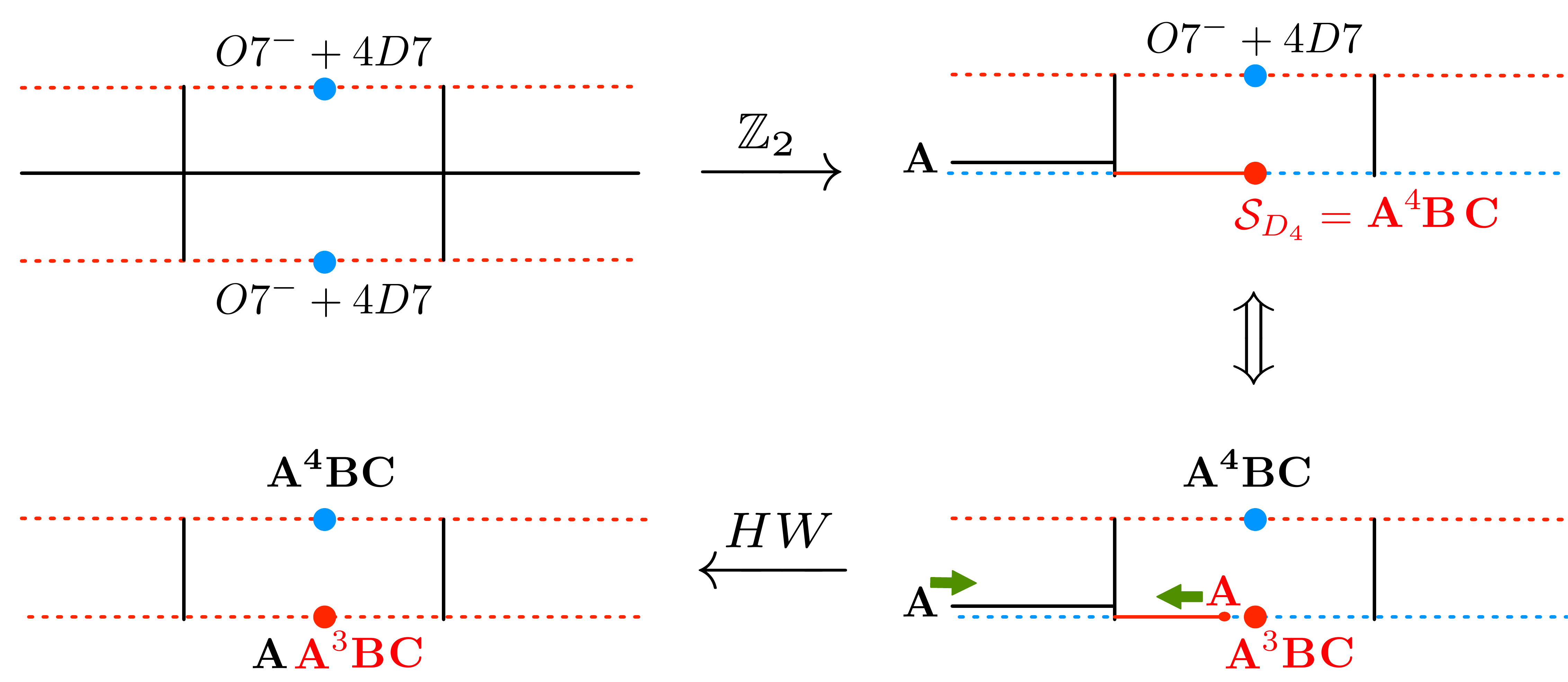}
     \caption{The $\mathbb{Z}_2$ quotient of 5d $Sp(2)+10\mathbf{F}$ theory leading to a brane for 5d $Sp(1)+8\mathbf{F}$, which is $E$-string on a circle.}
     \label{fig:Sp2+10F-Z2}
 \end{figure}

We note that such $D$-type conformal matters can be easily understood from 5-brane configurations with two O7$^-$-planes which show the $SU/Sp$ duality diagrammatically \cite{Hayashi:2015fsa}. From the perspective of 5-brane web with two O7$^-$-planes, the $SU(3)_0+10\mathbf{F}$ theory is realized as the $Sp(2)+10\mathbf{F}$ theory whose 5-brane configuration is given in Figure \ref{fig:Sp2+10F-Z2}. Applying the $\mathbb{Z}_2$ quotient, we obtain the second configuration in Figure \ref{fig:Sp2+10F-Z2}, where the $D_4$ type 7-branes are inserted at the $\mathbb{Z}_2$ fixed point such that a color D5-brane is attached. Expressed in terms of resolved 7-branes, one D7-brane $\bf A$ out of the $D_4$ 7-branes $\bf A^4BC$ is attached to the color D5-brane. By taking two Hanany-Witten moves, one with $\bf A$ attached to the color D5-brane and the other with $\bf A$ attached to the external flavor D5-brane in the opposite direction, one gets another 7-brane configuration of $D_4$ singularity of $\bf A^4BC$, given in the bottom-left diagram of Figure \ref{fig:Sp2+10F-Z2}. The resulting configuration is hence that of two $D_4$ singularities which is nothing but the brane configuration for the 6d $E$-string theory.

\begin{figure}[t]
     \centering
     \includegraphics[scale=.2]{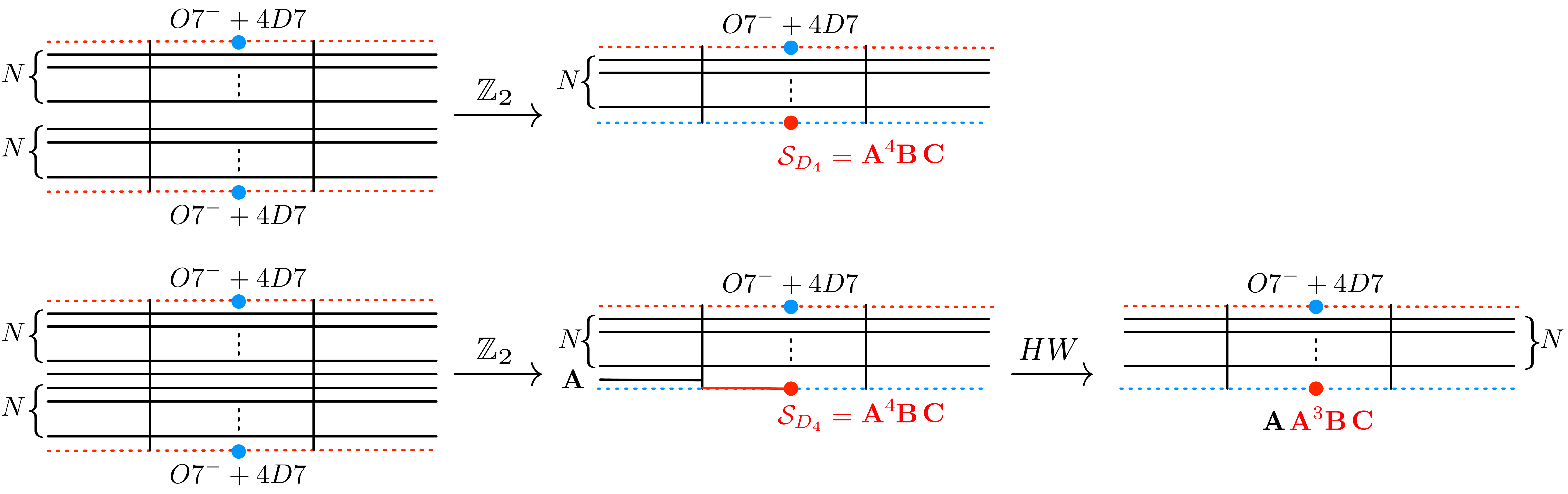}
     \caption{5-brane webs for 6d minimal $(2N+4, 2N+4)$ and $(2N+5, 2N+5)$ conformal matter, and their $\mathbb{Z}_2$ quotient leading to 6d $(N+4, N+4)$ conformal matter.}
     \label{fig:KKSp-Z2}
 \end{figure}

Based on a 5-brane web with two O7$^-$-planes, depicted in Figure \ref{fig:KKSp-Z2}, we see that the $\mathbb{Z}_2$ quotient of the 5d $SU(2N+3)_0+(4N+10) \mathbf{F}$  which is dual to the 6d $(2N+5, 2N+5)$ conformal matter theory on a circle ($N\ge 0$) leads to
\begin{align}
    5d~Sp(2N+2)+(4N+10) \mathbf{F} ~
    \xrightarrow[]{~\mathbb{Z}_2 \text{~}} ~ 5d~Sp(N+1)+(2N+8)\mathbf{F}\ ,
\end{align}
and equivalently,
\begin{align}
    5d~SU(2N+3)_0+(4N+10) \mathbf{F} ~
    \xrightarrow[]{~\mathbb{Z}_2 \text{~}} ~ 5d~SU(N+2)_0+(2N+8)\mathbf{F}\ .
    %: \quad \hat{D}_{2N+8}~{\rm symmetry},
\end{align}
In summary, the $\mathbb{Z}_2$ quotient of  the 6d $D$-type conformal matter theory realized by a 5-brane web with two O7$^-$-planes is given as follows:
\begin{align}
    6d~(2N+4, 2N+4) \text{~conf. mat.}~
    \xrightarrow[]{~\mathbb{Z}_2 \text{~}} ~ 6d~(N+4, N+4)\text{~conf. mat.}\ 
\end{align}
and 
\begin{align}
    6d~(2N+5, 2N+5) \text{~conf. mat.}~
    \xrightarrow[]{~\mathbb{Z}_2 \text{~}} ~ 6d~(N+4, N+4)\text{~conf. mat.}\ ,
\end{align}
which are illustrated in Figure \ref{fig:KKSp-Z2}.

\paragraph{SO-type.}
\begin{figure}[t]
     \centering
     \includegraphics[scale=.25]{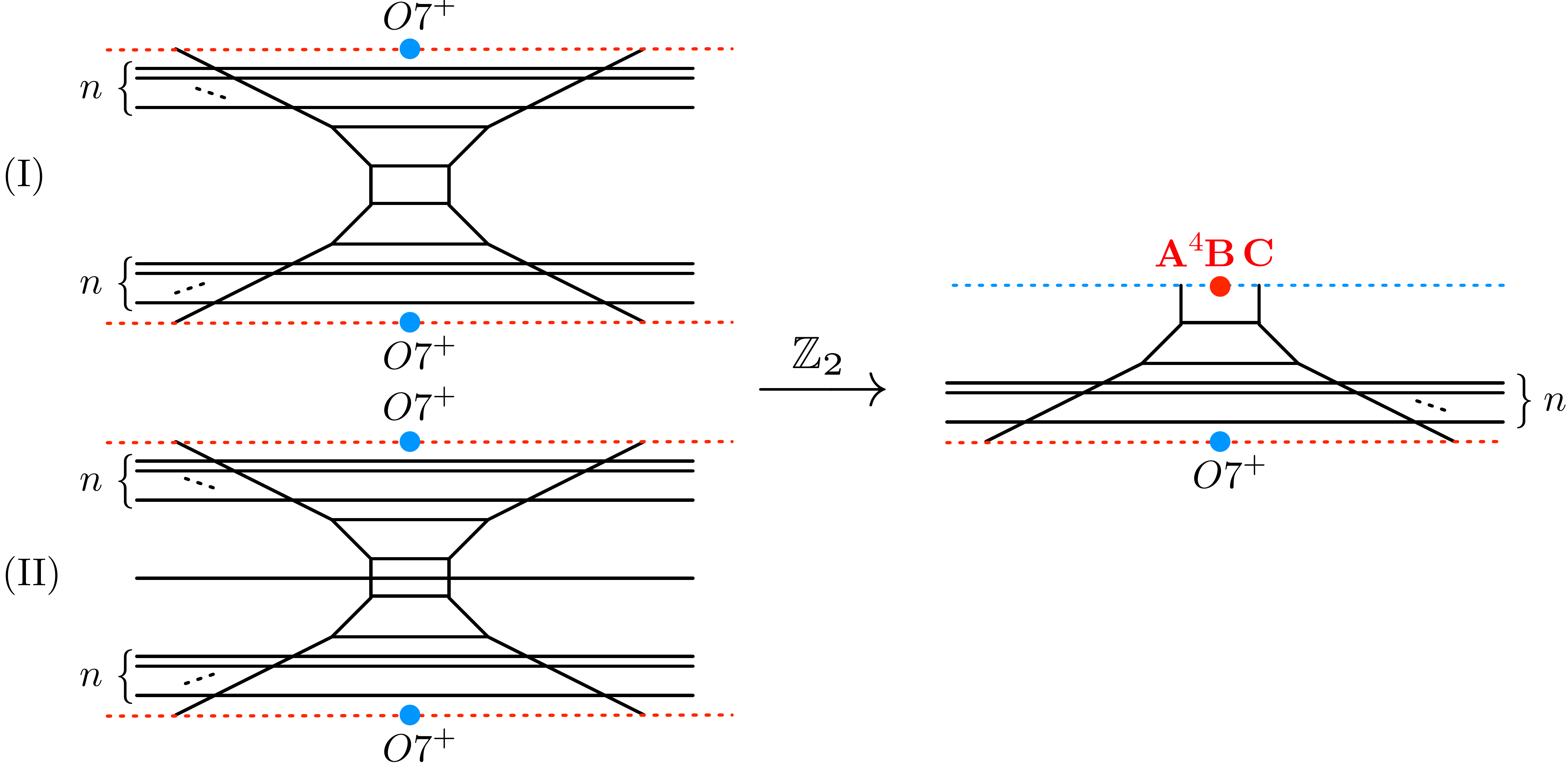}
     \caption{The $\mathbb{Z}_2$ quotient of 6d $SO(2N+8)+2N\mathbf{F}$ theory leading to 5d  $SO(2[N/2]+6)+(2[N/2]+4)\mathbf{F}$. Here, (I) is for $N=2n$ and  (II) is for $N=2n+1$.}
     \label{fig:6dSO-Z2}
 \end{figure}
The 6d $SO(2N+8)$ theory on a $-4$ curve shows a $\mathbb{Z}_2$ symmetry. When put on a circle, this theory is described by a 5-brane configuration with two O7$^+$-planes, as shown in Figure \ref{fig:6dSO-Z2}. The $\mathbb{Z}_2$ quotient of this theory yields a 5-brane web with both O7$^+$- and O7$^-$-planes, which corresponds to the 5d $SO(2[N/2]+6)+(2[N/2]+4)\mathbf{F}$ theory \cite{Jefferson:2017ahm, Hayashi:2015vhy}:
\begin{align}
    6d~SO(2N+8)+2N\mathbf{F} ~
    \xrightarrow[]{~\mathbb{Z}_2 \text{~}} ~ 5d~SO(2[N/2]+6)+(2[N/2]+4)\mathbf{F}\ .
\end{align}
For the $\mathbb{Z}_2$ quotient, one can consider two cases when $N$ is even ($N=2n$) and odd ($N=2n+1$) separately, as in Figure \ref{fig:6dSO-Z2}. The even case corresponds to the S-fold on a face, while the odd case corresponds to the S-fold on a line. Taking into account the Hanany-Witten transition effect for the S-fold on a line, one finds both even and odd cases after the $\mathbb{Z}_2$ quotient leads to the 5d $SO(2[N/2]+6)+(2[N/2]+4)\mathbf{F}$ theory. This is a  KK theory corresponding to the 6d $SU(2[N/2]+4)+(4[N/2]+8){\bf F}$ theory on a circle with $\mathbb{Z}_2$ twist \cite{Jefferson:2017ahm}.
Note that the resulting 5-brane configuration has two distinct orientifold planes which comes from a 6d SCFT on a circle with a twist \cite{Keurentjes:2000bs, Zafrir:2015rga,Hayashi:2015zka,Hayashi:2015vhy}.

%----------------------------------
\section{Conclusion}\label{sec:6}

In summary, we studied the S-foldings of discrete $\mathbb{Z}_n$ symmetries in 5d SCFTs which are generated by combinations of discrete rotations of associated 5-brane webs and ${\rm SL}(2,\mathbb{Z})$ action of Type IIB string theory. The brane configuration in the fundamental domain of the 5-brane web after the $\mathbb{Z}_n$ quotient puts strong constraints on the monodromy at the $\mathbb{Z}_n$ singularity. Based on the properties of the monodromy, we proposed additional 7-branes of type $D_4,E_6,E_7,E_8$ for $n=2,3,4,6$ respectively to be inserted into the singularity. In addition, when the fixed point of $\mathbb{Z}_n$ action is located at a vertex of intersecting $n$ 5-branes, the $n$ 5-branes are folded around the fixed point after the quotient and become a single 5-brane with one end attached to a 7-brane of the same charge at the singularity. The resulting 5-brane web together with the 7-branes inserted at the singularity allows us to read off the 5d theory arising from the $\mathbb{Z}_n$ quotient. We also suggested a systematic algorithm to calculate cubic prepotentials of the $\mathbb{Z}_n$ gauged theories and tested the results against the brane webs we obtained using our S-folding prescriptions.

In this article, we focused on a particular class of discrete symmetries related to rotations of 5-brane webs for 5d SCFTs. There are, however, other types of discrete symmetries in 5d field theories which are not associated with rotations of 5-brane webs. One such example is that of geometric symmetries which exchange identical compact surfaces in Calabi-Yau 3-folds for 5d theories. For example, the 5d $SO(8)$ gauge theory without matter is realized by gluing three $\mathbb{F}_2$ surfaces to a $\mathbb{F}_0$ surface \cite{DelZotto:2017pti,Hayashi:2017jze} and this geometry has a $\mathbb{Z}_3$ symmetry permuting the $\mathbb{F}_2$ surfaces. Based on geometric constructions, one may carry out concrete studies on the quotient of this type of discrete symmetries in 5d field theories. This would be a quotient of geometric automorphism of Calabi-Yau 3-folds from M-theory perspective which identifies geometric configurations related by the symmetry.

Another interesting extension of our results would be the study of discrete symmetries in 6d SCFTs and their quotients. We have a handful of   examples for $\mathbb{Z}_2$ and $\mathbb{Z}_3$ S-folds of 6d SCFTs which are visible in 5-brane webs possibly with orientifold planes. However, generic 6d SCFTs have no 5-brane web realization and therefore the study of their discrete symmetries and the corresponding quotients would require other techniques. Geometric constructions as well as prepotential considerations for some 6d SCFTs when compactified on a circle could provide controllable tools along that direction. For instance, the 6d minimal $SU(3)$ gauge theory has a geometric realization by three $\mathbb{F}_1$ surfaces embedded in a Calabi-Yau 3-fold, and the geometry and the associated prepotential exhibit a $\mathbb{Z}_3$ symmetry interchanging the three $\mathbb{F}_1$ surfaces. We observe that the $\mathbb{Z}_3$ S-folding gives rise to a 6d rank-1 SCFT with a trivial cubic prepotential, i.e. $\mathcal{F}=0$, when compactified on a circle. This provides a non-trivial clue to identify the $\mathbb{Z}_3$ gauged theory.

% The higher form symmetries~\cite{Gaiotto:2014kfa,Morrison:2020ool,Albertini:2020mdx,Bhardwaj:2020phs,BenettiGenolini:2020doj,Gukov:2020btk} appear in our study of the discrete gauging of the 5d SCFTs. The discrete global symmetries we have discussed are zero-form symmetries which act on the particle spectrum. We have studied the characteristics of gaugings or S-foldings of these 0-form symmetries. The new theory obtained after the S-folding would have naturally the dual discrete  global symmetry which is a 3-form symmetry and 
%whose charged states in the spectrum are co-dimension 1 defects or domain walls. 
%whose charged operators are co-dimension 2 operators or 3-plane operators.  
%The gauging of this dual 3-form symmetry or the un-gauging of the 0-form gauged theory  would lead back to the original theory with the discrete 0-form global symmetry. 
%A systematic study of the gauging of 3-form symmetries would be an interesting future task to pursue and this could provide another concrete check for our gauging prescription which relates one theory to another theory with 0-form and 3-form symmetries exchanged.

\acknowledgments
We would like to thank Hirotaka Hayashi,  Sungjay Lee, Kaiwen Sun, Xin Wang, and Futoshi Yagi for useful discussions. Hospitality at APCTP during the APCTP / KIAS / POSTECH program ``Frontiers in Theoretical Physics'' is kindly acknowledged. The research of HK is supported by Samsung Science and Technology Foundation under Project Number SSTF-BA2002-05 and by the National Research Foundation of Korea (NRF) grant funded by the Korea government (MSIT) (No. 2018R1D1A1B07042934). The research of SK is partially supported by the Fundamental Research Funds for the Central Universities 2682021ZTPY043. KL is supported in part by KIAS Individual Grant PG006904 and by the National Research Foundation of Korea (NRF) Grant funded by the Korea government (MSIT)  (No.  2017R1D1A1B06034369). 

%----------------------------------
%--------------------------------
%

\bibliographystyle{JHEP}
\bibliography{ref}
\end{document}